\documentclass[12pt, letterpaper, superscriptaddress, longbibliography]{revtex4-2}
\usepackage[none]{hyphenat} 

\usepackage{tocloft} 
\usepackage{fancyhdr}
\fancypagestyle{tocstyle}{
  \fancyhf{} 
  \fancyfoot[R]{\thepage~of~\pageref*{LastPage}}
 \fancyhead{} 
}

\usepackage{graphicx}
\usepackage{wrapfig}

\usepackage{caption}
\usepackage{subcaption}
\makeatletter
\renewcommand{\figurename}{\sectionfont Fig.}

\renewcommand{\thefigure}{\sectionfont \arabic{figure}}  
\makeatother

\usepackage{marvosym}

\usepackage[absolute,overlay]{textpos}

\usepackage{mathtools}
\usepackage{amsmath}

\usepackage{color}
\usepackage{xcolor} 
\definecolor{InstituteBlue}{HTML}{2e99a2}
\definecolor{InstituteYellow}{HTML}{ffd26d}
\usepackage{hyperref}
\hypersetup{
    colorlinks=true,
    citecolor  = InstituteBlue,
    urlcolor   = InstituteBlue,
    linkcolor  = InstituteBlue
}

\usepackage{csquotes}

\usepackage{media9}

\renewcommand{\thesection}{\arabic{section}}
\renewcommand{\thesubsection}{\thesection.\arabic{subsection}}

\makeatletter
\renewcommand{\p@subsection}{}
\renewcommand{\p@subsubsection}{}
\makeatother

\usepackage{fontspec}
\newfontfamily\sectionfont{ChapMedium.ttf}[ 
    Path = Fonts/ ,
    Extension = .ttf
]

\newfontfamily\sectionfontit{ChapMediumItalic.ttf}[ 
    Path = Fonts/ ,
    Extension = .ttf
]

\makeatletter
\def\@hangfrom@section#1#2#3{\@hangfrom{#1#2}{\large\sectionfont#3}}%
\def\@hangfroms@section#1#2{\@hangfroms{#1}{\large\sectionfont#2}}%
\makeatother

\usepackage{verbatim}

\usepackage{geometry}
\usepackage{enumitem}

\usepackage{lastpage} 
\usepackage{fancyhdr}
\usepackage{tabularx}  
\usepackage{ragged2e}  

\usepackage{media9}
\usepackage{animate}

\renewcommand{\thesection}{\sectionfont \arabic{section}}
\renewcommand{\thesubsection}{\sectionfont \arabic{section}.\arabic{subsection}}

\usepackage{amsmath}
\usepackage{amssymb}
\usepackage{physics}
\usepackage{tikz}
\usepackage{geometry}
\usepackage{booktabs}
\usepackage{xcolor}
\usepackage{tcolorbox}
\usetikzlibrary{arrows.meta, positioning, decorations.pathreplacing}

\usepackage{bbold}

\usepackage{float}

\usepackage{rotating}

\newcommand{\seclq}{{\rmfamily‘}}
\newcommand{\secrq}{{\rmfamily’}}

\definecolor{dred}{RGB}{139,0,0}

\usepackage{setspace}

\usepackage{media9}

\begin{document}

\title{\sectionfont\Large{A physicist-friendly primer \linebreak on the Hamiltonian for quantum sensing in proteins: \linebreak analytical expressions and insights \linebreak for a toy model of the radical-pair mechanism}}
\vspace*{-0.65cm}
\makeatletter
\renewcommand{\frontmatter@affiliationfont}{\small\itshape\selectfont}  
\makeatother

\makeatletter
\renewcommand{\frontmatter@affiliationfont}{\small\itshape\selectfont}
\renewcommand{\andname}{\ignorespaces}
\makeatother

\author{Clarice D.\ Aiello}
\email{clarice@quantumbiology.org}

\author{Brian L.\ Ross}
\author{Alessandro Lodesani}
\author{Morgan L.\ Sosa}

\affiliation{Quantum Biology Institute, Los Angeles, USA}

\date{\today}
\vspace*{-1cm}
\begin{abstract}
\begin{spacing}{0.95}
\noindent Electron spin-dependent chemical reactions in proteins, often discussed under the `radical-pair mechanism', have been studied for decades and remain the leading microscopic proposal for magnetic field sensing in biology. Yet the essential physics is often obscured by the complexity of realistic models. In this work, we present a physicist-friendly primer on the simplest radical-pair Hamiltonian that already captures many of the mechanism's best-known qualitative features. The contributions of this work are fourfold. \textbf{First}, we place on record a complete analytical solution of this toy model, which has previously been studied extensively, mostly through numerical and partial analytical approaches. Working in the experimentally relevant singlet--triplet basis, we derive closed-form expressions for the instantaneous singlet population and for two related time-averaged singlet yields. \textbf{Second}, we introduce a new interpretation of these results that makes several familiar features of radical-pair physics transparent. In particular, we show that the dynamics admit a bright--dark decomposition (in the sense of spin mixing), similar to structures widely studied in atomic and optical physics, for example in electromagnetically-induced transparency.
\textbf{Third}, through this bright--dark perspective, we clarify experimentally relevant features of the toy model. In particular, we show that the so-called `low-field effect' arises from a coherence term between bright and dark sectors, and that the special role of zero field is best understood as a phase-locking phenomenon rather than merely as enhanced mixing. The same framework also makes it possible to explicitly identify the `pathway that opens', as per the chemists' language, once a nonzero field is applied. \textbf{Fourth}, we import methods from quantum magnetometry, developed in the context of technological quantum sensing, to obtain further insight into the model. This allows us to clarify the role of initial state preparation and the trade-off between coherent phase accumulation and time-averaging penalties. The resulting toy model serves both as an analytically tractable benchmark and as a conceptual starting point for future work incorporating a true open quantum system treatment, unequal singlet and triplet decay rates, and fully directional magnetic field control.
\\

\noindent \scriptsize{A first version of the manuscript was deposited on \href{https://beta.dpid.org/1088}{https://beta.dpid.org/1088}
 on 2026/04/08 and was assigned a persistent digital identifier at that time.}
\end{spacing}
\end{abstract}

\maketitle
\vfill
\newpage
\clearpage
\section{Introduction}

\noindent How biological systems might sense weak magnetic fields remains one of the most intriguing open questions at the interface of quantum physics, chemistry, and biology. Among the proposed mechanisms, the radical-pair mechanism occupies a central position because it naturally links coherent spin dynamics to chemically amplified observables. In this picture, photoinduced charge separation, usually within a protein, creates a pair of electron spins whose singlet--triplet interconversion is modulated by magnetic interactions; spin-selective reaction channels then transduce that spin dynamics into a measurable chemical or \linebreak photophysical signal. Despite the apparent simplicity of this idea, the radical-pair literature often mixes together several distinct ingredients --- coherent Hamiltonian evolution, spin-\linebreak selective recombination, orientational anisotropy, and effective observables such as singlet yields --- in ways that can make the underlying physics harder to isolate than necessary.
\\

\noindent The goal of the present work is to develop a physicist-friendly primer on the simplest radical-pair Hamiltonian that already reproduces many of the best-known qualitative features of the mechanism. We consider two electron spins, one effectively free and one hyperfine-\linebreak coupled to a single nuclear spin, in a constant external magnetic field. We restrict our attention to an isotropic hyperfine interaction and neglect decoherence, recombination, and most of the nuclear dynamics, so that the problem can be solved analytically in closed form. This minimal setting is not intended to be chemically complete, nor can it account for some interesting features of the radical-pair dynamics such as directional magnetosensitivity. Its value lies instead in providing a fully solvable benchmark in which the coherent spin physics can be exposed as transparently as possible. To help achieve that, all of the main calculations can be followed along, step by step, using the accompanying SymPy code, which allows the reader to explore the exact expressions and their limiting cases interactively.
\\

\noindent Within this toy model, several familiar radical-pair characteristics can already be analyzed neatly. The singlet population can be obtained exactly as a function of time and field strength for a wide class of initial electron spin states, including generic incoherent mixtures in the singlet--triplet basis. In addition to the instantaneous singlet signal, we also obtain closed-form expressions for two commonly used observables obtained by post-processing the coherent dynamics: a finite-time average of the singlet population and a lifetime-\linebreak weighted singlet yield. These signals are not equivalent to incorporating Haberkorn-\linebreak type recombination directly into the dynamics, but they provide analytically tractable proxies that already display rich magnetic field dependence. 
\\

\noindent Significantly, we recognize that the Hamiltonian admits a bright--dark decomposition, \linebreak reminiscent of electromagnetically-induced transparency, in which one sector (`bright') actively participates in oscillatory spin mixing while another is symmetry-protected from it (`dark'). This decomposition clarifies that all the experimentally relevant signals contain three \linebreak qualitatively distinct pieces: a dark-sector contribution, a bright-sector contribution, and a bright--dark coherence term. This bright--dark framework also provides a clean physical interpretation for the so-called `low-field effect', which in the present model is traced to the survival of a coherence contribution that becomes phase-locked at exactly zero field. Relatedly, the same framework makes it possible to identify precisely which population-\linebreak transfer pathways are modified as the external field is increased from \(0\) to \(0^{+}\), thereby giving a concrete physical meaning to the literature's descriptions of `enhanced spin mixing at low field' or of a `pathway opening' once a nonzero field is applied.
\\

\noindent A further aim of this work is to connect the same solvable Hamiltonian to simple questions in quantum sensing. The magnetometric sensitivities of the two considered singlet yields can be obtained in closed form.
One of the central simplifications that emerges is that the sensitivities for a generic mixed initial state depend not on the four initial populations independently, but only on two particular combinations of them. The bright--dark framework clarifies that an optimal magnetometric state is one that balances dark-state phase memory against bright-state readout.
\\

\noindent The paper is organized as follows. We begin by defining the minimal Hamiltonian (Section~\ref{sec:H}) and solving it analytically in the experimentally relevant singlet--triplet basis (Section~\ref{sec:solution}). We then analyze the resulting coherent singlet dynamics, develop the bright--dark interpretation (Section~\ref{sec:analogy}), and derive closed expressions for the finite-time average and lifetime-weighted singlet yields (Section~\ref{sec:MFE}). We study finite-bias sensitivities for the singlet yield signals, and the dependence of magnetometric performance on the initial spin state (Section~\ref{sec:mag}). Short take-home messages are presented at the end of each main section. An extended discussion of each main section is provided in its corresponding supplementary section. We close by summarizing the main physical lessons of the toy model and outlining the natural next steps (Sections~\ref{sec:dis} and ~\ref{sec:con}): incorporating full open quantum system dynamics with distinct singlet and triplet recombination rates, and extending the field control beyond a single scalar Larmor frequency to explicit longitudinal and transverse components, thereby enabling the study of directional control and more realistic sensing protocols. Finally, we point readers to the notation table in Supplementary Section~\ref{not}, which summarizes the most important symbolic parameters used throughout the text for convenience.
\\

\clearpage
\section{Hamiltonian definition}
\label{sec:H}

\noindent The simplest Hamiltonian that reproduces the most well-studied and recognized spin and photophysics features of the radical pair mechanism~\cite{Hore2016, ZadehHaghighi2022MagneticField, Hore2025} describes a system in which:
\begin{enumerate}[leftmargin=0pt, label=\arabic*), itemsep=0pt]
\item two spin-$\frac{1}{2}$ electrons, which need not have any direct interaction (e.g., exchange or dipolar) between them, are left to evolve in time;
\item the first electron spin is assumed to be essentially `free', being only subject to the action of a constant external magnetic field of strength $B$ via the Zeeman interaction, with a corresponding Larmor frequency $\omega$;
\item the second electron spin is assumed to be isotropically hyperfine-coupled (with strength $a > 0$) to one spin-$\frac{1}{2}$ nucleus, and is therefore influenced both by the external magnetic field and by the magnetic field produced by the nearby nucleus; 
\item as will be shown explicitly in Supplementary Subsection~\ref{c5}, the electron spin states exhibit a degree of spin initialization; the term `spin initialization' is purposely used here by analogy with one of the DiVincenzo criteria for quantum computation, namely qubit initialization~\cite{DiVincenzo2000}. We restrict attention to initial electron spin states that are generic incoherent mixtures, excluding coherent superpositions;
\item there is no loss or decoherence.
\vspace*{0.25cm}
\end{enumerate}

\noindent Assuming the magnetic field $\hat{B}$ is along the $z$ direction, this minimal 3-spin Hamiltonian is thus of the following form:
\begin{equation}
    \mathcal{H} =
 \omega (\hat{S}_{z,1} + \hat{S}_{z,2}) + \hat{S}_2\cdot \hat{A} \cdot \hat{I} \ ,
\label{eq:H}
\end{equation}
\noindent where: the Larmor frequency of the electrons is given by $\omega \equiv \frac{\mu\,|g|\,B}{\hbar}$, with $\mu$ the Bohr magneton and $g$ the electron g-factor, the latter assumed to be the same for both electrons; the electron spin operators $\hat{S}_{i}$ along $z$ are given by $\hat{S}_{z,i}$, for $i = \{1, 2\}$; the nuclear spin operator is given by $\hat{I}$; and the hyperfine tensor is given by $\hat{A} = a \cdot \mathbb{1}_{3}$, where $\mathbb{1}_{3}$ is the $3 \times 3$ identity matrix. For simplicity, we neglect the small contribution of nuclear spin dynamics (for example, nuclear Zeeman interactions). Moreover, at room temperature, the nuclear spin population is assumed to be well-described by an incoherent, approximately equal mixture of spin states.
\\

\noindent All the symbolic calculations for the Hamiltonian can be followed along using the \linebreak accompanying SymPy code \href{https://bit.ly/Hamiltonian_SymPy}{\texttt{Hamiltonian\_SymPy.py}}. The step-by-step derivation of this Hamiltonian is found in Supplementary Section~\ref{sisec:hamiltonian}. Without loss of generality, we set $\  \hbar = 1$, so that energies are expressed in angular frequency units. 
\\

\noindent The radical pair mechanism can be probed through an experimentally relevant quantity (e.g., fluorescence or chemical product yield) that is effectively a proxy for the decayed population into the electron spin singlet state (equivalently, into the electron spin triplet states manifold), regardless of the final nuclear spin state. We are not aware of conventional radical-pair reactions in which the products associated with the individual triplet sublevels can be readily distinguished as different stable chemical species. For nuclear spins, however, this statement requires qualification: in chemically-induced dynamic
nuclear polarization~\cite{Nagakura1998DynamicSpinChemistry}, different nuclear spin states can influence reaction pathways and produce detectable nonequilibrium nuclear spin populations in the products, even though the products are not usually distinct chemical species solely by virtue of their nuclear spin state.
\\

\noindent  Hence, the experimentally relevant (and accessible) basis for the Hamiltonian in question is composed of the following eight states:
\begin{eqnarray}
\{|S\rangle, |T_{0}\rangle, |T_{+}\rangle, |T_{-}\rangle\} \otimes \{ \ket{\uparrow}, \ket{\downarrow} \} \ ,
\end{eqnarray}
\noindent where: the electron spins can be in a singlet $|S\rangle$ or in any one of the triplet $\{|T_{0}\rangle, |T_{+}\rangle, |T_{-}\rangle\}$ states; and the nuclear spin can be `up' or `down' given an arbitrary quantization direction (here, along $z$ in the lab frame). 
\\

\begin{figure}[H]
  \centering

  \pgfmathsetmacro{\Scale}{1.3}
  \pgfmathsetlengthmacro{\LevelLW}{\Scale*1.5pt}

  \makebox[\textwidth][c]{%
  \begin{tikzpicture}[
      scale=\Scale,
      y=1.2cm,
      every node/.style={transform shape},
      level/.style={draw, line width=1.5pt, color=black},
      energy/.style={font=\large, color=black},
      state/.style={font=\large, color=black},
  ]

  \def\levelwidth{4}
  \def\labeloffset{3.2}



  \coordinate (E2) at (0, 4.5);
  \draw[level] (-\levelwidth/2, 4.5) -- (\levelwidth/2, 4.5);
  \node[state, left] at (-\levelwidth/2-0.3, 4.5) {$\ket{T_+} \otimes \ket{\uparrow}$};
  \node[energy, right] at (\levelwidth/2+0.1, 4.5) {$+\ a/4 + \omega$};

  \coordinate (E7) at (0, 3.0); 
  \draw[level] (-\levelwidth/2, 3.0) -- (\levelwidth/2, 3.0);
  \node[state, left] at (-\levelwidth/2-0.3, 3.0) {$\ket{T_-} \otimes \ket{\downarrow}$};
  \node[energy, right] at (\levelwidth/2+0.1, 3.0) {$+\ a/4 - \omega $};


\coordinate (E4) at (0, 0.8);
\draw[level] (-\levelwidth/2, 0.8) -- (\levelwidth/2, 0.8);
\node[state, left]  at (-\levelwidth/2-0.3, 0.8) {$\ket{T_0} \otimes \ket{\uparrow}$};
\node[energy, right] at (\levelwidth/2+0.1, 0.8) {$0$};

\coordinate (E1) at (0, 0.3);
\draw[level] (-\levelwidth/2, 0.3) -- (\levelwidth/2, 0.3);
\node[state, left]  at (-\levelwidth/2-0.3, 0.3) {$\ket{S} \otimes \ket{\uparrow}$};
\node[energy, right] at (\levelwidth/2+0.1, 0.3) {$0$};

\coordinate (E5) at (0, -0.3);
\draw[level] (-\levelwidth/2, -0.3) -- (\levelwidth/2, -0.3);
\node[state, left]  at (-\levelwidth/2-0.3, -0.3) {$\ket{S} \otimes \ket{\downarrow}$};
\node[energy, right] at (\levelwidth/2+0.1, -0.3) {$0$};

\coordinate (E8) at (0, -0.8);
\draw[level] (-\levelwidth/2, -0.8) -- (\levelwidth/2, -0.8);
\node[state, left]  at (-\levelwidth/2-0.3, -0.8) {$\ket{T_0} \otimes \ket{\downarrow}$};
\node[energy, right] at (\levelwidth/2+0.1, -0.8) {$0$};

  \coordinate (E6) at (0, -3.0);
  \draw[level] (-\levelwidth/2, -3.0) -- (\levelwidth/2, -3.0);
  \node[state, left] at (-\levelwidth/2-0.3, -3.0) {$\ket{T_+} \otimes \ket{\downarrow}$};
  \node[energy, right] at (\levelwidth/2+0.1, -3.0) {$-\ a/4 + \omega $};

  \coordinate (E3) at (0, -4.5);
  \draw[level] (-\levelwidth/2, -4.5) -- (\levelwidth/2, -4.5);
  \node[state, left] at (-\levelwidth/2-0.3, -4.5) {$\ket{T_-} \otimes \ket{\uparrow}$};
  \node[energy, right] at (\levelwidth/2+0.1, -4.5) {$ -\ a/4 - \omega$};

  \draw[{Stealth[length=4mm]}-{Stealth[length=4mm]}, thick, black, densely dashed]
    (-\levelwidth/2-1.8-0.75-1, 4.5) -- (-\levelwidth/2-1.8-0.75-1, 3.0);
  \node[font=\large, black, fill=white, inner sep=2pt]
      at (-\levelwidth/2-1.8-0.75-1, 3.75) {$2\omega$};

  \draw[-{Stealth[length=4mm]}, thick, black]
      (\levelwidth/2 + 6 - 1+0.9 + 0.4, -5.5) -- (\levelwidth/2 + 6 -1+0.9 +0.4, 5.5);
  \node[font=\large, rotate=90, black]
      at (\levelwidth/2 + 6.8 -1+0.9, 0) {Energy, low external field regime ($\omega \ll a/4$)};

  \end{tikzpicture}%
  }

  \captionsetup[figure]{aboveskip=4pt,belowskip=8pt}
  \captionof{figure}{\justifying \textbf{Energy level diagram (diagonal terms of the Hamiltonian $\mathcal{H}$ of Eq.~\eqref{eq:H}) in the low external magnetic field regime ($\omega \ll a/4$; typically, e.g., geomagnetic conditions).}
  Quantum states are labeled in the basis $\ket{\text{electron spins}} \otimes \ket{\text{nuclear spin}}$.  The exact Zeeman splitting between the $\ket{T_{-}}$ and $\ket{T_{+}}$ manifolds is $2 \omega$; this term does not dominate the energy structure and is a perturbation on top of the hyperfine coupling $a$. 
  The hyperfine coupling $a$ dominates the energy structure, creating a clear separation between nuclear spin manifolds.}
   \label{fig:energy_low}
  \end{figure}
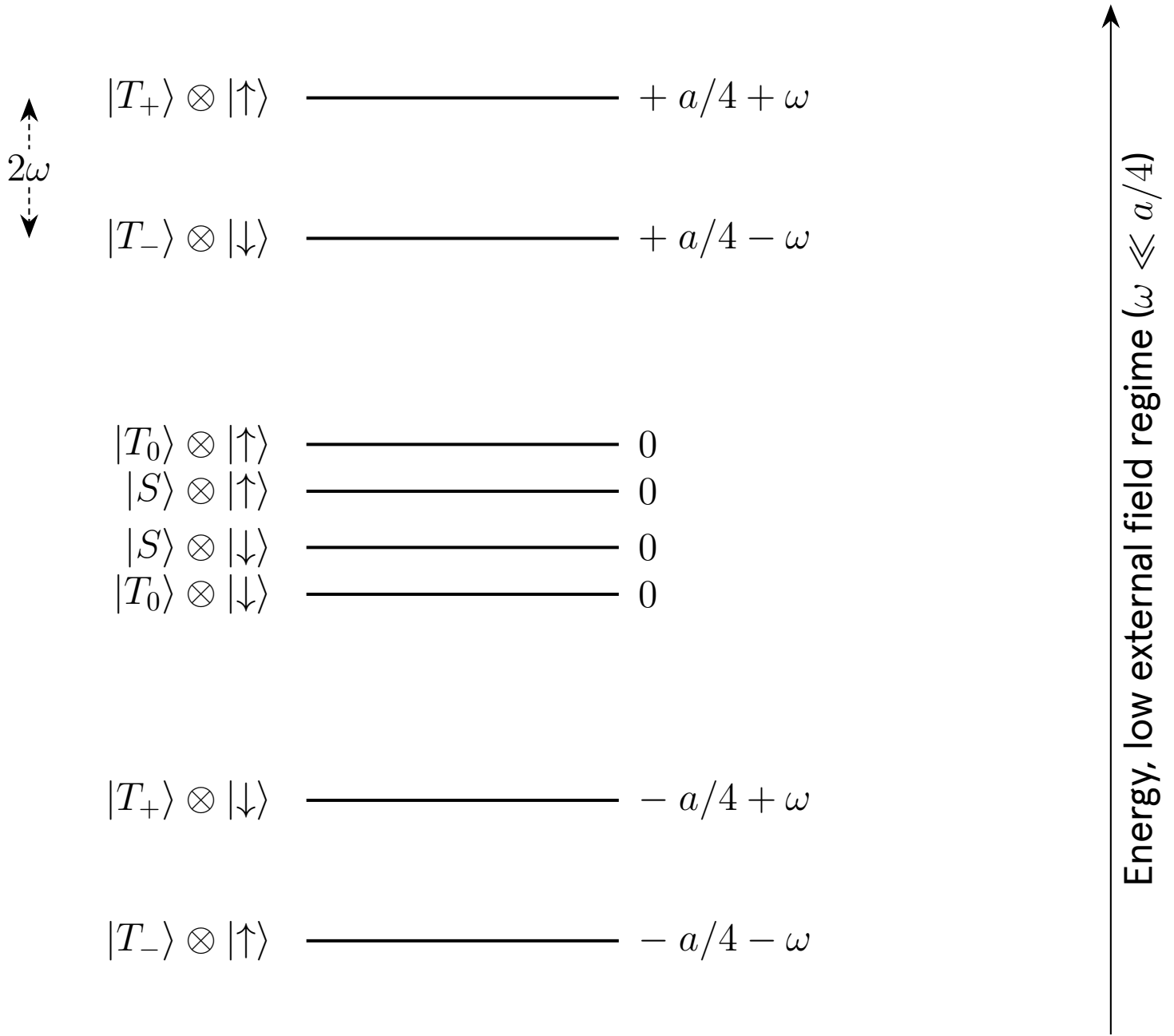

\begin{figure}[H]
  \centering

  \pgfmathsetmacro{\Scale}{1.3} 
  \pgfmathsetlengthmacro{\LevelLW}{\Scale*1.5pt} 

  \makebox[\textwidth][c]{%
  \begin{tikzpicture}[
      scale=\Scale,
      y=1.2cm, 
      every node/.style={transform shape},
      level/.style={draw, line width=1.5pt, color=black},
      energy/.style={font=\large, color=black},
      state/.style={font=\large, color=black},
  ]

  \def\levelwidth{4}
  \def\labeloffset{3.2}

  \coordinate (E1) at (0, 0.3);
  \draw[level] (-\levelwidth/2, 0.3) -- (\levelwidth/2, 0.3);
  \node[state, left] at (-\levelwidth/2-0.3, 0.3) {$\ket{S} \otimes \ket{\uparrow}$};
  \node[energy, right] at (\levelwidth/2+0.1, 0.3) {$0$};

  \coordinate (E2) at (0, 4.5);
  \draw[level] (-\levelwidth/2, 4.5) -- (\levelwidth/2, 4.5);
  \node[state, left] at (-\levelwidth/2-0.3, 4.5) {$\ket{T_+} \otimes \ket{\uparrow}$};
  \node[energy, right] at (\levelwidth/2+0.1, 4.5) {$+\ \omega + a/4$};

  \coordinate (E3) at (0, -4.5);
  \draw[level] (-\levelwidth/2, -4.5) -- (\levelwidth/2, -4.5);
  \node[state, left] at (-\levelwidth/2-0.3, -4.5) {$\ket{T_-} \otimes \ket{\uparrow}$};
  \node[energy, right] at (\levelwidth/2+0.1, -4.5) {$-\ \omega - a/4$};

  \coordinate (E4) at (0, 0.8);
  \draw[level] (-\levelwidth/2, 0.8) -- (\levelwidth/2, 0.8);
  \node[state, left] at (-\levelwidth/2-0.3, 0.8) {$\ket{T_0} \otimes \ket{\uparrow}$};
  \node[energy, right] at (\levelwidth/2+0.1, 0.8) {$0$};

  \coordinate (E5) at (0, -0.3);
  \draw[level] (-\levelwidth/2, -0.3) -- (\levelwidth/2, -0.3);
  \node[state, left] at (-\levelwidth/2-0.3, -0.3) {$\ket{S} \otimes \ket{\downarrow}$};
  \node[energy, right] at (\levelwidth/2+0.1, -0.3) {$0$};

  \coordinate (E6) at (0, 4.0);
  \draw[level] (-\levelwidth/2, 4.0) -- (\levelwidth/2, 4.0);
  \node[state, left] at (-\levelwidth/2-0.3, 4.0) {$\ket{T_+} \otimes \ket{\downarrow}$};
  \node[energy, right] at (\levelwidth/2+0.1, 4.0) {$+\ \omega - a/4$};

  \coordinate (E7) at (0, -4.0);
  \draw[level] (-\levelwidth/2, -4.0) -- (\levelwidth/2, -4.0);
  \node[state, left] at (-\levelwidth/2-0.3, -4.0) {$\ket{T_-} \otimes \ket{\downarrow}$};
  \node[energy, right] at (\levelwidth/2+0.1, -4.0) {$-\ \omega + a/4$};

  \coordinate (E8) at (0, -0.8);
  \draw[level] (-\levelwidth/2, -0.8) -- (\levelwidth/2, -0.8);
  \node[state, left] at (-\levelwidth/2-0.3, -0.8) {$\ket{T_0} \otimes \ket{\downarrow}$};
  \node[energy, right] at (\levelwidth/2+0.1, -0.8) {$0$};

  \draw[{Stealth[length=4mm]}-{Stealth[length=4mm]}, thick, black, densely dashed]
    (-\levelwidth/2-1.8-0.75-1, 4.5) -- (-\levelwidth/2-1.8-0.75-1, -4.0);
  \node[font=\large, black, fill=white, inner sep=2pt]
      at (-\levelwidth/2-1.8-0.75-1, 0.25) {$2\omega$};

  \draw[-{Stealth[length=4mm]}, thick, black]
      (\levelwidth/2 + 6 - 1+0.9 + 0.4, -5.5) -- (\levelwidth/2 + 6 -1+0.9 + 0.4, 5.5);
  \node[font=\large, rotate=90, black]
      at (\levelwidth/2 + 6.8 -1+0.9, 0) {Energy, high external field regime ($\omega \gg a/4$)};

  \end{tikzpicture}%
  }
  \captionsetup[figure]{aboveskip=4pt,belowskip=8pt}
  \captionof{figure}{\justifying \textbf{Energy level diagram (diagonal terms of the Hamiltonian $\mathcal{H}$ of Eq.~\eqref{eq:H}) in the high external magnetic field regime ($\omega \gg a/4$).}
  Quantum states are labeled in the basis $\ket{\text{electron spins}} \otimes \ket{\text{nuclear spin}}$.  The exact Zeeman splitting between the $\ket{T_{-}}$ and $\ket{T_{+}}$ manifolds is $2 \omega$; this term dominates the energy structure.}
   \label{fig:energy_high}
  
  \end{figure}
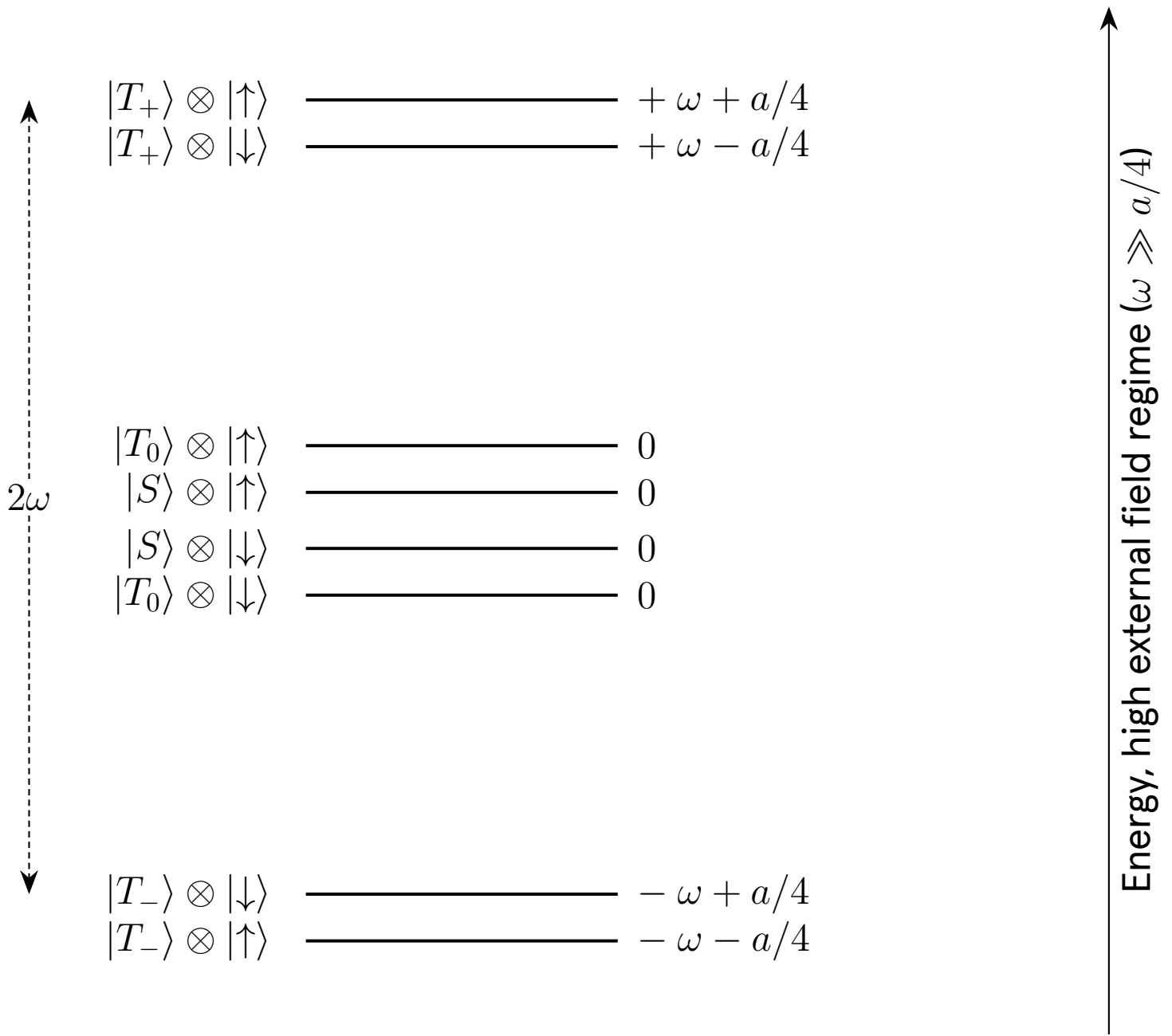

\noindent Solving for the energies in the above basis yields the equivalent diagrams of Figs.~\ref{fig:energy_low} and ~\ref{fig:energy_high}, depicting regimes in which the externally applied magnetic field is, respectively, much smaller or much larger than the considered hyperfine interaction strength.
\\

\begin{figure}[H]
\vspace*{-2cm}
\centering

\begin{tikzpicture}[
    state/.style={draw, circle, minimum size=2cm, font=\Large\bfseries},
    arrow/.style={-{Stealth[length=2.5mm]}, thick},
    label/.style={fill=white, inner sep=1pt, opacity=0.95, text opacity=1},
] 

\def\colsep{9}
\def\rowsep{2.8}

\node[state] (S_up)  at (0,  1*\rowsep) {$\scriptstyle\ket{S} \otimes \ket{\uparrow}$};
\node[state] (Tp_up) at (0,  3.25*\rowsep) {$\scriptstyle\ket{T_+} \otimes \ket{\uparrow}$};
\node[state] (Tm_up) at (0, -3.25*\rowsep) {$\scriptstyle\ket{T_-} \otimes \ket{\uparrow}$};
\node[state] (T0_up) at (0,  0*\rowsep) {$\scriptstyle\ket{T_0} \otimes \ket{\uparrow}$};

\node[state] (S_down)  at (\colsep,  1*\rowsep) {$\scriptstyle\ket{S} \otimes \ket{\downarrow}$};
\node[state] (Tp_down) at (\colsep,  3.25*\rowsep) {$\scriptstyle\ket{T_+} \otimes \ket{\downarrow}$};
\node[state] (Tm_down) at (\colsep, -3.25*\rowsep) {$\scriptstyle\ket{T_-} \otimes \ket{\downarrow}$};
\node[state] (T0_down) at (\colsep,  0*\rowsep) {$\scriptstyle\ket{T_0} \otimes \ket{\downarrow}$};


\draw[arrow, <->, blue!80, line width=1.2pt]
  (S_up) -- (T0_up)
  node[label, midway, left] {$-\frac{a}{4}$};






\draw[arrow, <->, cyan!70!blue, line width=1.2pt]
  (S_down) -- (T0_down)
  node[label, midway, right] {$+\frac{a}{4}$};

\draw[arrow, <->, purple!70, densely dotted, line width=1pt]
  (T0_up) to[bend left=6]
  node[label, pos=0.88, below right] {$\frac{a}{2\sqrt{2}}$} (Tp_down);

\draw[arrow, <->, brown!70, dash dot, line width=1pt]
  (Tm_up) to[bend right=2]
  node[label, pos=0.12, above left] {-$\frac{a}{2\sqrt{2}}$} (S_down);


\draw[arrow, <->, orange!80!red, densely dotted, line width=1pt]
  (Tm_up) to[bend right=6]
  node[label, pos=0.12, below right] {$\frac{a}{2\sqrt{2}}$} (T0_down);



\draw[arrow, <->, magenta!70, dash dot dot, line width=1pt]
  (Tp_down) to[bend right=3]
  node[label, pos=0.20, above left] {$\frac{a}{2\sqrt{2}}$} (S_up);

\node[font=\Large, blue!80] at (0, 11) {$\ket{\uparrow}$};
\node[font=\Large, cyan!70!blue] at (\colsep, 11) {$\ket{\downarrow}$};

\node[anchor=north west, align=left, font=\footnotesize] at (11, 4) {   
  \textcolor{blue!80}{$\blacksquare$} within nuclear $\ket{\uparrow}$ manifold \\
  \textcolor{cyan!70!blue}{$\blacksquare$} within nuclear $\ket{\downarrow}$ manifold\\
  \textcolor{magenta!70}{$\blacksquare$} $\ket{S}\leftrightarrow \ket{T_{+}}$ coupling\\
  \textcolor{brown!70}{$\blacksquare$} $\ket{S} \leftrightarrow \ket{T_{-}}$ coupling\\
  \textcolor{purple!70}{$\blacksquare$} $\ket{T_{0}} \leftrightarrow \ket{T_{+}}$ coupling\\
  \textcolor{orange!80!red}{$\blacksquare$} $\ket{T_{0}} \leftrightarrow \ket{T_{-}}$ coupling\\
};

\end{tikzpicture}

\captionsetup[figure]{aboveskip=4pt,belowskip=8pt}
  \captionof{figure}{\justifying \textbf{Couplings diagram (off-diagonal terms of the Hamiltonian $\mathcal{H}$ of Eq.~\eqref{eq:H}) for a minimal isotropic hyperfine tensor of strength $a$. This toy model will be analytically solved.}
  Quantum states are labeled in the basis $\ket{\text{electron spins}} \otimes \ket{\text{nuclear spin}}$.  States $\ket{T_{+}}\otimes\ket{\uparrow}$ and $\ket{T_{-}}\otimes\ket{\downarrow}$ are completely decoupled. The independent subspaces $\{ \ket{S}\otimes\ket{\uparrow}, \ket{T_{0}}\otimes\ket{\uparrow},  \ket{T_{+}}\otimes\ket{\downarrow}\}$ and $\{ \ket{S}\otimes\ket{\downarrow},  \ket{T_{0}}\otimes\ket{\downarrow},  \ket{T_{-}}\otimes\ket{\uparrow}\}$ are completely symmetric in their coupling strengths.}
   \label{fig:couplings}
\end{figure}
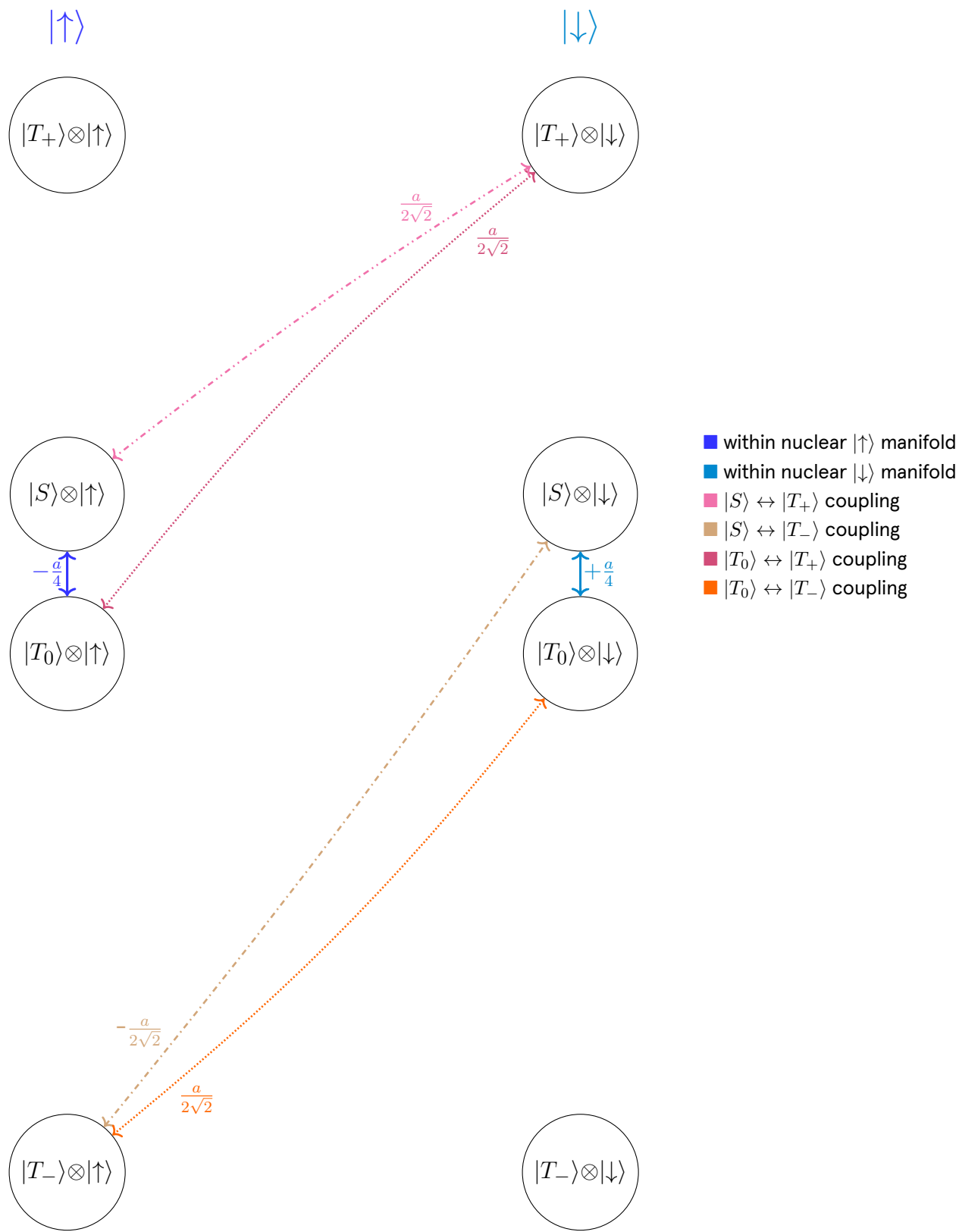

\noindent In turn, solving for the couplings in the above basis results in the diagram of Fig.~\ref{fig:couplings}.
\\

\noindent In the next section, we solve the dynamics generated by this Hamiltonian.
\vfill
\newpage
\newpage
\vspace*{-2cm}

\vspace*{-2cm}


  \newpage

\newpage
\begin{center}
\sectionfont{Take-home messages}
\end{center}

\begin{enumerate}
    \item This toy model for the radical pair mechanism assumes two spin-$\frac{1}{2}$ electrons, one of which is free, whereas the other is hyperfine-coupled to one spin-$\frac{1}{2}$ nucleus. The electrons otherwise do not interact among themselves. The hyperfine interaction is assumed to be isotropic with strength $a$. An external DC magnetic field is applied, yielding Larmor frequency $\omega$. Most of the nuclear dynamics is neglected. There is no decoherence.
   \item We will analytically solve the Hamiltonian in an experimentally relevant basis, namely $\{|S\rangle, |T_{0}\rangle, |T_{+}\rangle, |T_{-}\rangle\} \otimes \{ \ket{\uparrow}, \ket{\downarrow} \}$, where: the electron spins can be in a singlet $|S\rangle$ or in any one of the triplet $\{|T_{0}\rangle, |T_{+}\rangle, |T_{-}\rangle\}$ states; and the nuclear spin can be `up' or `down'.
   \item Energies in this basis are shown in Figs.~\ref{fig:energy_low} and ~\ref{fig:energy_high}, which are equivalent. The ratio $\frac{\omega}{a}$ determines which term dominates the energy structure: the hyperfine interaction or the external field.
   \item Couplings in this basis are shown in Fig.~\ref{fig:couplings}. Of the eight states, two are completely decoupled from the dynamics, while the remaining six split into two subspaces with symmetric coupling structure.
\end{enumerate}         
\clearpage
\section{Solution for the singlet population and density matrix}
\label{sec:solution}

\noindent 

\noindent Given the couplings of Fig.~\ref{fig:couplings}, we first note that the $\ket{T_{+}}\otimes\ket{\uparrow}$ and the $\ket{T_{-}}\otimes\ket{\downarrow}$ states are individually decoupled from any other state; their dynamics are thus trivial: any initial spin population in those states stays trapped in them. Secondly, we note that the subspaces $\{ \ket{S}\otimes\ket{\uparrow}, \ket{T_{0}}\otimes\ket{\uparrow},  \ket{T_{+}}\otimes\ket{\downarrow}\}$ (corresponding to $m_J = \frac{1}{2}$ in the notation of~\cite{brocklehurst1996freeradical, timmel1998weakfields, lewis2018lowfield}) and $\{ \ket{S}\otimes\ket{\downarrow},  \ket{T_{0}}\otimes\ket{\downarrow},  \ket{T_{-}}\otimes\ket{\uparrow}\}$ (corresponding to $m_J = -\frac{1}{2}$) are each decoupled from the rest of the Hilbert space and are related by a simple symmetry. In particular, the second subspace is obtained from the first by a nuclear spin flip, the swap $\ket{T_{+}} \rightarrow \ket{T_{-}}$, and the sign change $\omega \rightarrow - \omega$. It is therefore sufficient to calculate the dynamics of only one of the two three-state subspaces, for instance $\{ \ket{S}\otimes\ket{\uparrow}, \ket{T_{0}}\otimes\ket{\uparrow},  \ket{T_{+}}\otimes\ket{\downarrow}\}$. We show explicitly in Supplementary Subsection~\ref{explanation} how the results obtained for this subspace can be repurposed to describe both the dynamics of the second subspace and those of the full eight-state system. In the following, for simplicity, we drop the nuclear spin state ket in the description of the quantum state. 
\\

\noindent All the symbolic calculations for the time evolution of one of the three-state \linebreak subspaces  can be followed along using the accompanying SymPy code \linebreak \href{https://bit.ly/SingletPopulationEvolution_SymPy}{\texttt{SingletPopulationEvolution\_SymPy.py}}. 
\\

\noindent In the first subspace, we can define the basis vectors:
\begin{equation}
\ket{S} = 
\begin{pmatrix}
  1 \\
  0 \\
  0 
\end{pmatrix} \ , \qquad 
\ket{T_{0}} =
\begin{pmatrix}
  0 \\
  1 \\
  0 
\end{pmatrix} \ , \qquad 
\ket{T_{+}} = 
\begin{pmatrix}
  0 \\
  0 \\
  1 
\end{pmatrix} \ , 
\end{equation}
\noindent and their corresponding Hamiltonian:
\begin{equation}
\mathcal{H}_{1} = 
\left(
\begin{array}{ccc}
 0 & -\frac{a}{4} & \frac{a}{2 \sqrt{2}} \\
 -\frac{a}{4} & 0 & \frac{a}{2 \sqrt{2}} \\
 \frac{a}{2 \sqrt{2}} & \frac{a}{2 \sqrt{2}} & -\frac{a}{4}+\omega  \\
\end{array}
\right) \ . 
\label{H1}
\end{equation}

\noindent Because $\mathcal{H}_{1}$ is symmetric under the exchange of the labels $\ket{S}$ and $\ket{T_{0}}$ (i.e., in the way $\ket{T_{+}}$ sees them), we can define a symmetric basis:
\begin{equation}
\ket{D} \equiv \frac{\ket{T_{0}} - \ket{S}}{\sqrt{2}} \ , \qquad \ket{B} \equiv \frac{\ket{T_{0}} + \ket{S}}{\sqrt{2}} \ ,  \qquad \ket{T_{+}} \ .
\end{equation}

\noindent Conversely, 
\begin{equation}
    \ket{S} = \frac{\ket{B} - \ket{D}}{\sqrt{2}} \ , \qquad \ket{T_{0}} = \frac{\ket{B} + \ket{D}}{\sqrt{2}} \ . 
\end{equation}

\noindent In this new $\{\ket{D}, \ket{B}, \ket{T_{+}} \}$ basis, the Hamiltonian becomes:
\begin{equation}
  \mathcal{H}_{2} = 
\left(
\begin{array}{ccc}
 \frac{a}{4} & 0 & 0 \\
 0 & -\frac{a}{4} & \frac{a}{2} \\
 0 & \frac{a}{2} & -\frac{a}{4}+\omega  \\
\end{array}
\right) \ .
\label{H2}
\end{equation}

\noindent Note that $\ket{D}$ does not couple to either $\ket{B}$ or $\ket{T_{+}}$. In physics jargon (especially that inspired by the field of electromagnetically-induced transparency~\cite{Fleischhauer2005EIT}, see Supplementary Subsection~\ref{subsec:EIT}), this means that $\ket{D}$ is a `dark' state: destructive interference removes $\ket{D}$ from the dynamics. More precisely, $\ket{D}$ is dark to spin mixing. To the best of our knowledge, only one previous work~\cite{Xu2016} explicitly makes a similar decomposition and analysis in the context of radical pairs. Put simply, state $\ket{D}$ evolves as:
\begin{equation}
   \ket{D(t)} = e^{-i t\frac{a}{4}}\ket{D(0)} \ . 
\end{equation}

\noindent Throughout, the notation $\ket{X(t)}$ is used as shorthand for the component of a quantum state $\ket{\psi(t)}$ along $\ket{X}$, i.e., $|X(t)\rangle \equiv |X\rangle\langle X|\psi(t)\rangle$. Accordingly, $\left|\ket{X(t)}\right|^2 \equiv \langle\psi(t) | X \rangle \langle X | \psi(t)\rangle$ denotes the population in state $\ket{X}$ at time $t$; and $\rho_{XY}(t) \equiv \langle X | \rho(t) | Y \rangle$ denotes the density matrix element between states $\ket{X}$ and $\ket{Y}$ at time $t$ (a coherence, for $\ket{X} \neq \ket{Y}$) .
\\

\noindent The remaining Hamiltonian for the $\{\ket{B}$, $\ket{T_{+}}\}$ subspace is then:
\begin{equation}
  \mathcal{H}_{3} = 
\left(
\begin{array}{cc}
-\frac{a}{4} & \frac{a}{2} \\
 \frac{a}{2} & -\frac{a}{4}+\omega  \\
\end{array}
\right)
 \ = \ -\frac{a}{4} \mathbb{1}_{2} \ + \ 
\left(
\begin{array}{cc}
0 & \frac{a}{2} \\
 \frac{a}{2} & \omega  \\
\end{array}
\right)\ \equiv \ -\frac{a}{4} \mathbb{1}_{2} \ + \ \mathcal{H}_{4} \ ,
\end{equation}
\noindent where $\mathbb{1}_{2}$ is the $2\times2$ identity matrix. Note that the term proportional to $\mathbb{1}_{2}$ does not alter the dynamics of $\{\ket{B}, \ket{T_{+}}\}$, only shifting their energy zero; it does, however, contribute a global phase $e^{+ i t \frac{a}{4}}$ which will need to be taken into account. This phase becomes physically relevant when this block is recombined with the dark state $\ket{D}$: it then enters as a relative phase in $\ket{S} = \frac{\ket{B}  - \ket{D}}{\sqrt{2}}$, and we will argue shortly that our experimental observable is a function of $\ket{S(t)}$.
\\

\noindent In the absence of decoherence, and using
\begin{equation}
\ket{B} = 
\begin{pmatrix}
  1 \\
  0 
\end{pmatrix} \ , \qquad
\ket{T_{+}} =
\begin{pmatrix}
  0 \\
  1 
\end{pmatrix} \ ,  
\end{equation}

\noindent the dynamics of $\{\ket{B}, \ket{T_{+}}\}$ are given by:
\begin{equation}
\begin{pmatrix}
   B(t) \\
   T_{+}(t)
\end{pmatrix} = e^{+ i t\frac{a}{4}} \ \cdot \  e^{- i t \mathcal{H}_{4}} \ \cdot \ \begin{pmatrix}
  B(0) \\
  T_{+}(0)
\end{pmatrix} \ , 
\label{17}
\end{equation}

\noindent where the exact expression for the exponentiation of $\mathcal{H}_{4}$ is given below:
\begin{equation}
    e^{- i t \mathcal{H}_{4}} = \frac{e^{-i t\frac{\omega}{2}}}{\Omega} 
\left(
\begin{array}{cc}
  \Omega  \cos(\theta)+i \omega  \sin(\theta) & - i a  \sin(\theta)\\
- i a  \sin(\theta) & \Omega  \cos(\theta)-i \omega  \sin(\theta) \\
\end{array}
\right) \ ,
\label{18}
\end{equation}
\noindent where we defined $\Omega \equiv \sqrt{a^2 + \omega^2}$ and $\theta \equiv \left(  \frac{\Omega t }{2} \right)$. 
\\

\noindent We previously claimed that, experimentally, observables in this type of spin-dependent chemical reactions are a proxy for the electron spin population projected into the singlet state (or in all of the triplet states). Here, decoherence is neglected, and the population in the singlet state as a function of time, $\left|\ket{S(t)}\right|^2$, is analytically calculated below for a generic incoherent initial electron spin state; $\left|\ket{S(t)}\right|^2$ is thus referred to as our `signal'.
\\

\noindent The signal $\left|\ket{S(t)}\right|^2$ will be a function of the initial populations in the $\{ \ket{S}, \ket{T_{0}},  \ket{T_{+}}\}$ states. If the initial spin population has a nonzero $\ket{T_{-}}$ component, then, depending on the nuclear spin state, that fraction may either remain indefinitely trapped in the uncoupled state $\ket{T_{-}}\otimes\ket{\downarrow}$ or participate in the dynamics of the subspace $\{ \ket{S}\otimes\ket{\downarrow},  \ket{T_{0}}\otimes\ket{\downarrow},  \ket{T_{-}}\otimes\ket{\uparrow}\}$. With a nonzero $\ket{T_{-}}$ component, the expressions presented for $\left|\ket{S(t)}\right|^2$ will need to be scaled by an appropriate factor. 
\\

\noindent As previously mentioned, the singlet population dynamics for the second, not explicitly considered three-state subspace are obtained from the results presented in the following pages by making the replacements $\ket{T_{+}} \rightarrow \ket{T_{-}}$ and $\omega \rightarrow  -\omega$ (plus a nuclear spin flip; other dynamics of the nuclear spin are being neglected); details can be found in 
Supplementary Subsection~\ref{explanation}.
\\

\noindent Predicting the distribution of initial electron spin states in a protein that supports spin-\linebreak dependent chemical reactions is often challenging~\cite{schweiger2001principles}. Whether a radical pair is formed in a singlet or triplet state is best understood as the outcome of a kinetic competition between electron transfer and intersystem crossing; this competition may be further complicated by other factors, such as spin relaxation rates and symmetry conservation arguments. As a rule of thumb, if electron transfer occurs directly from an excited singlet precursor before intersystem crossing takes place, the radical pair is typically singlet-born; if intersystem crossing to a triplet state is faster, the radical pair is typically triplet-born. Even in the latter case, however, the distribution among the triplet sublevels is not straightforward to predict. We therefore base all calculations here on a generic incoherent mixture of the four possible initial spin states.
\\

\noindent Hence, the signal resulting from the true eight-level dynamics, for a generic incoherent mixture of spins starting in  $\alpha  \ket{T_{+}}$, $\beta \ket{T_{0}}$, $\gamma  \ket{T_{-}}$, and $(1-\alpha-\beta-\gamma)\ket{S}$, and a maximal incoherent mixture of initial nuclear spin states, is in reality given by
\begin{equation}
\begin{split}
\textcolor{InstituteBlue}{%
\underline{\left|\ket{S(t)}\right|^2_{\substack{
\alpha \ket{T_{+}},\;
\beta \ket{T_{0}},\;
\gamma \ket{T_{-}},\\
(1-\alpha-\beta-\gamma)\ket{S}\ @\ t=0
}}}}
={}&\phantom{+ \ \ }
\frac{\alpha}{2}
\left|\ket{S(t,\omega)}\right|^2_{100\% \ket{T_{+}} \ @ \ t=0}
+
\frac{\gamma}{2}
\left|\ket{S(t,-\omega)}\right|^2_{100\% \ket{T_{+}} \ @ \ t=0}
\\
&+
\frac{\beta}{2}\Big(
\left|\ket{S(t,\omega)}\right|^2_{100\% \ket{T_{0}} \ @ \ t=0}
+
\left|\ket{S(t,-\omega)}\right|^2_{100\% \ket{T_{0}} \ @ \ t=0}
\Big)
\\
&+
\frac{(1-\alpha-\beta-\gamma)}{2}\Big(
\left|\ket{S(t,\omega)}\right|^2_{100\% \ket{S} \ @ \ t=0}
+
\left|\ket{S(t,-\omega)}\right|^2_{100\% \ket{S} \ @ \ t=0}
\Big)\ .
\end{split}
\label{signalform}
\end{equation}

\noindent To distinguish the signal derived for the considered three-state subspace, $\left|\ket{S(t)}\right|^2$, from the signal of the full eight-level system, \textcolor{InstituteBlue}{\underline{$\left|\ket{S(t)}\right|^2$}}, we represent the latter in \textcolor{InstituteBlue}{blue} and \underline{underlined}. 
\\

\noindent The full signal is:
\begin{equation}
\begin{split}
\textcolor{InstituteBlue}{%
\underline{\left|\ket{S(t)}\right|^2_{\substack{
\alpha \ket{T_{+}},\;
\beta \ket{T_{0}},\;
\gamma \ket{T_{-}},\\
(1-\alpha-\beta-\gamma)\ket{S}\ @\ t=0
}}}}
={}&\phantom{ \ + \ }
\frac{1-\alpha-\gamma}{2}
-(1-2\alpha-2\gamma)\frac{a^2}{4\Omega^2}\sin(\theta)^2
\\
&+
\frac{((1-\alpha-\beta-\gamma)-\beta)}{2}
\cos\!\left(\frac{at}{2}\right)
\left[
\cos(\theta) \cos\!\left(\frac{\omega t}{2}\right)
+\frac{\omega}{\Omega}\sin(\theta) \sin\!\left(\frac{\omega t}{2}\right)
\right]\ .
\end{split}
\label{gen}
\end{equation}

\noindent In many formulas henceforth, a useful structural simplification will emerge: many expressions depend only on three combinations of the initial populations, rather than on the four populations separately. In particular, they often depend on the total initial population in the $\{\ket{S},\ket{T_0}\}$ sector,

\begin{equation}
  \Sigma \equiv ((1-\alpha-\beta-\gamma)+\beta) = 1-\alpha-\gamma \ ,
\end{equation}

\noindent and on the initial population imbalance within that sector,

\begin{equation}
\Delta \equiv ((1-\alpha-\beta-\gamma)-\beta) \ .
\end{equation}

\noindent To parametrize the full problem, it is then convenient to introduce a third variable measuring the imbalance between the polarized triplet states,

\begin{equation}
\zeta \equiv \alpha-\gamma  \ .
\end{equation}

\noindent We will therefore also present our results in terms of the three parameters $\Sigma$, $\Delta$, and $\zeta$.
\\

\noindent In the new parametrization, the full signal is:
\begin{equation}
\begin{split}
\textcolor{InstituteBlue}{%
\underline{\left|\ket{S(t)}\right|^2_{\substack{
\alpha \ket{T_{+}},\;
\beta \ket{T_{0}},\;
\gamma \ket{T_{-}},\\
(1-\alpha-\beta-\gamma)\ket{S}\ @\ t=0
}}}}
={}&\phantom{ \ + \ }
\frac{\Sigma}{2}
-\frac{(2\Sigma-1)a^2}{4\Omega^2}\sin(\theta)^2
\\
&+
\frac{\Delta}{2}\cos\!\left(\frac{at}{2}\right)
\left[
\cos(\theta) \cos\!\left(\frac{\omega t}{2}\right)
+\frac{\omega}{\Omega}\sin(\theta) \sin\!\left(\frac{\omega t}{2}\right)
\right]\ .
\end{split}
\label{gennew}
\end{equation}

\noindent Note that: the first term is DC; the second term oscillates with $\Omega$; and the third term has a multifrequency content with components $\frac{|a \pm \omega \pm \Omega|}{2}$.
\\

\noindent In the absence of a magnetic field ($\omega \rightarrow 0$), 
\begin{equation}
\begin{split}
\lim_{\omega\to 0}\,
\textcolor{InstituteBlue}{%
\underline{\left|\ket{S(t)}\right|^2_{\substack{
\alpha \ket{T_{+}},\;
\beta \ket{T_{0}},\;
\gamma \ket{T_{-}},\\
(1-\alpha-\beta-\gamma)\ket{S}\ @\ t=0
}}}}
={}&
\frac{\Sigma+\Delta}{2}
-\left(\frac{\Sigma+\Delta}{2} - \frac{1}{4}\right)\sin\!\left(\frac{at}{2}\right)^2\ ,
\end{split}
\end{equation}

\noindent and, for very large magnetic fields ($\omega \rightarrow \infty$), 
\begin{equation}
\begin{split}
\lim_{\omega\to\infty}\,
\textcolor{InstituteBlue}{%
\underline{\left|\ket{S(t)}\right|^2_{\substack{
\alpha \ket{T_{+}},\;
\beta \ket{T_{0}},\;
\gamma \ket{T_{-}},\\
(1-\alpha-\beta-\gamma)\ket{S}\ @\ t=0
}}}}
={}&
\frac{\Sigma+\Delta}{2}
-\Delta \sin\!\left(\frac{at}{4}\right)^2 \ .
\end{split}
\end{equation}

\noindent Comparing the two limiting expressions reveals that the magnetic field changes both the effective oscillation timescale and the part of the initial state that controls the signal. In the zero-field limit, the singlet population oscillates as $\sin\left(\frac{a t}{2}\right)^2$, whereas in the large-field limit it oscillates as $\sin\left(\frac{a t}{4}\right)^2$, so the dynamics become slower by a factor of $2$ at high field. Such a slower dynamics at high field will reappear several times; we offer an explanation for it later in the text. Moreover, the two considered limits probe different population imbalances. For $\omega\to0$, the oscillation amplitude is set by $\left(\frac{\Sigma+\Delta}{2} -\frac{1}{4}\right)$, that is, by the deviation of the initial singlet population from the equal-sharing value \(\frac{1}{4}\); in this sense the zero-field dynamics depend only on the total singlet weight and not on how the population in the three triplets is distributed internally. In turn, for $\omega\to\infty$ the oscillation amplitude is $\Delta$, so the signal is controlled specifically by the imbalance between the initial \(\ket{S}\) and \(\ket{T_0}\) populations, while \(\ket{T_\pm}\) effectively decouple due to the large Zeeman splitting. Thus the field drives a crossover from collective singlet--triplet mixing at low field to an effective \(\ket{S}\leftrightarrow \ket{T_0}\) problem at high field.
\\

\noindent The signal can be artificially time-averaged; the assumption is that we perform an average for $t \rightarrow \infty$, which is equivalent to letting terms in $\sin(\cdot) \to 0$ and terms in $\sin(\cdot)^2 \to \frac{1}{2}$. It yields:
\begin{align}
\langle \textcolor{InstituteBlue}{\underline{\left|\ket{S(t)}\right|^2_{
  \alpha  \ket{T_{+}},\;
  \beta  \ket{T_{0}},\;
  \gamma  \ket{T_{-}},\;
  (1-\alpha-\beta-\gamma)\ket{S}\ @\ t=0
}}}  \rangle_{t\to\infty}
&=
\frac{1+2\Sigma+2\Delta}{8} \ ,
\qquad \omega=0 \ ; 
\\[1ex]
\langle \textcolor{InstituteBlue}{\underline{\left|\ket{S(t)}\right|^2_{
  \alpha  \ket{T_{+}},\;
  \beta  \ket{T_{0}},\;
  \gamma  \ket{T_{-}},\;
  (1-\alpha-\beta-\gamma)\ket{S}\ @\ t=0
}}}  \rangle_{t\to\infty}
&=
\frac{(1+2\Sigma)a^2+4\Sigma\omega^2}{8\Omega^2} \ ,
\qquad \omega\neq0 \ ;
\end{align}

\noindent Note that this long-time average of the signal is discontinuous between $\omega = 0$ and $\omega = 0^{+}$ by a quantity $\frac{\Delta}{4}$. Mathematically, this discontinuity is due to the fact that the limits $\omega \to 0$ and $t \to \infty$ do not commute; this is true in a class of problems where the $\omega\to0$ limit turns an oscillatory term into a non-oscillatory one, so the long-time operation `sees' a DC piece only at exactly zero frequency. This will become explicit in Eq.~(\ref{27}) of Section~\ref{sec:analogy}. With any finite reaction lifetime, dephasing, or finite observation window (see Section~\ref{sec:MFE}), that discontinuity is washed away.
\\

\noindent Also, the time-averaged signal, for $\omega \neq 0$, depends only on the total initial populations in the $\left\{\ket{S}, \ket{T_{0}}\right\}$ manifold. $\ket{S}$ and $\ket{T_{0}}$ are the two states that mix directly in the considered dynamics since they have the same energy. In other words, the long-time average for $\omega \neq 0$ `forgets' the distinction between them. In contrast, the long-time average at exactly $\omega = 0$ `remembers' the imbalance between $\ket{S}$ and $\ket{T_{0}}$ because the zero-field triplet degeneracy leaves behind a DC contribution of $\frac{\Delta}{4}$ that survives this average. 
\\

\noindent For high fields, we find $\lim_{\omega\to\infty}
\langle  \textcolor{InstituteBlue}{\underline{\left|\ket{S(t)}\right|^2_{
  \alpha  \ket{T_{+}},\;
  \beta  \ket{T_{0}},\;
  \gamma  \ket{T_{-}},\;
  (1-\alpha-\beta-\gamma)\ket{S}\ @\ t=0
}}} \rangle_{t\to\infty}
=
\frac{\Sigma}{2}$, \linebreak reflecting the fact that, in the absence of decoherence, the fractions initially in $\ket{T_{+}}$ and $\ket{T_{-}}$ are effectively trapped by the large Zeeman splitting, whereas the remaining fraction $\Sigma$, initially in the $\{\ket{S},\ket{T_{0}}\}$ manifold, continues to oscillate coherently between $\ket{S}$ and $\ket{T_{0}}$. Since the long-time singlet population average of that coherent $\ket{S}\leftrightarrow\ket{T_{0}}$ oscillation is $\frac{1}{2}$, the overall high-field time-averaged signal is indeed $ \frac{1}{2} \cdot \Sigma = \frac{\Sigma}{2}$. In other words, the high-field plateau level only depends on the total initial weight in the $\{\ket{S}, \ket{T_{0}}\}$ manifold.
\\

\noindent All the generic terms derived above have been checked against the particular initial spin states studied in Supplementary Subsections~\ref{c1} through~\ref{c5}, namely: pure in $\ket{T_{+}}$; pure in $\ket{T_{0}}$; equiprobable for all triplets, with no population weight in $\ket{S}$; pure in $\ket{S}$; and equiprobable for all four states. The results are self-consistent. The complete signal, as well as its low- and high-field limits, are reported in Table~\ref{Table1} for these particular cases. Note that all signals are even in $\omega$, and reduce smoothly from $\omega = 0$ to $\omega = 0^{+}$.
\\

\noindent One note about the fifth data row in Table~\ref{Table1}, for an equiprobable initial spin polarization (i.e., $\alpha = \beta = \gamma = \frac{1}{4}$): note that the signal is constant. In this model, thus, no magnetic field sensing arises in the absence of some degree of initial spin polarization. An equiprobable (maximal-mixture) initial electron spin state does not give rise to modulation of the singlet population as a function of time.
\\

\begin{figure}[H]
  \centering
  \includegraphics[width=0.9\textwidth]{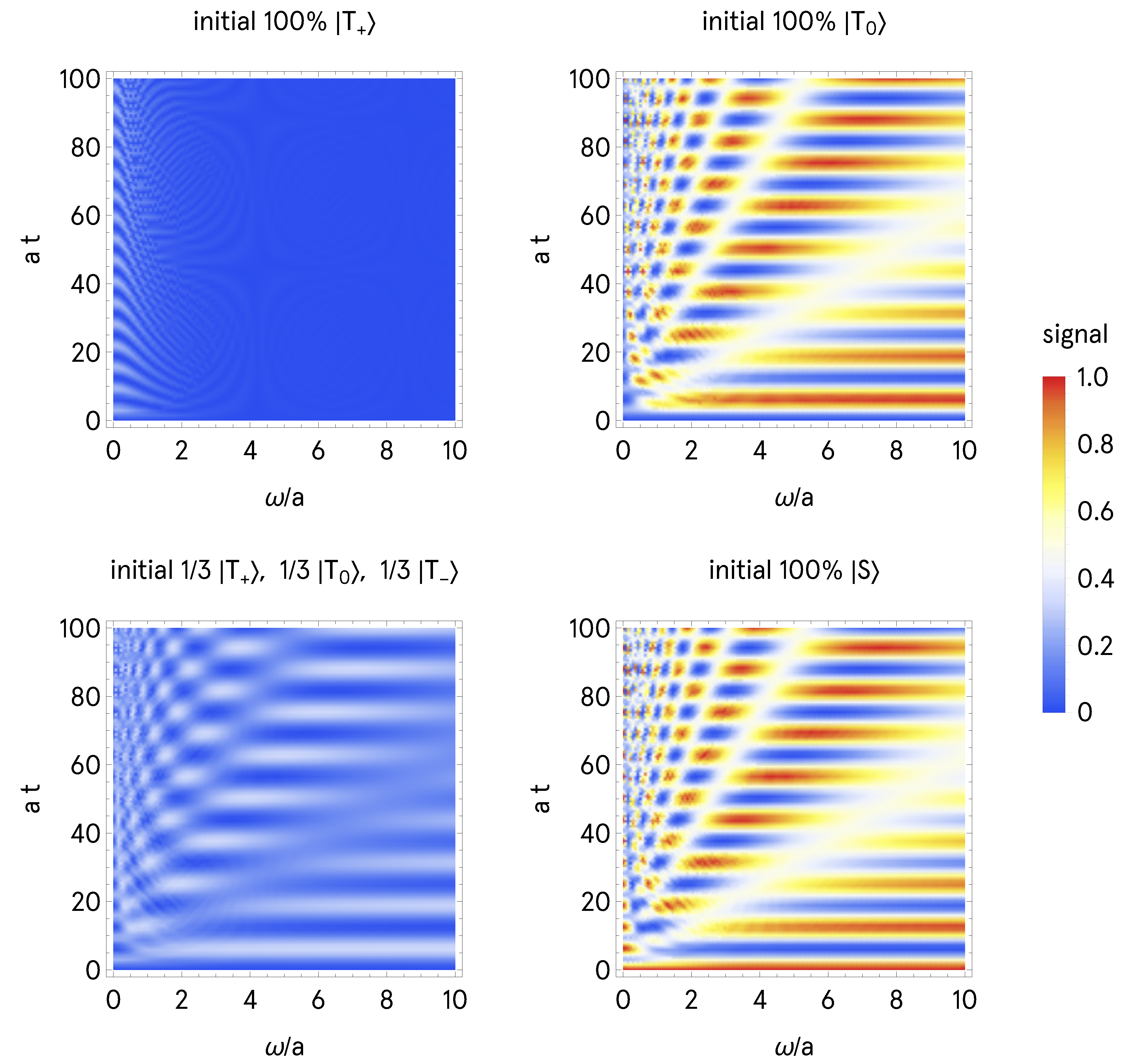}
  \captionsetup[figure]{aboveskip=4pt,belowskip=8pt}
  \captionof{figure}{\justifying \textbf{Density plot of the signal (time-evolution of the singlet population), expressed for different initial spin polarizations.}
The signal is plotted as a function of the dimensionless parameters \(at\) and \(\frac{\omega}{a}\), shown for four representative initial states: pure \(\ket{T_{+}}\), pure \(\ket{T_{0}}\), an equiprobable triplet mixture with no singlet population, and pure \(\ket{S}\). The axes are shown in dimensionless form: \(at\) is time measured in units of the hyperfine timescale \(\frac{1}{a}\), while \(\frac{\omega}{a}\) measures the Larmor frequency relative to the hyperfine coupling strength. The plots illustrate the rich multi-frequency structure of the dynamics across field strength, while also showing the smooth crossover from the low-field to the high-field regime and the progressive slowing down of the oscillations as \(\frac{\omega}{a}\) increases. The plot for an initial \(\ket{T_{+}}\) has fewer features because \(\ket{T_{+}}\) decouples at high field.}
   \label{P4}
  \end{figure}

\noindent In Fig.~\ref{P4}, we plot the dynamics of the four nontrivial initial states using the dimensionless variables \(a t\) and \(\frac{\omega}{a}\). This choice is natural because both \(a\) and \(\omega\) have units of angular frequency, so \(\frac{\omega}{a}\) measures the external field strength relative to the hyperfine coupling, while \(at\) rescales time by the intrinsic hyperfine timescale \(\frac{1}{a}\). With this rescaling, the signal depends only on dimensionless combinations of the parameters, e.g., \(\Omega t = (at)\sqrt{1+(\frac{\omega}{a})^2}\). Thus \(a\) sets the `internal clock' of the dynamics and is the natural field scale of the problem. Although the dynamics display a rich multi-frequency dependence on \(\omega\), the corresponding plots are visually busy and therefore are more useful as a compact overview of the parameter dependence than as a transparent guide to the underlying mechanisms. What is nevertheless visible is the smooth crossover from the low-field regime to the high-field regime, together with the progressive slowing down of the oscillations at large \(\frac{\omega}{a}\). The initial \(\ket{T_0}\) and initial \(\ket{S}\) cases are governed by the same characteristic scales and therefore exhibit similar patterning in the \((\frac{\omega}{a},at)\) plane; they differ by the sign with which the \(\Delta\)-controlled term enters on top of a common background, see Suppl.~Eqs.~(\ref{t0si}) and~(\ref{s}). The initially polarized \(\ket{T_{+}}\) case remains comparatively featureless because the polarized triplet states become effectively decoupled at high field. On a separate note, we will be able to confirm with the example of Fig.~\ref{100} of Section~\ref{sec:MFE} that, for a time parameter value of
$a t \sim 100$, results are already very close to those of a proper long-time average $t \to \infty$.
\\

\vfill
\newpage

\begin{table}[H]
\hspace*{-0.3cm}
    \centering
\begin{tabular}{|c|c|c|c|}
\hline
initial state & signal \textcolor{InstituteBlue}{\underline{$\left|\ket{S(t)}\right|^2$}} & $\lim_{\omega \to 0}	\textcolor{InstituteBlue}{\underline{\left|\ket{S(t)}\right|^2}}$ & $\lim_{\omega \to \infty}	\textcolor{InstituteBlue}{\underline{\left|\ket{S(t)}\right|^2}}$  \\ \hline
100\% $\ket{T_{+}}$                    & Suppl.~Eq.~(\ref{tpsi}) & $\frac{1}{4}\sin(\frac{a t}{2})^2$ & 0   \\ \hline
100\% $\ket{T_{0}}$                    & Suppl.~Eq.~(\ref{t0si}) & $\frac{1}{4}\sin(\frac{a t}{2})^2$ & $\sin(\frac{a t}{4})^2$   \\ \hline
$\frac{1}{3} \ket{T_{+}}$, \ $\frac{1}{3} \ket{T_{0}}$, \ $\frac{1}{3} \ket{T_{-}}$                                      & Suppl.~Eq.~(\ref{eq})& $\frac{1}{4}\sin(\frac{a t}{2})^2$ & $\frac{1}{3}\sin(\frac{a t}{4})^2$   \\ \hline
100\% $\ket{S}$                                                                   & Suppl.~Eq.~(\ref{s}) & $1 - \frac{3}{4}\sin(\frac{a t}{2})^2$ & $1 - \sin(\frac{a t}{4})^2$   \\ \hline
$\frac{1}{4} \ket{T_{+}}, \ \frac{1}{4} \ket{T_{0}}, \ \frac{1}{4} \ket{T_{-}}, \ \frac{1}{4} \ket{S}$ & $\frac{1}{4}$ & $\frac{1}{4}$ & $\frac{1}{4}$  \\ \hline
\scriptsize{$\alpha \ket{T_{+}}, \ \beta  \ket{T_{0}}, \ \gamma  \ket{T_{-}}, \ (1-\alpha-\beta-\gamma) \ket{S}$} & Eq.~(\ref{gennew}) & 
$\frac{\Sigma+\Delta}{2}-\left(\frac{\Sigma+\Delta}{2}- \frac{1}{4}\right)\sin\left(\frac{a t}{2}\right)^2
$ & $\frac{\Sigma+\Delta}{2}-\Delta\sin\left(\frac{a t}{4}\right)^2$ \\ \hline
\end{tabular}
   \captionsetup[table]{aboveskip=4pt,belowskip=8pt}
  \captionof{table}{\justifying \textbf{Signal (time-evolution of the singlet population) for the full eight-level system dynamics and its low- and high-magnetic field limits, expressed for different initial spin polarizations.} All results are self-consistent. Closed expressions can be found in the accompanying SymPy code. The initial spin states appear in the signals only via two parameters, $\Sigma$ and $\Delta$, which measure the total initial weight in the $\{ \ket{S}, \ket{T_0}\}$ sector, and the initial difference between the populations in these states. All signals are smoothly continuous for $\omega$ going from $0$ to $0^{+}$.}
    \label{Table1}
\end{table}

\vfill\newpage

\vfill
\newpage

\noindent In Table~\ref{Table2}, we present the long-time averaged signals. Note the discontinuity from $\omega = 0$ to $\omega = 0^{+}$ when the initial $\ket{S}$ and $\ket{T_{0}}$ populations are unequal (i.e., when $\Delta \neq 0$). 
\\

\begin{figure}[H]
  \centering
  \includegraphics[width=\textwidth]{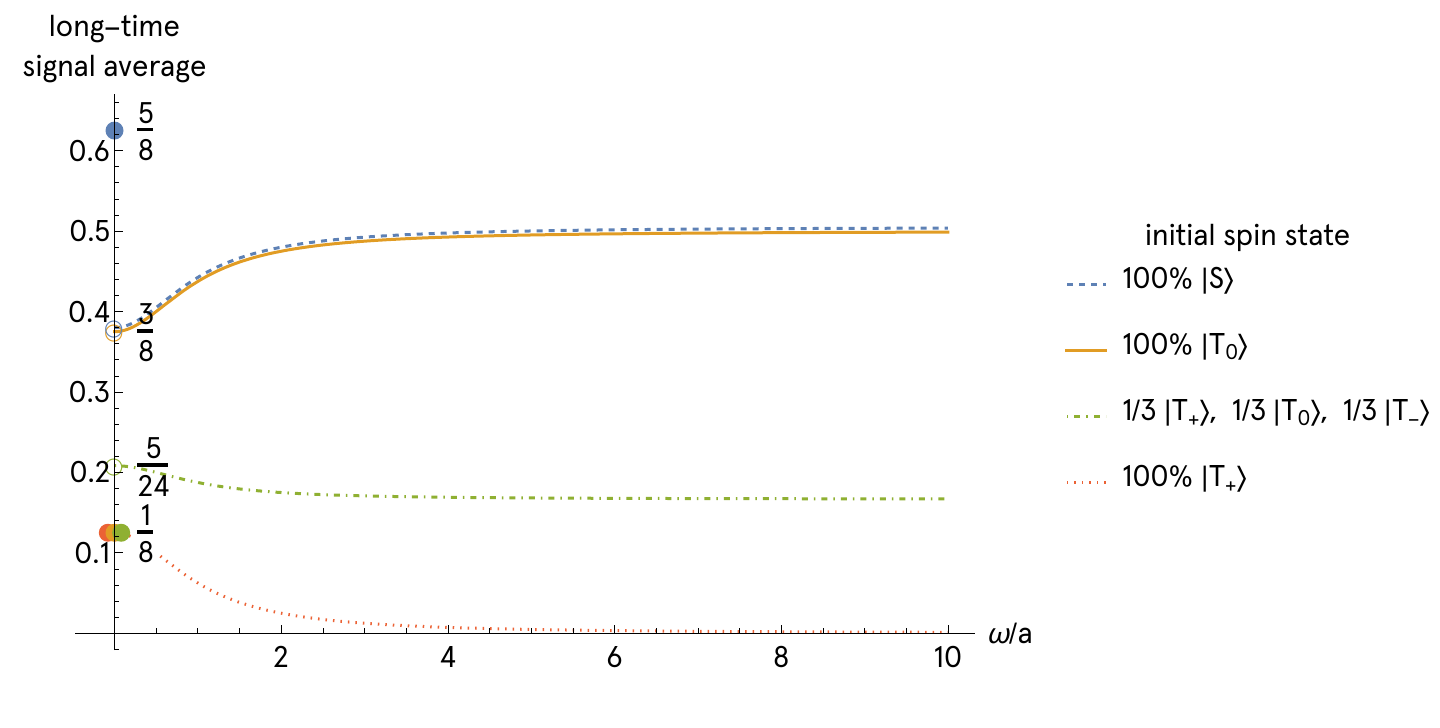}
  \captionsetup[figure]{aboveskip=4pt,belowskip=8pt}
  \captionof{figure}{\justifying \textbf{Long-time signal average expressed for different initial spin polarizations, as a function of field strength.}
The signal is plotted as a function of the dimensionless parameter \(\frac{\omega}{a}\), for four representative initial states: pure \(\ket{S}\), pure \(\ket{T_{0}}\), an equiprobable triplet mixture with no singlet population, and pure \(\ket{T_{+}}\). Unlike the instantaneous dynamics of Fig.~\ref{P4}, the long-time average removes the oscillatory structure and reveals the persistent population background. For any \(\omega \neq 0\), the average depends only on \(\Sigma\), the total initial weight in the \(\{\ket{S},\ket{T_{0}}\}\) manifold, so the pure \(\ket{S}\) and pure \(\ket{T_{0}}\) cases coincide. The curves approach the high-field plateau values \(\Sigma/2\). A discontinuity arises for cases with $\Delta \neq 0$.}
   \label{P1}
  \end{figure}

\noindent Fig.~\ref{P1} shows the long-time average of the signal as a function of the dimensionless field strength \(\frac{\omega}{a}\) for the same representative initial spin states considered in Fig.~\ref{P4}. In contrast to the instantaneous dynamics, the long-time average suppresses the oscillatory multi-\linebreak frequency structure and isolates the persistent population background. For any nonzero field, this average depends only on \(\Sigma\), the total initial population in the \(\{\ket{S},\ket{T_{0}}\}\) manifold, and not on \(\Delta\), the imbalance between \(\ket{S}\) and \(\ket{T_{0}}\). As a result, the initially pure \(\ket{S}\) and initially pure \(\ket{T_{0}}\) cases become identical for all \(\omega \neq 0\), even though their instantaneous signals differ. The curves then interpolate smoothly toward their high-field plateaus. At exactly \(\omega=0\), however, the long-time average can retain a memory of \(\Delta\), producing the singular zero-field behavior discussed above for all cases with \(\Delta \neq 0\).

\vfill
\newpage

\begin{table}[H]
    \centering
\begin{tabular}{|c|c|}
\hline
initial state & $\langle  \textcolor{InstituteBlue}{\underline{\left|\ket{S(t)}\right|^2}} \rangle_{t\to\infty}$ for $\omega = 0 / \omega \neq 0/ \omega \rightarrow \infty$ \\ \hline
100\% $\ket{T_{+}}$                    &  $\frac{1}{8} / \frac{a^2}{8\Omega^2} / 0$  \\ \hline
100\% $\ket{T_{0}}$                    & $ \frac{1}{8} / \frac{3a^2 + 4\omega^2}{8\Omega^2} / \frac{1}{2}$  \\ \hline
$\frac{1}{3} \ket{T_{+}}$, \ $\frac{1}{3} \ket{T_{0}}$, \ $\frac{1}{3} \ket{T_{-}}$                                  & $\frac{1}{8} / \frac{5a^2 + 4\omega^2}{24\Omega^2} / \frac{1}{6}$  \\ \hline
100\% $\ket{S}$                                                                   & $\frac{5}{8} / \frac{3a^2 + 4\omega^2}{8\Omega^2} / \frac{1}{2}$  \\ \hline
$\frac{1}{4} \ket{T_{+}}, \ \frac{1}{4} \ket{T_{0}}, \ \frac{1}{4} \ket{T_{-}}, \ \frac{1}{4} \ket{S}$ & $\frac{1}{4} / \frac{1}{4} / \frac{1}{4}$  \\ \hline
$\alpha \ket{T_{+}}, \ \beta  \ket{T_{0}}, \ \gamma  \ket{T_{-}}, \ (1-\alpha-\beta-\gamma) \ket{S}$ & $ \frac{1+2\Sigma+2\Delta}{8}
/ \frac{(1+2\Sigma)a^{2}+4\Sigma\omega^{2}}{8\Omega^{2}}/ \frac{\Sigma}{2} $ \\ \hline

\end{tabular}
   \captionsetup[table]{aboveskip=4pt,belowskip=8pt}
  \captionof{table}{\justifying \textbf{Long-time signal average for the full eight-level system dynamics and its low- and high-magnetic field limits, expressed for different initial spin polarizations.} All results are self-consistent.
  The model assumes no decoherence, so the average is for $t \to \infty$. The initial spin states appear in the signals only via two parameters, $\Sigma$ and $\Delta$, which measure the total initial weight in the $\{ \ket{S}, \ket{T_0}\}$ sector, and the initial difference between the populations in these states. Long-time averaged signals are discontinuous for $\omega$ going from $0$ to $0^{+}$ if $\Delta \neq 0$.}
    \label{Table2}
\end{table}

\newpage

\noindent For completeness, we give here the signals (i.e., the population dynamics) for all the electron spin states, as well as the nonzero coherences for the full $8\times8$ density matrix.
\\

\noindent The signals are:
\begin{align}
\textcolor{InstituteBlue}{\underline{\left|\ket{T_{0}(t)}\right|^2_{\substack{
  \alpha  \ket{T_{+}},\;
  \beta  \ket{T_{0}},\;
  \gamma  \ket{T_{-}},\\
  (1-\alpha-\beta-\gamma)\ket{S}\ @\ t=0
}}}}
=&\phantom{ \ + \ }
\frac{\Sigma}{2}
-\frac{(2\Sigma-1)a^{2}}{4\Omega^{2}}\sin(\theta)^2
\nonumber\\
&
-
\frac{\Delta}{2}\cos\left(\frac{at}{2}\right)
\left[
\cos(\theta)\cos\left(\frac{\omega t}{2}\right)
+\frac{\omega}{\Omega}\sin(\theta)\sin\left(\frac{\omega t}{2}\right)
\right]\ ,
\\
\textcolor{InstituteBlue}{\underline{\left|\ket{T_{+}(t)}\right|^2_{\substack{
  \alpha  \ket{T_{+}},\;
  \beta  \ket{T_{0}},\;
  \gamma  \ket{T_{-}},\\
  (1-\alpha-\beta-\gamma)\ket{S}\ @\ t=0
}}}}
=&\phantom{ \ + \ }
\frac{1-\Sigma+\zeta}{2}
+\frac{(2\Sigma-1-\zeta)a^{2}}{4\Omega^{2}}\sin(\theta)^2\ ,
\\
\textcolor{InstituteBlue}{\underline{\left|\ket{T_{-}(t)}\right|^2_{\substack{
  \alpha  \ket{T_{+}},\;
  \beta  \ket{T_{0}},\;
  \gamma  \ket{T_{-}},\\
  (1-\alpha-\beta-\gamma)\ket{S}\ @\ t=0
}}}}
=&\phantom{ \ + \ }
\frac{1-\Sigma-\zeta}{2}
+\frac{(2\Sigma-1+\zeta)a^{2}}{4\Omega^{2}}\sin(\theta)^2\ .
\end{align}

\noindent Note that \textcolor{InstituteBlue}{\underline{$\left|\ket{S(t)}\right|^2$}} $+$ \textcolor{InstituteBlue}{\underline{$\left|\ket{T_{0}(t)}\right|^2$}} $+$ \textcolor{InstituteBlue}{\underline{$\left|\ket{T_{+}(t)}\right|^2$}} $+$ \textcolor{InstituteBlue}{\underline{$\left|\ket{T_{-}(t)}\right|^2$}} $ = 1$ for any $t$, as expected. In addition, all full signal expressions are smoothly continuous for $\omega$ going from $0$ to $0^{+}$. Note also that the $\ket{S}$ and $\ket{T_{0}}$ signals differ in sign only in their last term, which we will argue in Section~\ref{sec:analogy} comes from a bright--dark coherence term; its strength is proportional to the initial population imbalance between $\ket{S}$ and $\ket{T_{0}}$, $\Delta$.
\\

\noindent As previously, we might take the limits $\omega \to 0$, $\omega \to \infty$ for all the signals --- the results are displayed in Table~\ref{tab:sig_limits}. Here, the main message is that the magnetic field changes both the active subspace and the dynamics oscillation timescale (as previously, the oscillation period doubles at high fields). In the zero-field limit, all four populations participate in the dynamics. The oscillation amplitudes of the \(\ket S\) and \(\ket{T_0}\) signals are set by the deviation of their initial populations from the equal-sharing value \(\frac{1}{4}\). On the other hand, for \(\zeta\neq0\), the oscillation amplitudes of the \(\ket{T_\pm}\) signals are not determined solely by their own deviation from the equal-sharing value \(\frac{1}{4}\). Rather, they contain an additional correction proportional to the initial population asymmetry in \(\ket{T_+}\) and \(\ket{T_-}\), \(\zeta\). Physically, this reflects the fact that at zero field the two states belong to the same degenerate triplet manifold, so the oscillation of each \(\ket{T_\pm}\) population depends not only on its own initial weight, but also on how the total \(\ket{T_\pm}\) weight is distributed between the two channels. In the special case $\alpha = \gamma$, i.e., $\zeta = 0$, the oscillating amplitudes of the \(\ket{T_\pm}\) signals do reduce to their deviation from $\frac{1}{4}$. In the high-field limit the dynamics collapse to an effective \(\ket{S}\leftrightarrow\ket{T_0}\) problem as expected.
\\

\noindent Results for the long-time averaged signals are found in Table~\ref{tab:sig_ta}. This averaging removes much of the detailed dynamical information and exposes which initial imbalances are truly retained. A discontinuity at zero field is present for the signals $\textcolor{InstituteBlue}{\underline{\left|\ket{S(t)}\right|^2}}$ and $\textcolor{InstituteBlue}{\underline{\left|\ket{T_{0}(t)}\right|^2}}$ whenever $\Delta\neq0$: for $\omega = 0$, these signals still depend on the initial imbalance \(\Delta\). However, for any nonzero field this dependence disappears from their long-time averages, making the zero-field point singular. 
In the high-field limit one finds them equally sharing the population on average. The $\textcolor{InstituteBlue}{\underline{\left|\ket{T_{\pm}(t)}\right|^2}}$ averages continue to retain memory of the initial asymmetry through \(\zeta\), so no discontinuities arise.
\vfill
\newpage
 
\begin{table}[H]
    \centering
\begin{tabular}{|c|c|c|}
\hline
signal & $\lim_{\omega\to0}$ signal & $\lim_{\omega\to\infty}$ signal\\ \hline
$\textcolor{InstituteBlue}{\underline{\left|\ket{S(t)}\right|^2_{
  \alpha  \ket{T_{+}},\;
  \beta  \ket{T_{0}},\;
  \gamma  \ket{T_{-}},\;
  (1-\alpha-\beta-\gamma)\ket{S}\ @\ t=0
}}}$
&
$\frac{\Sigma+\Delta}{2}-\left(\frac{\Sigma+\Delta}{2}-\frac{1}{4}\right)\sin\left(\frac{a t}{2}\right)^2$
&
$\frac{\Sigma+\Delta}{2}-\Delta\sin\left(\frac{a t}{4}\right)^2$
\\ \hline

$\textcolor{InstituteBlue}{\underline{\left|\ket{T_{0}(t)}\right|^2_{
  \alpha  \ket{T_{+}},\;
  \beta  \ket{T_{0}},\;
  \gamma  \ket{T_{-}},\;
  (1-\alpha-\beta-\gamma)\ket{S}\ @\ t=0
}}}$
&
$\frac{\Sigma-\Delta}{2}-\left(\frac{\Sigma-\Delta}{2}-\frac{1}{4}\right)\sin\left(\frac{a t}{2}\right)^2$
&
$\frac{\Sigma-\Delta}{2}+\Delta\sin\left(\frac{a t}{4}\right)^2$
\\ \hline

$\textcolor{InstituteBlue}{\underline{\left|\ket{T_{+}(t)}\right|^2_{
  \alpha  \ket{T_{+}},\;
  \beta  \ket{T_{0}},\;
  \gamma  \ket{T_{-}},\;
  (1-\alpha-\beta-\gamma)\ket{S}\ @\ t=0
}}}$
&
$\frac{1-\Sigma+\zeta}{2}+\left(\frac{2\Sigma-1-\zeta}{4}\right)\sin\left(\frac{a t}{2}\right)^2$
&
$\frac{1-\Sigma+\zeta}{2}$
\\ \hline

$\textcolor{InstituteBlue}{\underline{\left|\ket{T_{-}(t)}\right|^2_{
  \alpha  \ket{T_{+}},\;
  \beta  \ket{T_{0}},\;
  \gamma  \ket{T_{-}},\;
  (1-\alpha-\beta-\gamma)\ket{S}\ @\ t=0
}}}$
&
$\frac{1-\Sigma-\zeta}{2}+\left(\frac{2\Sigma-1+\zeta}{4}\right)\sin\left(\frac{a t}{2}\right)^2$
&
$\frac{1-\Sigma-\zeta}{2}$
\\ \hline
\end{tabular}
   \captionsetup[table]{aboveskip=4pt,belowskip=8pt}
  \captionof{table}{\justifying \textbf{Signal (time-evolution of the population) for the full eight-level system dynamics and its low- and high-magnetic field limits, expressed for different spin observables.} All results are self-consistent. Closed expressions can be found in the accompanying SymPy code. The initial spin states appear in the signals via three parameters: $\Sigma$ and $\Delta$, which measure the total initial weight in the $\{ \ket{S}, \ket{T_0}\}$ sector, and the initial difference between the populations in these states; and $\zeta$, which measures the $\ket{T_{\pm}}$ initial population imbalance. All signals are smoothly continuous for $\omega$ going from $0$ to $0^{+}$.}
    \label{tab:sig_limits}
\end{table}

\begin{table}[H]
    \centering
\begin{tabular}{|c|c|}
\hline
signal & long-time signal average for $\omega = 0 / \omega \neq 0/ \omega \rightarrow \infty$ \\ \hline
$\textcolor{InstituteBlue}{\underline{\left|\ket{S(t)}\right|^2_{
  \alpha  \ket{T_{+}},\;
  \beta  \ket{T_{0}},\;
  \gamma  \ket{T_{-}},\;
  (1-\alpha-\beta-\gamma)\ket{S}\ @\ t=0
}}}$                
&  $\frac{1+2\Sigma+2\Delta}{8} \ / \ \frac{(1+2\Sigma)a^{2}+4\Sigma\omega^{2}}{8\Omega^{2}} \ / \ \frac{\Sigma}{2}$  \\ \hline
$\textcolor{InstituteBlue}{\underline{\left|\ket{T_{0}(t)}\right|^2_{
  \alpha  \ket{T_{+}},\;
  \beta  \ket{T_{0}},\;
  \gamma  \ket{T_{-}},\;
  (1-\alpha-\beta-\gamma)\ket{S}\ @\ t=0
}}}$                 
&  $\frac{1+2\Sigma-2\Delta}{8} \ / \ \frac{(1+2\Sigma)a^2 + 4\Sigma\omega^2}{8\Omega^2} \ / \ \frac{\Sigma}{2}$ \\ \hline
$\textcolor{InstituteBlue}{\underline{\left|\ket{T_{+}(t)}\right|^2_{
  \alpha  \ket{T_{+}},\;
  \beta  \ket{T_{0}},\;
  \gamma  \ket{T_{-}},\;
  (1-\alpha-\beta-\gamma)\ket{S}\ @\ t=0
}}}$
& $\frac{3-2\Sigma+3\zeta}{8} \ / \ \frac{1-\Sigma+\zeta}{2} + \frac{(2\Sigma-1-\zeta)a^2}{8\Omega^2} \ / \ \frac{1-\Sigma+\zeta}{2}$\\ \hline
$\textcolor{InstituteBlue}{\underline{\left|\ket{T_{-}(t)}\right|^2_{
  \alpha  \ket{T_{+}},\;
  \beta  \ket{T_{0}},\;
  \gamma  \ket{T_{-}},\;
  (1-\alpha-\beta-\gamma)\ket{S}\ @\ t=0
}}}$                                      
& $\frac{3-2\Sigma-3\zeta}{8} \ / \ \frac{1-\Sigma-\zeta}{2} + \frac{(2\Sigma-1+\zeta)a^2}{8\Omega^2} \ / \ \frac{1-\Sigma-\zeta}{2}$  \\ \hline
\end{tabular}
   \captionsetup[table]{aboveskip=4pt,belowskip=8pt}
   \captionof{table}{\justifying \textbf{Long-time signal average for the full eight-level system dynamics and its low- and high-magnetic field limits, expressed for different spin observables.} 
All results are self-consistent.
  Closed expressions can be found in the accompanying SymPy code. The model assumes no decoherence, so the average is for $t \to \infty$. 
The initial spin states appear in the curves via three parameters: $\Sigma$ and $\Delta$, which measure the total initial weight in the $\{ \ket{S}, \ket{T_0}\}$ sector, and the initial difference between the populations in these states; and $\zeta$, which measures the $\ket{T_{\pm}}$ initial population imbalance. Long-time averaged signals for the $\ket{S}$ and $\ket{T_{0}}$, but not the $\ket{T_{\pm}}$, observables are discontinuous for $\omega$ going from $0$ to $0^{+}$ if $\Delta \neq 0$.}
    \label{tab:sig_ta}
\end{table}

\vfill
\newpage

\noindent We now turn to the calculation of the coherence terms. These are useful because they reveal the mechanism behind the population signals: they show which states are actually connected by the Hamiltonian, make the bright--dark interference structure explicit (see Section~\ref{sec:analogy}), and identify which terms are responsible for the magnetic field dependence. 
\\

\noindent The nonzero coherences are, for the first subspace sector of the full $8\times8$ density matrix:
\begin{align}
\rho_{ST_0}(t)
=&
-\frac{a^2}{8\Omega^2}(2\Sigma-1-\zeta)\sin(\theta)^2
\nonumber\\
&
-\frac{i\Delta}{4}
\left[
\cos(\theta)\,\sin\!\left(\frac{(a-\omega)t}{2}\right)
+\frac{\omega}{\Omega}\sin(\theta)\,\cos\!\left(\frac{(a-\omega)t}{2}\right)
\right] \ , \label{25}
\\
\rho_{ST_+}(t)
=&\phantom{ \ + \ }
\frac{a}{4\sqrt{2}\,\Omega}\sin(\theta)
\left[
\phantom{-}\Delta \sin\!\left(\frac{(a-\omega)t}{2}\right)
-
\frac{\omega}{\Omega}(2\Sigma-1-\zeta)\sin(\theta)
\right]
\nonumber\\
&
+
\frac{ia}{4\sqrt{2}\,\Omega}\sin(\theta)
\left[
\phantom{-}\Delta \cos\!\left(\frac{(a-\omega)t}{2}\right)
+\phantom{\hspace*{0.12cm}}(2\Sigma-1-\zeta)\cos(\theta)
\right] \  , \label{rhoST+}
\\
\rho_{T_0T_+}(t)
=&-
\frac{a}{4\sqrt{2}\,\Omega}\sin(\theta)
\left[
\phantom{-}\Delta \sin\!\left(\frac{(a-\omega)t}{2}\right)
+
\frac{\omega}{\Omega}(2\Sigma-1-\zeta)\sin(\theta)
\right]
\nonumber\\
&
+
\frac{ia}{4\sqrt{2}\,\Omega}\sin(\theta)
\left[-
\Delta \cos\!\left(\frac{(a-\omega)t}{2}\right)+\phantom{\hspace*{0.12cm}}
(2\Sigma-1-\zeta)\cos(\theta)
\right]  \ ;
\end{align}

\noindent and for the second subspace sector of the full $8\times8$ density matrix:
\begin{align}
\rho_{ST_0}(t)
=&\phantom{ \ + \ }
\frac{a^2}{8\Omega^2}(2\Sigma-1+\zeta)\sin(\theta)^2
\nonumber\\
&
+\frac{i\Delta}{4}
\left[
\cos(\theta)\,\sin\!\left(\frac{(a+\omega)t}{2}\right)
-\frac{\omega}{\Omega}\sin(\theta)\,\cos\!\left(\frac{(a+\omega)t}{2}\right)
\right]\ ,
\\
\rho_{ST_-}(t)
=&-\frac{a}{4\sqrt{2}\,\Omega}\sin(\theta)
\left[
\phantom{-}\Delta \sin\!\left(\frac{(a+\omega)t}{2}\right)
+
\frac{\omega}{\Omega}(2\Sigma-1+\zeta)\sin(\theta)
\right]
\nonumber\\
&
-
\frac{ia}{4\sqrt{2}\,\Omega}\sin(\theta)
\left[
\phantom{-}\Delta \cos\!\left(\frac{(a+\omega)t}{2}\right)
+\phantom{\hspace*{0.12cm}}
(2\Sigma-1+\zeta)\cos(\theta)
\right] \ ,
\\
\rho_{T_0T_-}(t)
=&\phantom{ \ - \ }\frac{a}{4\sqrt{2}\,\Omega}\sin(\theta)
\left[-
\Delta \sin\!\left(\frac{(a+\omega)t}{2}\right) + 
\frac{\omega}{\Omega}(2\Sigma-1+\zeta)\sin(\theta)
\right]
\nonumber\\
&
+
\frac{ia}{4\sqrt{2}\,\Omega}\sin(\theta)
\left[-
\Delta \cos\!\left(\frac{(a+\omega)t}{2}\right) + \phantom{\hspace*{0.12cm}}
(2\Sigma-1+\zeta)\cos(\theta)
\right]\ \label{30}.
\end{align}

\noindent All coherences are smoothly continuous for $\omega$ going from $0$ to $0^{+}$.
\\

\noindent The last term of the coherence $\rho_{ST_{0}}(t)$ will be shown in Section~\ref{sec:analogy} to be a signature of the bright--dark recombination \(\ket{B}\propto \ket{S}+\ket{T_0}\), \(\ket{D}\propto \ket{T_0}-\ket{S}\).
\\

\noindent A note on the real and imaginary parts of the coherences. We show in Supplementary Subsection~\ref{cohexpl} that, for two coupled states, population transfer is controlled by the imaginary part of the corresponding coherence; thus, for example, \(\mathrm{Im}(\rho_{ST_0}(t))\) is the \linebreak component associated with back-and-forth population exchange between \(\ket S\) and \(\ket{T_0}\). By contrast, the real part of the coherence describes phase correlation between the two amplitudes without, by itself, representing an instantaneous population current. 
\\

\noindent Generally, we can thus see that population transfer between $\ket{S}$ and $\ket{T_{0}}$ only occurs for $\Delta \neq 0$; and that population transfer between $\ket{S}$ and $\ket{T_{\pm}}$ depend on both $\Delta \neq 0$ and $(2\Sigma -1 \mp\zeta) \neq 0$. Note that $\Delta + (2\Sigma - 1 - \zeta) = ((1 - \alpha - \beta - \gamma) - \alpha)$ and $\Delta + (2\Sigma - 1 + \zeta) = ((1 - \alpha - \beta - \gamma) - \gamma)$ --- in other words, the population transfer $\ket{S} \leftrightarrow \ket{T_{m}}$ vanishes if $\ket{S}$ and $\ket{T_{m}}$, for $m = \{0, +, - \}$, are initially equally populated. The same is valid for coherences linking two triplet states. Hence, the instantaneous population transfer in a given coherence channel vanishes when the two states involved are initially equally populated, as expected for incoherent initial states. 
\\

\noindent Restricting attention to the first three-state subspace, the relevant population transfer channels between the singlet and triplet manifolds are governed by the imaginary parts of the corresponding coherences. From Eqs.~(\ref{25}) and~(\ref{rhoST+}), these are
\begin{equation}
\Im (\rho_{ST_0}(t))
=
-\frac{\Delta}{4}
\left[
\cos(\theta)\sin\!\left(\frac{(a-\omega)t}{2}\right)
+
\frac{\omega}{\Omega}\sin(\theta)\cos\!\left(\frac{(a-\omega)t}{2}\right)
\right],
\end{equation}
and
\begin{equation}
\Im (\rho_{ST_+}(t))
=
\frac{a}{4\sqrt{2}\,\Omega}\sin(\theta)
\left[
\Delta \cos\!\left(\frac{(a-\omega)t}{2}\right)
+
(2\Sigma-1-\zeta)\cos(\theta)
\right].
\end{equation}
At exactly zero field, these reduce to
\begin{equation}
\Im (\rho_{ST_0}(t))
=
-\frac{\Delta}{8}\sin(at) \ 
\end{equation}
\noindent and
\begin{equation}
\Im (\rho_{ST_+}(t))
=
\frac{(2\Sigma+\Delta-1-\zeta)}{8\sqrt{2}}\sin(at) \ ,
\end{equation}
so both channels already have generically nonzero \(O(1)\) contributions to the population transfer at \(\omega=0\). In particular, at zero field \(\ket{S}\leftrightarrow\ket{T_0}\) transfer is present whenever \(\Delta\neq 0\), while \(\ket{S}\leftrightarrow\ket{T_+}\) transfer is present whenever \((2\Sigma+\Delta-1-\zeta)\neq 0\); note that \((2\Sigma+\Delta-1-\zeta) = 2((1-\alpha -\beta -\gamma) -\alpha)\) is proportional to the difference between the initial populations in $\ket{S}$ and $\ket{T_{+}}$. Thus, in physics parlance, no new Hamiltonian pathway appears when going from \(\omega=0\) to \(\omega=0^{+}\): the \(\ket{S}\leftrightarrow\ket{T_0}\) and \(\ket{S}\leftrightarrow\ket{T_+}\) channels are already dynamically active at zero field. However, for \(\omega\neq 0\), an additional term
\begin{equation}
-\frac{\Delta}{4}\frac{\omega}{\Omega}\sin(\theta)\cos\!\left(\frac{(a-\omega)t}{2}\right)
\end{equation}
\noindent appears only in \(\Im (\rho_{ST_0}(t))\). It vanishes identically at \(\omega=0\) and is linear in \(\omega\) for small fields. There is no analogous additive term in \(\Im (\rho_{ST_+}(t))\); \(\Im (\rho_{ST_+}(t))\)'s field dependence enters only through the replacement \(\Omega=\sqrt{a^2+\omega^2}\), the mixing angle \(\theta=\frac{\Omega t}{2}\), and the phase shift in \(\cos\left(\frac{(a-\omega)t}{2}\right)\).
Thus, while \(\Im(\rho_{ST_+}(t))\) is also modified at small nonzero field, its change arises from perturbing a term already present at \(\omega=0\), rather than from the appearance of a new, explicitly `field-activated' contribution. In this precise sense, the chemistry literature's statement that a `pathway opens' at low field is best understood not as the appearance of a new coupling, but as the emergence in \(\Im(\rho_{ST_0}(t))\) of a distinct additive term $\propto \omega + O(\omega^3)$, which singles out the \(\ket{S}\leftrightarrow\ket{T_0}\) coherence once the exact triplet degeneracy is lifted.
\\

\noindent Note that a small nonzero field can generate nonzero imaginary parts in coherence channels that vanish at exactly zero field for specially chosen initial populations, but this should not be interpreted as the opening of a new Hamiltonian pathway. Rather, it reflects the lifting of a zero-field cancellation that depends on state preparation, while the underlying coupling network remains unchanged. An example of such a particular case is developed in Supplementary Subsection~\ref{particular}.
\\

\noindent The limits \(\omega\to0\) and \(\omega\to\infty\) of all coherences are given in Table~\ref{tab:coh}. The first and last three rows correspond to the first and second subspaces, respectively. This table sharpens the distinction between the two field regimes. In the zero-field limit, coherences are generically nonzero and contain both real and imaginary parts. The imaginary parts indicate ongoing coherent population exchange among the three states within a given subspace, while the real parts show that the dynamics also retain nontrivial interference between amplitudes. For high fields, there is only population transfer between $\ket{S} \leftrightarrow \ket{T_0}$, as expected; the dynamics reduces to that of an effective two-level resonant system: the \(\ket{T_\pm}\) states become both population-trapped and coherence-decoupled. The surviving coherence also oscillates more slowly: whereas the \(\omega\to0\) coherences vary as $\sin(at)$, the \(\omega\to\infty\) coherence varies as \(\sin(\frac{at}{2})\), corresponding to a doubling of the oscillation period. The slowing down by a factor of $2$ in the high-field limit, which has already appeared several times, can be understood as a reduction of the effective mixing scale. In a two-level system, the population oscillation frequency is set by the off-diagonal coupling. Comparing $\sin\!\left(\frac{at}{2}\right)^2$ at zero field with $\sin\!\left(\frac{at}{4}\right)^2$ in the high-field limit shows that the relevant coupling is reduced by a factor of $2$ in the latter regime. 
We will now explicitly derive this factor in the following way. Take for instance the first three-state subspace, where the Hamiltonian $\mathcal{H}_1$ (Eq.~(\ref{H1})) can be written in the bright--dark basis \(\{\ket D,\ket B,\ket{T_+}\}\) to yield $\mathcal{H}_2$ (Eq.~(\ref{H2})); \(\ket{D}\) decouples and the only active off-diagonal matrix element is \(\langle B|\mathcal{H}_2|T_+\rangle=\frac{a}{2}\). In the high-field limit, the \(\ket{T_\pm}\) states are Zeeman-detuned and the dynamics reduces to the direct \(\ket S\leftrightarrow\ket{T_0}\) channel, with matrix element \(\langle S|\mathcal{H}_1|T_0\rangle=- \frac{a}{4}\). The effective mixing scale is therefore reduced from \(\frac{a}{2}\) to \(\frac{a}{4}\), which explains the calculated slowdown in the oscillation period.
\\

\noindent The long-time averaged values of the coherences are displayed in Table~\ref{tab:coh_ta}. The first and last three rows correspond to the first and second subspaces, respectively. A nonzero real-valued long-time averaged coherence reflects not persistent population transfer, but persistent phase correlation in the \(\{\ket S,\ket{T_m}\}\) basis: it indicates that the long-time averaged state is not diagonal in that basis and retains a stationary coherent admixture of the \linebreak corresponding states (a `memory' that they were coherently linked). In this sense, the long-time averages probe which interference terms do not dephase away under the unitary dynamics, or equivalently, which contributions survive the \(t\to\infty\) averaging procedure. In addition, these averages reveal two qualitatively different field dependences. The \linebreak coherences between \(\ket S\leftrightarrow\ket{T_0}\) scale as \(\frac{a^2}{\Omega^2}=\frac{a^2}{a^2+\omega^2}\), so they are continuous through \(\omega=0\), largest at zero field, and monotonically suppressed as the field increases. In turn, the coherences involving \(\ket{T_\pm}\) are discontinuous at zero field: at exactly \(\omega=0\) they retain finite values proportional to the initial \(\{\ket S,\ket{T_0}\}\) imbalance \(\Delta\), whereas for any nonzero field they scale as \(\frac{a\omega}{\Omega^2}=\frac{a\omega}{a^2+\omega^2}\), and therefore vanish both as \(\omega\to0^+\) and as \(\omega\to\infty\). Hence the \(\ket{T_\pm}\)-related long-time coherences are not only discontinuous at \(\omega=0\), but also non-monotonic in field strength, reaching their largest values at intermediate fields \(\omega\sim a\) (indeed, \(\frac{d}{d\omega}\frac{a\omega}{a^2+\omega^2}=0\) at \(\omega=a\)). The discontinuity at \(\omega=0\) reflects the singular nature of the long-time average at exact triplet degeneracy rather than sustained population current. At zero field, the triplet manifold is exactly degenerate, so certain contributions to the \(\ket{T_\pm}\) coherences are non-oscillatory and survive the long-time average. For any nonzero field, however small, the Zeeman splitting lifts this degeneracy, these terms acquire relative phases oscillating at frequencies \(\sim \frac{|a\pm\omega\pm\Omega|}{2}\), and their infinite-time averages vanish. Thus the discontinuity is not an abrupt change in the finite-time dynamics, but a consequence of degeneracy lifting combined with the \(t\to\infty\) averaging procedure.
\\

\begin{figure}[H]
  \centering
  \includegraphics[width=\textwidth]{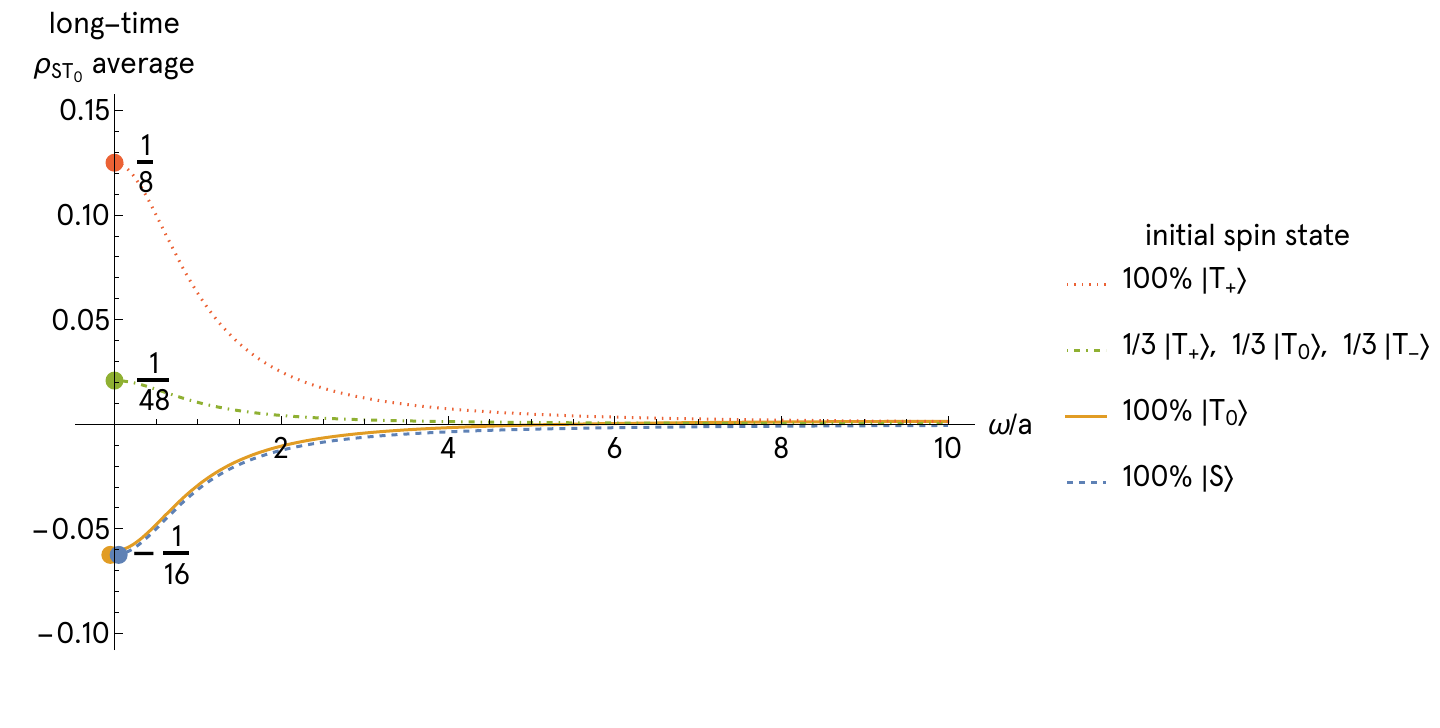}
  \captionsetup[figure]{aboveskip=4pt,belowskip=8pt}
  \captionof{figure}{\justifying \textbf{Long-time average of coherence $\rho_{ST_0}$ expressed for different initial spin polarizations, as a function of field strength.}
The coherence is plotted as a function of the dimensionless parameter \(\frac{\omega}{a}\), for the same representative initial spin states used in the previous figures. Unlike coherences involving \(\ket{T_{\pm}}\), this averaged coherence is continuous through \(\omega=0\), is largest at zero field, and is monotonically suppressed as the field increases. 
Physically, a monotonically decreasing coherence indicates that the applied field steadily suppresses the stationary phase correlation associated with the effective $\ket{S} \leftrightarrow \ket{T_0}$ channel.}
   \label{P2}
  \end{figure}

\begin{figure}[H]
  \centering
  \includegraphics[width=\textwidth]{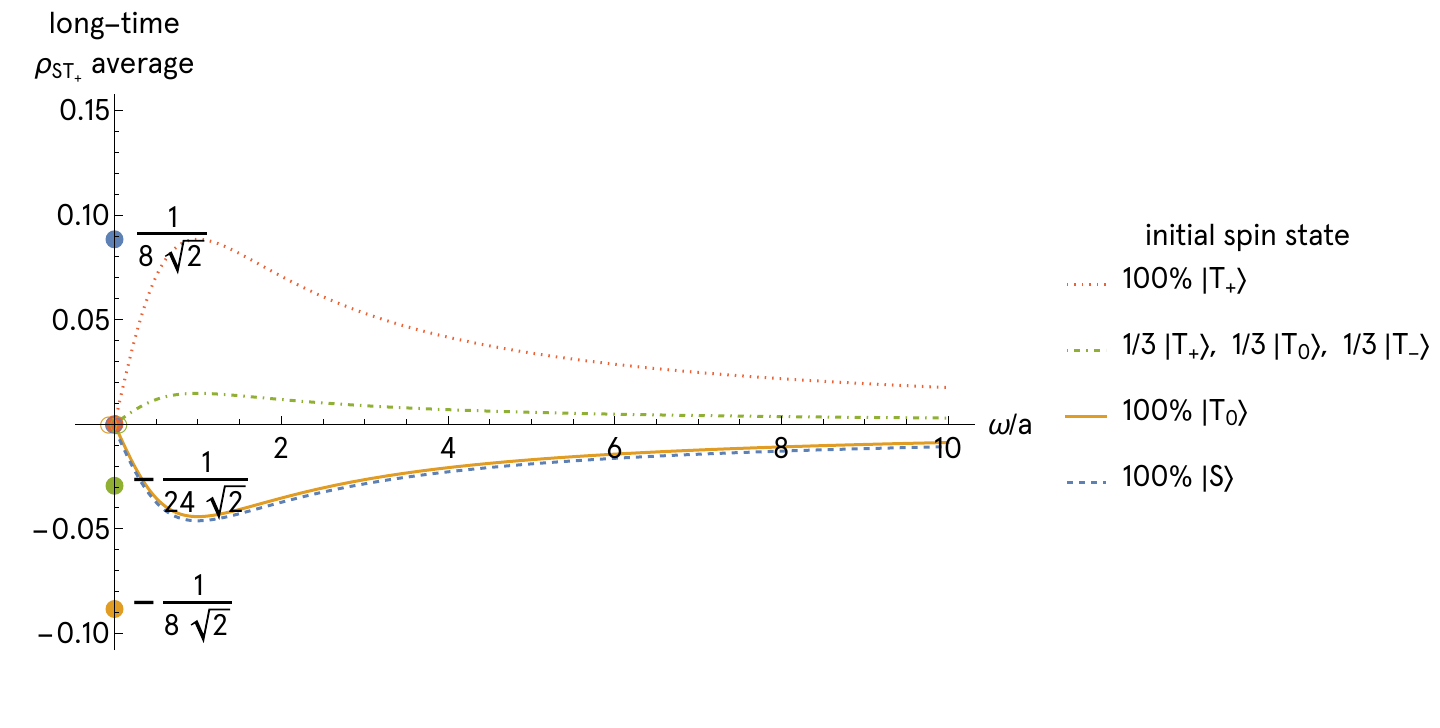}
  \captionsetup[figure]{aboveskip=4pt,belowskip=8pt}
  \captionof{figure}{\justifying \textbf{Long-time average of coherence $\rho_{ST_+}$ expressed for different initial spin polarizations, as a function of field strength.}
The coherence is plotted as a function of the dimensionless parameter \(\frac{\omega}{a}\), for the same representative initial spin states used in the previous figures. This averaged coherence is singular at \(\omega=0\) for all cases with $\Delta \neq 0$: at exact zero field it can retain a finite value which, for any \(\omega\neq 0\), is suppressed by the lifting of the triplet degeneracy. As a result, it vanishes both as \(\omega\to 0^{+}\) and as \(\omega\to\infty\), and is largest only at intermediate fields \(\omega\sim a\). Physically, a non-monotonic behavior indicates that this coherence is only a low- and intermediate-field interference remnant tied to exact zero-field triplet degeneracy, rather than a stationary coherence that survives the crossover to the high-field regime.}
   \label{P3}
  \end{figure}

\noindent Figs.~\ref{P2} and~\ref{P3} show the long-time averages of representative coherences $\rho_{ST_0}$ and $\rho_{ST_+}$ as functions of the dimensionless field strength \(\frac{\omega}{a}\). The two figures illustrate the two qualitatively distinct field dependences found analytically. 
In this way, Figs.~\ref{P2} and~\ref{P3} make visible the difference between a coherence that is continuously weakened by the field and one whose long-time average survives only because of exact zero-field degeneracy.
\vfill
\newpage 
\begin{table}[H]
    \centering
\begin{tabular}{|c|c|c|}
\hline
coherence & $\lim_{\omega\to0}$ coherence & $\lim_{\omega\to\infty}$ coherence\\ \hline
$\rho_{ST_0}(t)$ & $-\frac{(2\Sigma-1-\zeta)}{8}\sin^2\!\left(\frac{at}{2}\right)
-\frac{i\Delta}{8}\sin(at)$ & $-\frac{i\Delta}{4}\sin\!\left(\frac{at}{2}\right)$\\ \hline
$\rho_{ST_+}(t)$  &  $\frac{\Delta}{4\sqrt2}\sin^2\!\left(\frac{at}{2}\right)
+\frac{i(2\Sigma+\Delta-\zeta-1)}{8\sqrt2}\sin(at)$& $0$ \\ \hline
$\rho_{T_0T_+}(t)$  & $-\frac{\Delta}{4\sqrt2}\sin^2\!\left(\frac{at}{2}\right)
+\frac{i(2\Sigma-\Delta-\zeta-1)}{8\sqrt2}\sin(at)$ & $0$\\ \hline
$\rho_{ST_0}(t)$  & $\frac{(2\Sigma-1+\zeta)}{8}\sin^2\!\left(\frac{at}{2}\right)
+\frac{i\Delta}{8}\sin(at)$ & $ \frac{i\Delta}{4}\sin\!\left(\frac{at}{2}\right)  $\\ \hline
$\rho_{ST_-}(t)$  & $-\frac{\Delta}{4\sqrt2}\sin^2\!\left(\frac{at}{2}\right)
-\frac{i(2\Sigma+\Delta+\zeta-1)}{8\sqrt2}\sin(at)$& $0$\\ \hline
$\rho_{T_0T_-}(t)$  & $-\frac{\Delta}{4\sqrt2}\sin^2\!\left(\frac{at}{2}\right)
+\frac{i(2\Sigma-\Delta+\zeta-1)}{8\sqrt2}\sin(at)$& $  0 $ \\ \hline
\end{tabular}
   \captionsetup[table]{aboveskip=4pt,belowskip=8pt}
 \captionof{table}{\justifying \textbf{Coherences (time-evolution of the off-diagonal terms in the density matrix) for the full eight-level system dynamics and its low- and high-magnetic field limits.} Closed expressions can be found for all the derived coherences in the accompanying SymPy code. The first (last) three lines are for the first (second) three-state subspace. The initial spin states appear in the coherences via three parameters: $\Sigma$ and $\Delta$, which measure the total initial weight in the $\{ \ket{S}, \ket{T_0}\}$ sector, and the initial difference between the populations in these states; and $\zeta$, which measures the $\ket{T_{\pm}}$ initial population imbalance. All coherences are smoothly continuous for $\omega$ going from $0$ to $0^{+}$.}
    \label{tab:coh}
\end{table}

\begin{table}[H]
    \centering
\begin{tabular}{|c|c|}
\hline
coherence & long-time coherence average for $\omega = 0/\omega \neq 0/\omega \to \infty$\\ \hline
$\rho_{ST_0}(t)$ & $-\frac{(2\Sigma-1-\zeta)}{16}/ -\frac{(2\Sigma-1-\zeta)a^2}{16\Omega^2} / 0 $\\ \hline
$\rho_{ST_+}(t)$   &  $ \frac{\Delta}{8\sqrt2} / -\frac{(2\Sigma-1-\zeta)a\omega}{8\sqrt2\,\Omega^2} / 0 $\\ \hline
$\rho_{T_0T_+}(t)$   & $-\frac{\Delta}{8\sqrt2}/-\frac{(2\Sigma-1-\zeta)a\omega}{8\sqrt2\,\Omega^2} /0 $\\ \hline
$\rho_{ST_0}(t)$   &  $ \frac{(2\Sigma-1+\zeta)}{16}  / \frac{(2\Sigma-1+\zeta)a^2}{16\Omega^2}/ 0 $\\ \hline
$\rho_{ST_-}(t)$   & $ -\frac{\Delta}{8\sqrt2} / -\frac{(2\Sigma-1+\zeta)a\omega}{8\sqrt2\,\Omega^2}  / 0 $\\ \hline
$\rho_{T_0T_-}(t)$   & $ -\frac{\Delta}{8\sqrt2}  / \frac{(2\Sigma-1+\zeta)a\omega}{8\sqrt2\,\Omega^2}  / 0  $\\ \hline
\end{tabular}
   \captionsetup[table]{aboveskip=4pt,belowskip=8pt}
 \captionof{table}{\justifying \textbf{Long-time coherence average for the full eight-level system dynamics and its low- and high-magnetic field limits.} The first (last) three lines are for the first (second) three-state subspace. This assumes no decoherence, so that the average is for $t \to \infty$. The initial spin states appear in the coherences via three parameters: $\Sigma$ and $\Delta$, which measure the total initial weight in the $\{ \ket{S}, \ket{T_0}\}$ sector, and the initial difference between the populations in these states; and $\zeta$, which measures the $\ket{T_{\pm}}$ initial population imbalance. Long-time averaged coherences involving $\ket{T_{\pm}}$, but not the ones between $\ket{S} \leftrightarrow \ket{T_{0}}$, are discontinuous for $\omega$ going from $0$ to $0^{+}$ if $\Delta \neq 0$.}
    \label{tab:coh_ta}
\end{table}

\vfill
\newpage

\vfill
\newpage

\vfill
\newpage

\noindent In summary, the preceding results help clarify the often-invoked statement in the literature that a `new pathway opens' when going from \(\omega=0\) to \(\omega=0^+\). As our explicit population and coherence formulas show, this statement is ambiguous unless one specifies what is meant by `pathway'. If one means a bare Hamiltonian coupling, then no new pathway opens at \(\omega=0^+\): the coupling structure of the model is fixed. If instead one means dynamical participation, then all three $\ket{S}\leftrightarrow\ket{T_m}$ channels are already active at \(\omega=0\), as seen from their nontrivial population dynamics and the corresponding coherences. Moreover, whether a given channel carries instantaneous population transfer depends not only on the Hamiltonian, but also on the initial spin populations, since the imaginary part of a coherence vanishes whenever the two states involved are initially equally populated. This means that analyses based on particular initial states can make a channel appear effectively absent, even though the corresponding coupling is present and would become dynamically active for a different preparation. 
\\

\noindent What changes singularly at \(\omega=0^+\) is therefore not the appearance of a new coupling, but the lifting of the exact zero-field triplet degeneracy and the consequent loss of the stationary, `phase-locked' interference term that exists only at \(\omega=0\) (see Section~\ref{sec:analogy}). 
In addition, a qualitatively distinct contribution unique to the \(\rho_{ST_0}(t)\) coherence does emerge for \(\omega>0\): its imaginary part contains a term that vanishes at \(\omega=0\) and is linear in \(\omega\) for small fields. We believe this is the feature often described phenomenologically as an `opened pathway'. More precisely, once the triplet degeneracy is lifted, \(\ket{T_0}\) is dynamically singled out \linebreak relative to \(\ket{T_\pm}\) and increased \(\ket{S}\leftrightarrow\ket{T_0}\) mixing follows.
\\





\newpage
\begin{center}
\sectionfont{Take-home messages}
\end{center}

\begin{enumerate}
 \item The Hamiltonian can be decomposed into a `bright' and a `dark' sector; in this lossless model, the dark sector is exactly symmetry-protected and decoupled from the spin-\linebreak mixing channel, so any population initially prepared in it remains trapped and does not participate in the dynamics.
  \item The experimentally relevant `signal', namely the singlet population as a function of time, is obtained by recombining the bright and dark components.
It has a DC component, a component oscillating with $\Omega$, and a component oscillating at frequencies $\frac{|a \pm \omega \pm \Omega|}{2}$.
   The signal is calculated in full, but also at the relevant limits given by $\omega \to 0$, $\omega \to \infty$, and $t \to \infty$. 
 \item We explicitly present all the terms of the density matrix, for different initial spin preparations, in full but also at the relevant limits given by $\omega \to 0$, $\omega \to \infty$, and $t \to \infty$. The population transfer pathways are determined by the imaginary parts of the relevant coherences, with amplitudes set by the initial spin preparation.
\item Most features of the dynamics seem to depend heavily on two quantities: the difference between the initial populations in $\ket{S}$ and $\ket{T_0}$, $\Delta$, and their sum, $\Sigma$.
\item In this lossless toy model, at zero field, the special role of \(\omega=0\) is an interference effect associated with exact triplet degeneracy. 
At $\omega = 0$, exact triplet degeneracy makes some bright--dark interference frequency components, for example the one $\sim e^{-i\frac{(a - \omega - \Omega)t}{2}} = e^{-i\frac{(a - 0 - a)t}{2}} = e^{0} = 1$, collapse to zero frequency, leaving a stationary DC contribution. For any $\omega > 0$, the degeneracy is lifted and that contribution acquires nonzero relative phases
$\sim e^{-i\frac{|a \pm \omega \pm \Omega|t}{2}}$, so it becomes purely oscillatory and averages away over long times.
In that regime, the \(\ket S\leftrightarrow\ket{T_0}\) coherence provides the dominant contribution to the population transfer through a term linear in \(\omega\) at low fields. This gives a precise interpretation of the chemistry literature's common statement that a `pathway opens up' once a nonzero field is applied.

 
 \end{enumerate}

\clearpage
\section{\seclq Bright-dark\secrq \ interpretation of the Hamiltonian}
\label{sec:analogy}

\noindent In this section, we explore the analogy between the Hamiltonian considered here and Hamiltonians that arise in another area of physics. To our knowledge, the bright--dark \linebreak perspective that we will uncover has only once been previously emphasized in the radical-pair literature~\cite{Xu2016}. Framing the dynamics in a language familiar from electromagnetically-induced transparency~\cite{Fleischhauer2005EIT} and related interference phenomena offers a genuinely new way of looking at radical-pair physics: rather than viewing the model solely as a spin-chemistry problem, one can interpret it as an interference problem with bright and dark sectors, destructive interference, and coherence transfer. For details on this analogy, see Supplementary Subsection~\ref{subsec:EIT}. This viewpoint brings into a common framework several key features of the model that were hitherto left disconnected, including: the emergence of a decoupled dark sector; the importance of bright--dark coherence; the low-field effect and the singular nature of the zero-field point; and the crossover at high field to an effective \(\ket{S}\leftrightarrow\ket{T_0}\) description.
\\

\noindent We showed in Section~\ref{sec:solution} that the Hamiltonian for the first subspace $\mathcal{H}_2$ (Eq.~(\ref{H2}); also the Hamiltonian for the second subspace, $\mathcal{H}_6$, found in Supplementary Section~\ref{explanation}, \linebreak Suppl.~Eq.~(\ref{eq:Hdown2})) is simple in a basis $\{\ket{D}, \ket{B}, \ket{T_{+}}\}$ that contains the dark and bright states. In turn, our signal, or experimental readout, is the singlet population $\left|\ket{S(t)}\right|^2$ obtained using the projector $|S\rangle\langle S|$, which is not diagonal in that basis:
\begin{equation}
    \ket{S}=\frac{\ket{B}-\ket{D}}{\sqrt2} \Rightarrow |S\rangle\langle S| = \frac12\Big(|B\rangle\langle B|+|D\rangle\langle D|-|B\rangle\langle D|-|D\rangle\langle B|\Big) \ . 
\end{equation}

\noindent In that basis, the signal is immediately
\begin{equation}
\left|\ket{S(t)}\right|^2 
= \frac12\Big(\left|\ket{B(t)}\right|^2+\left|\ket{D(t)}\right|^2\Big)- \text{Re}\big[\rho_{BD}(t)\big] \ .
\label{sigdb}
\end{equation}

\noindent The instantaneous singlet population is the average of the bright and dark populations, minus the real part of their interference term, the density matrix coherence element $\rho_{BD}(t)$. Thus, it receives contributions from: the bright state population, which mixes with $\ket{T_{+}}$ and hence oscillates in time; the dark state population, which is constant in time; and also from the bright--dark coherence term, which encodes the relative phase between the two pathways that recombine into $\ket{S}$. This is why $\ket{S}$ and $\ket{T_{0}}$, although they have the same initial bright and dark populations, produce different signals: they differ only by the sign of the bright--dark coherence (as was explicitly mentioned in Section~\ref{sec:solution}). 
\\

\noindent Calculating the expression for $\rho_{BD}(t)$, we obtain for the first subspace:
\begin{equation}
    \text{Re}\big[\rho_{BD}(t)\big] = \frac{(\beta'-(1 - \alpha' -\beta'))}{2}
\left[
\cos(\theta)\cos\left(\frac{(a-\omega)t}{2}\right)
-
\frac{\omega}{\Omega}\sin(\theta)\sin\left(\frac{(a-\omega)t}{2}\right)
\right] \ ,
\label{cohsub}
\end{equation}

\noindent where the initial populations within that subspace are $\alpha'  \ket{T_{+}}$, $\beta' \ket{T_{0}}$, and $(1-\alpha'-\beta')\ket{S}$. Note that the strength of the interference term is proportional to $(\beta'-(1 - \alpha' -\beta'))$ -- namely, to the difference between the initial populations in $\ket{T_{0}}$ and $\ket{S}$. 
\\

\noindent Some limits of the above expression for the coherence merit being considered:
\begin{align}
    \lim_{\omega \to 0} \text{Re}\big[\rho_{BD}(t)\big] &=  \frac{(\beta'-(1 - \alpha' -\beta'))}{4}\big(1 + \cos(at)\big)   \ , \label{27} \\
    \lim_{\omega \to \infty} \text{Re}\big[\rho_{BD}(t)\big] &=  \frac{(\beta'-(1 - \alpha' -\beta'))}{2}\cos\left(\frac{a t}{2}\right)  \ . 
\end{align}

\noindent At exactly \(\omega=0\), the bright--dark coherence contribution contains not only an oscillatory part but also a genuine DC term of strength \(\frac{\Delta}{4}\), with $\Delta = ((1 - \alpha' -\beta') - \beta')$, which shifts the baseline of the singlet signal and survives time averaging.
\\

\noindent We can also take the limit of $\text{Re}\big[\rho_{BD}(t)\big]$ for the long time-average. This yields:
\begin{align}
\langle \text{Re}\big[\rho_{BD}(t)\big] \rangle_{t \to \infty} &= \frac{(\beta'-(1 - \alpha' -\beta'))}{4} \ , \quad \omega = 0 \ , \\
    \langle \text{Re}\big[\rho_{BD}(t)\big] \rangle_{t \to \infty} &= 0 \ , \quad \omega \neq 0  \ .
    \label{new37}
\end{align}

\noindent Perhaps the most important insight is that \(\mathrm{Re}[\rho_{BD}(t)]\) changes not only in magnitude but in qualitative character at \(\omega=0\). At exactly zero field, the bright--dark coherence is sign-definite up to the sign of \((\beta'-(1-\alpha'-\beta'))\), contains a genuine DC component, and never changes sign during the oscillation: it is effectively a rectified, `phase-locked' interference term. For any \(\omega\neq 0\), in turn, this stationary part is lost and the coherence becomes purely oscillatory with zero long-time average; in the high-field limit it also oscillates more slowly, with doubled period, similarly to what was already observed in Section~\ref{sec:solution}. In this sense, a magnetic field converts the bright--dark coherence from a DC+AC interference term at \(\omega=0\) into a purely AC coherence for any nonzero field. Equivalently, the long-time average of \(\mathrm{Re}[\rho_{BD}(t)]\) is singular at \(\omega=0\): an arbitrarily small field destroys the stationary interference contribution. 
Moreover, \(\mathrm{Re}[\rho_{BD}(t)]\) vanishes identically when \(\beta' = (1-\alpha'-\beta')\), showing that it is controlled entirely by the initial population imbalance between \(\ket{S}\) and \(\ket{T_0}\). 
\\

\noindent For the eight-level dynamics, we show explicitly in Supplementary Section~\ref{sigcontr} that the full signal can be decomposed into different contributions:
\begin{equation}
\begin{split}
\textcolor{InstituteBlue}{%
\underline{\left|\ket{S(t)}\right|^2_{\substack{
\alpha \ket{T_{+}},\;
\beta \ket{T_{0}},\;
\gamma \ket{T_{-}},\\
(1-\alpha-\beta-\gamma)\ket{S}\ @\ t=0
}}}}
={}&
\underbrace{\frac{\Sigma}{4}}_{\text{dark-state population}}
\\
&\quad+
\underbrace{
\left[
\frac{\Sigma}{4}
-\frac{(2\Sigma-1)a^2}{4\Omega^2}\sin(\theta)^2
\right]
}_{\text{bright-state population}}
\\
&\quad+
\underbrace{
\frac{\Delta}{2}
\cos\!\left(\frac{at}{2}\right)
\left[
\cos(\theta) \cos\!\left(\frac{\omega t}{2}\right)
+\frac{\omega}{\Omega}\sin(\theta) \sin\!\left(\frac{\omega t}{2}\right)
\right]
}_{\text{bright--dark coherence}}
\ .
\end{split}
\label{full}
\end{equation}

\noindent As expected, the magnitude of the bright--dark coherence term is proportional to the initial population difference \(((1-\alpha-\beta-\gamma)-\beta)=\Delta\) between \(\ket{S}\) and \(\ket{T_0}\). In turn, the oscillatory part of the bright-state population contribution depends on the total initial population in the \(\{\ket{S},\ket{T_0}\}\) manifold, namely \(\Sigma\). Thus the decomposition cleanly separates two kinds of information about the initial state: \(\Sigma\) controls how much weight is available in the active bright sector and how much is trapped in the dark sector, whereas \(\Delta\) controls the interference between bright and dark components. In Section~\ref{sec:MFE}, we will show that the initial imbalance in \(\{\ket{S},\ket{T_0}\}\) controls a coherence-driven low-field structure, whereas the total weight in that manifold controls the amount of actual population redistribution generated by the dynamics and the high-field plateau of the signal and of the yield curves.
\\

\noindent Also noteworthy is that the bright--dark coherence contribution to the signal is resolved into sidebands at angular frequencies \(\frac{|a\pm\omega\pm\Omega|}{2}\), arising from interference between the dark state and the two states that compose the bright block. In turn, the bright population contribution oscillates at angular frequency \(\Omega\), reflecting the beating between the two eigenstates within the bright block. Both the signal and the singlet yields (see Section~\ref{sec:MFE} for the latter) therefore contain oscillations originating from these two distinct mechanisms.
\\

\noindent Moreover, the bright--dark viewpoint is most useful in the low-field, unresolved-triplet regime, and the crossover to high field can be understood as the breakdown of this collective mixing picture. In the low-field regime, the singlet couples not to a single triplet eigenstate but to a bright superposition within the triplet manifold, while the orthogonal dark \linebreak combination is decoupled from the spin-mixing Hamiltonian. The mixing dynamics are therefore governed by the bright block (although the reconstructed singlet signal can still contain interference between bright and dark contributions, see Eq.~(\ref{full})). As the magnetic field increases, the Zeeman splitting resolves the triplet manifold: the \(\ket{T_\pm}\) states are shifted off resonance and effectively decouple from the singlet dynamics, leaving an effective \(\ket{S}\leftrightarrow\ket{T_0}\) mixing channel in the high-field limit.
\\

\noindent Finally, the considered simple toy model, in the absence of loss, yields an idealized dark state sector that is completely isolated from the spin-mixing dynamics. In a more realistic radical-pair model, decoherence and recombination, anisotropic hyperfine terms, exchange and dipolar couplings will turn the dark state into only `an approximate dark state'. 
\vfill

\newpage
\begin{center}
\sectionfont{Take-home messages}
\end{center}
\begin{enumerate}
    \item We uncover the origin of the signal's three distinct contributions: the dark-sector contribution is time-independent, the bright-sector contribution oscillates at \linebreak frequency \(\Omega\), and the bright--dark coherence contribution gives rise to sidebands at frequencies \(\frac{|a\pm\omega\pm\Omega|}{2}\).
    \item The bright--dark coherence behaves qualitatively differently at \(\omega=0\): exactly at zero field it contains a constant, `phase-locked' component, whereas for any \(\omega\neq 0\) it becomes purely oscillatory. In this sense, \(\omega=0\) is a singularity.
    \item The strength of the bright--dark coherence contribution depends directly on \(\Delta\), the initial population difference between \(\ket S\) and \(\ket{T_0}\). Because this coherence changes character at \(\omega=0\), \(\Delta\) controls the low-field structure of the signal.
    \item The bright- and dark-sector population contributions depend directly on \(\Sigma\), the initial total population in the \(\{\ket S,\ket{T_0}\}\) manifold. Accordingly, \(\Sigma\) controls the high-field and long-time structure of the signal, including for example the plateau value in the high-field limit.
\end{enumerate}
\vfill
\clearpage
\section{Solution for the singlet yield}
\label{sec:MFE}

\noindent Historically, the radical pair mechanism has been probed experimentally in setups in which electron spins are not initialized simultaneously into the coherent dynamics, but are instead generated continuously, and in which singlet population yields are averaged over timescales far exceeding the coherence time of the spin dynamics --- i.e., in what a physicist would describe as a `steady-state regime'. Plots of this steady-state integrated singlet population, measured as a function of magnetic field strength, for continuously generated radical pairs rather than a single synchronized initial ensemble, are referred to in the literature as `magnetic field effect curves' for the singlet yield. `Singlet yield' and `magnetic field effect curve' will thus be used interchangeably in what follows.
\\

\noindent All the symbolic calculations for the singlet yield expressions can be followed along using the accompanying SymPy code \href{https://bit.ly/SingletYield_SymPy}{\texttt{SingletYield\_SymPy.py}}.
\\

\noindent The quantity in the present toy model that most naturally corresponds to the experimentally measured magnetic field effect curves is thus not the instantaneous singlet population (our signal $\textcolor{InstituteBlue}{\underline{\left|\ket{S(t)}\right|^2}}$), but rather the singlet yield, obtained, for example, using a normalized finite-time integral,
\begin{equation}
    \langle \textcolor{InstituteBlue}{\underline{\left|\ket{S(t,\omega)}\right|^2}}\rangle_T \equiv \frac{1}{T}\int_0^T \textcolor{InstituteBlue}{\underline{\left|\ket{S(t)}\right|^2}} \ dt \ , 
\end{equation}
or an exponentially-weighted integral taking into account $k$, a radical pair recombination rate, 
\begin{equation}
\Phi_k(\omega)\equiv k\int_0^\infty e^{-kt}\, \textcolor{InstituteBlue}{\underline{\left|\ket{S(t)}\right|^2}} \ dt \ .
\label{eq:weighted_yield_def}
\end{equation}

\noindent The first yield represents the time-averaged singlet population over an observation window of duration $T$;  $\langle \textcolor{InstituteBlue}{\underline{\left|\ket{S(t,\omega)}\right|^2}}\rangle_T$ should be understood as a function of $\omega$ for fixed $T$.
This average weights all times equally between $0$ and $T$ and corresponds to a simplified scenario with continuous radical-pair production and a uniform age distribution on the interval \([0,T]\); equivalently, it describes a no-loss ensemble observed over a finite time window, or a continuously replenished ensemble in which each radical pair is removed after a fixed maximum lifetime \(T\). In this sense, the second expression for the yield is more standard in that times are not weighted equally: if radical pair recombination occurs, late-time dynamics contribute less because fewer radical pairs remain. Note that Eq.~(\ref{eq:weighted_yield_def}) is exactly like a spin-independent survival process for the radical pair as a whole, with mean survival time $k^{-1}$; it is not a singlet-selective or triplet-selective recombination model. In this sense, $k$ is not a proper decay rate inside the spin dynamics but a weighting parameter in the yield definition, prescribing how it emphasizes early times versus late times, regardless of spin state.
Similarly to the first yield, the lifetime-weighted yield should then be understood as a function of \(\omega\) for fixed recombination rate \(k\), so that the magnetic field dependence is examined under an exponential survival-time distribution set by the radical pair lifetime \(k^{-1}\). However, neither of these yields is intended to replace a full Haberkorn approach~\cite{Fay2018SpinSelective}. In that formalism, spin-selective recombination is built into the master equation itself, leading to a time-dependent loss of radical-pair population and yields defined through the recombination dynamics. 
\\

\noindent The expressions for the analytical integrations $ \langle \textcolor{InstituteBlue}{\underline{\left|\ket{S(t, \omega)}\right|^2}}\rangle_T$ (henceforth `finite-$T$ average singlet yield') for different spin initializations are given in Table~\ref{tab:mfe_T}, where
\begin{equation}
\begin{split}
\mathcal K(T)
\equiv{}&
\phantom{ \ + \ }\frac{1}{4}\left(1+\frac{\omega}{\Omega}\right)
\left(
\text{sinc}\!\left(\frac{(a+\omega-\Omega)T}{2}\right)
+
\text{sinc}\!\left(\frac{(a-\omega+\Omega)T}{2}\right)
\right)
\\
&+
\frac{1}{4}\left(1-\frac{\omega}{\Omega}\right)
\left(
\text{sinc}\!\left(\frac{(a-\omega-\Omega)T}{2}\right)
+
\text{sinc}\!\left(\frac{(a+\omega+\Omega)T}{2}\right)
\right) \ ;
\end{split}
\label{Kappaterms}
\end{equation}

\noindent in turn, the corresponding values for $\Phi_k(\omega)$ (henceforth `exponentially-weighted singlet yield') are given in Table~\ref{tab:mfe_k}, with
\begin{equation}
\begin{split}
K(\omega)
\equiv{}&\phantom{ \ + \ }
\frac14\left(1+\frac{\omega}{\Omega}\right)
\Big[
\Lambda_k(a+\omega-\Omega)+\Lambda_k(a-\omega+\Omega)
\Big]
\\
&+
\frac14\left(1-\frac{\omega}{\Omega}\right)
\Big[
\Lambda_k(a-\omega-\Omega)+\Lambda_k(a+\omega+\Omega)
\Big]\ ,
\end{split}
\label{eq:Kk_def}
\end{equation}
\noindent where we introduced
\begin{equation}
    \Lambda_k(\lambda)\equiv \frac{4k^2}{4k^2+\lambda^2} \ .
\end{equation}

\noindent These expressions define our toy model's singlet yield curves. All singlet yields are smooth going from $\omega = 0$ to $0^{+}$.
\\

\noindent For a generic initial state for the spins, the coefficient accompanying \(\mathcal K(T)\) and \(K(\omega)\) in the singlet yield expressions is proportional to the difference between the initial $\ket{S}$ and $\ket{T_{0}}$ populations, $\Delta$. This term is the contribution to the singlet yield that originates from the bright--dark coherence term. Any low-field feature is thus controlled entirely by the initial \(\ket{S}\leftrightarrow\ket{T_0}\) imbalance. By contrast, \(\Sigma\) governs the smooth population background through the bright- and dark-sector contributions. Indeed, the dark-state contribution to the singlet yield is $\frac{\Sigma}{4}$, while the bright-state contribution is
\begin{align}
 \frac{\Sigma}{4}
-\frac{(2\Sigma-1)a^2}{8\Omega^2} + \frac{(2\Sigma-1)a^2}{8\Omega^3 T}\sin(\Omega T) \quad &\text{ for }   \langle \textcolor{InstituteBlue}{\underline{\left|\ket{S(t,\omega)}\right|^2}}\rangle_T; \\
\frac{\Sigma}{4} - \frac{(2\Sigma-1)a^2}{8(k^2+\Omega^2)} \quad &\text{ for } \Phi_k(\omega) \ .
\end{align}

\noindent Note that the limit \(T\to\infty\) is analogous to \(k\to0\). For any fixed \(\omega\neq0\), the correction terms proportional to \(\mathcal{K}(T)\) and \(K(\omega)\) vanish as \(T\to\infty\) and \(k\to0\), respectively, so the normalized integrated signals of Tables~\ref{tab:mfe_T} and~\ref{tab:mfe_k} reduce to the corresponding \(\omega\neq0\) long-time averages of Table~\ref{Table2}. At exactly \(\omega=0\), however, \(\mathcal{K}(T)\to\tfrac12\) as \(T\to\infty\), and \(K(\omega)\to\tfrac12\) as \(k\to0\), thereby reproducing instead the zero-field entries of Table~\ref{Table2}. The surviving constant factor at exactly \(\omega=0\) reflects the fact that the limits \(T\to\infty\) (\(k\to0\)) and \(\omega\to0\) do not commute.
Thus, in this toy model, the finite-$T$ average and the exponentially-weighted singlet yields can be viewed as two different ways to get rid of the $\omega = 0$ discontinuity in the long-time averaged signals of Table~\ref{Table2}: a multiple sinc-type smoothing of the long-time averaged singlet yield curve through $\mathcal{K}(T)$, and a multiple Lorentzian-type smoothing through $K(\omega)$. Strictly speaking, however, the yields are obtained by integrating the full instantaneous singlet signal, not by applying a smoothing kernel directly to the long-time average; the smoothing interpretation emerges only after the exact integrated expressions are derived.
\\
\noindent It should by now be clear that, in general, the singlet yield expressions exhibit a nontrivial oscillatory structure as a function of $\omega$ rather than simple monotonic behavior. Because these expressions contain oscillatory contributions depending nonlinearly on $\omega$, the \linebreak corresponding magnetic field effect curves can exhibit bends, ripples, and local extrema as the magnetic field is scanned. This oscillatory magnetic field dependence is not usually explicitly emphasized in the existing literature. Perhaps this is because current experimental capabilities are not sophisticated enough to resolve it; or because experiments operate in a long time-average regime (effectively, approaching $T \to \infty$ or $k \to 0$) where most of those oscillating features are washed out (i.e., in a regime in which magnetic field effect curves are well-described by the long-time signal averages of Table~\ref{Table2}).
\\

\begin{figure}[H]
  \centering
  \includegraphics[width=\textwidth]{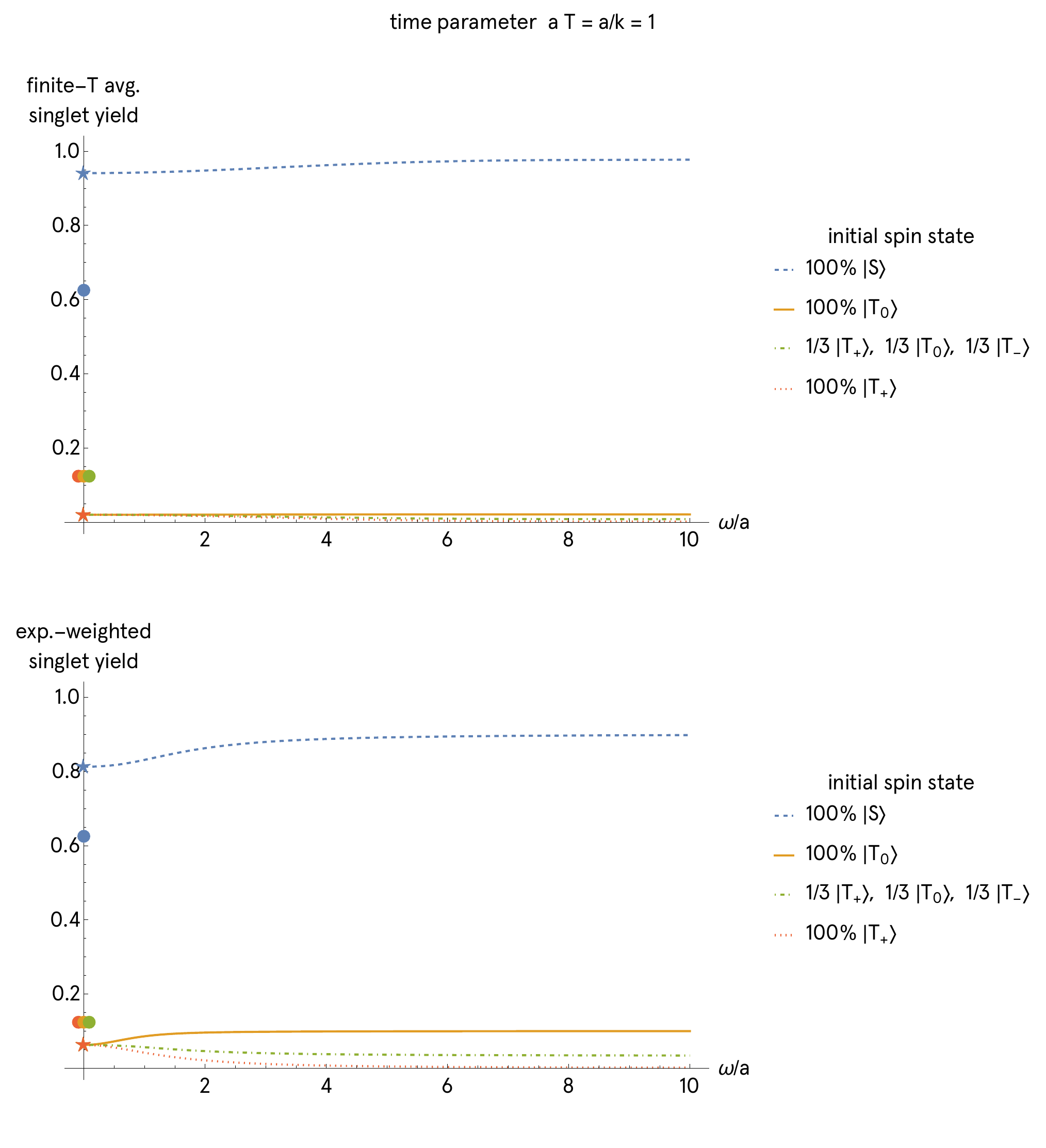}
  \captionsetup[figure]{aboveskip=4pt,belowskip=8pt}
  \captionof{figure}{\justifying \textbf{Singlet yields for time parameter \(aT = \frac{a}{k} =1\).}
  The observation window is short, and appreciable spin-mixing has not yet occurred. The filled circles are the values of the yields for $\omega = 0$ in the limit $t \to \infty$, namely $\frac{1}{8}$ and $\frac{5}{8}$. The stars are the values of the yields for $\omega = 0$ for this given time parameter. At zero field, the curves are still far from their $t\to \infty$ values. }
   \label{1}
  \end{figure}

\begin{figure}[H]
  \centering
  \includegraphics[width=\textwidth]{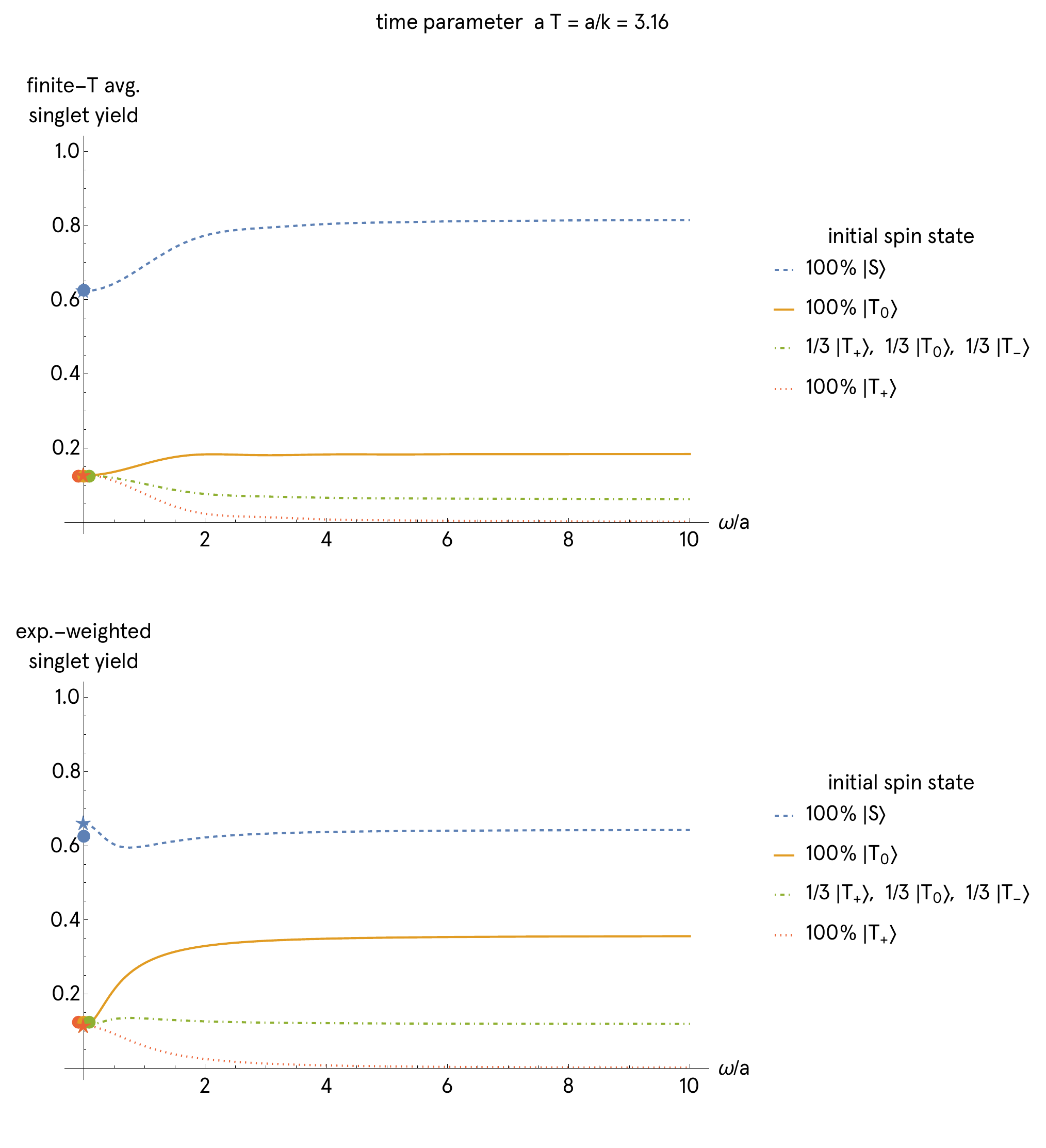}
  \captionsetup[figure]{aboveskip=4pt,belowskip=8pt}
  \captionof{figure}{\justifying \textbf{Singlet yields for time parameter \(aT = \frac{a}{k} =3.16\).}
  The singlet yield curves start to develop appreciable bends. The filled circles are the values of the yields for $\omega = 0$ in the limit $t \to \infty$, namely $\frac{1}{8}$ and $\frac{5}{8}$. The stars are the values of the yields for $\omega = 0$ for this given time parameter. At zero field, the curves are already close to their $t\to \infty$ values.}
   \label{3}
  \end{figure}

\begin{figure}[H]
  \centering
  \includegraphics[width=\textwidth]{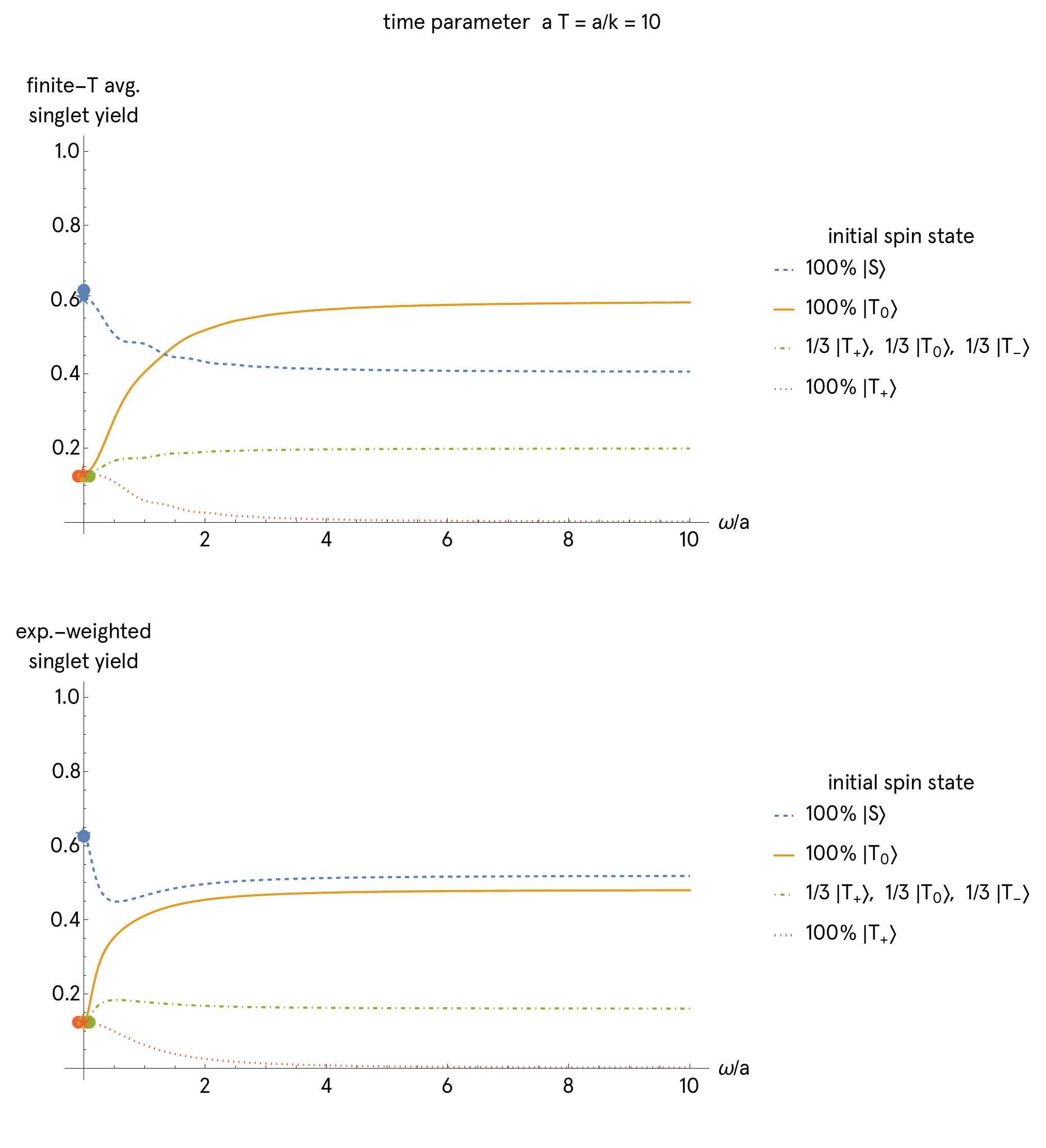}
  \captionsetup[figure]{aboveskip=4pt,belowskip=8pt}
  \captionof{figure}{\justifying \textbf{Singlet yields for time parameter \(aT=10\).}
  The oscillatory character of the finite-$T$ average singlet yield starts to be pronounced, and a low-field feature in the exponentially-weighted singlet yield starts to develop. The filled circles are the values of the yields for $\omega = 0$ in the limit $t \to \infty$, namely $\frac{1}{8}$ and $\frac{5}{8}$. The stars are the values of the yields for $\omega = 0$ for this given time parameter.}  
   \label{10}
  \end{figure}

\begin{figure}[H]
  \centering
  \includegraphics[width=\textwidth]{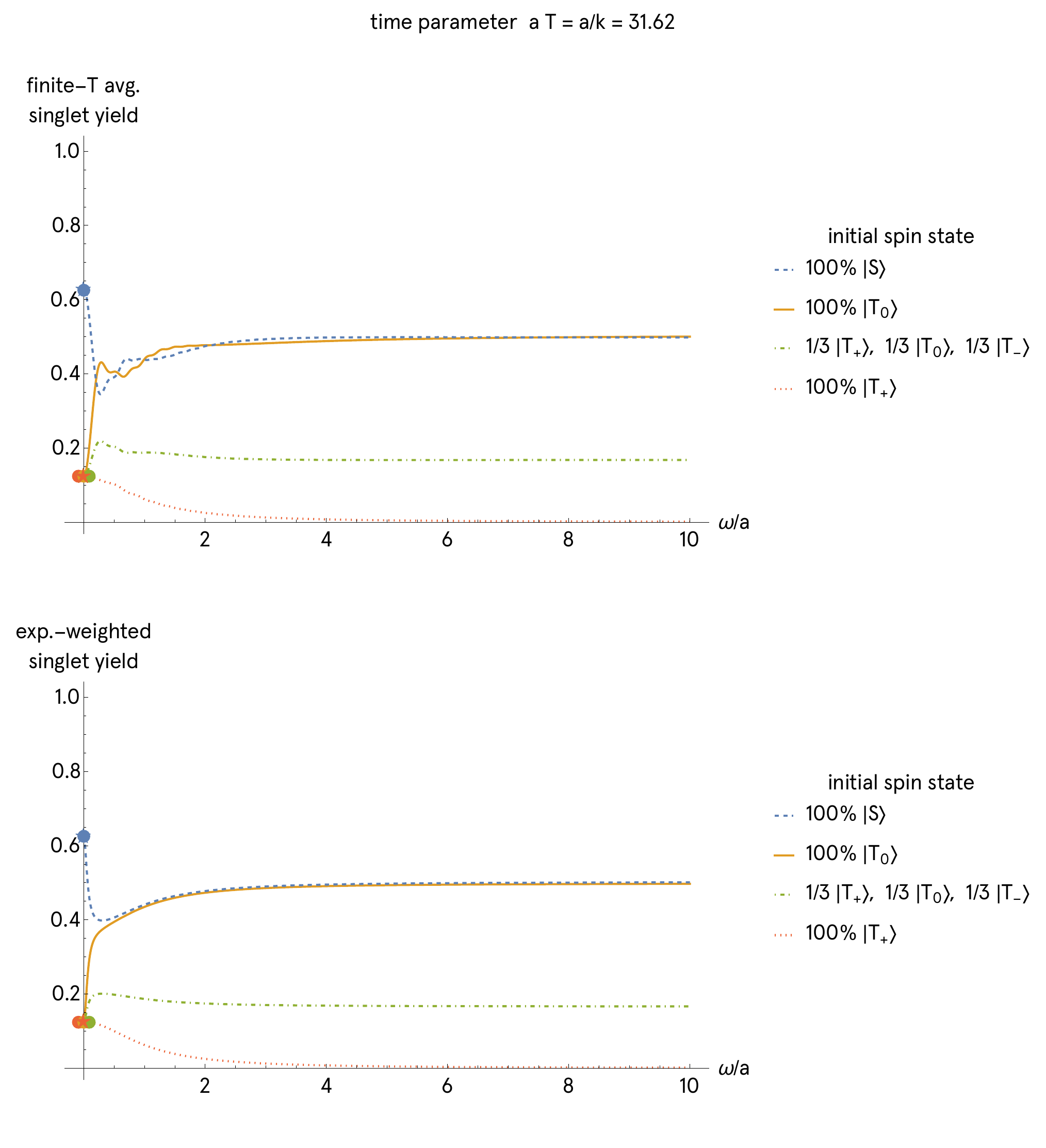}
  \captionsetup[figure]{aboveskip=4pt,belowskip=8pt}
  \captionof{figure}{\justifying \textbf{Singlet yields for time parameter \(aT=31.62\).}
  The oscillatory character of the finite-$T$ average singlet yield is very clear, and the low-field feature in the exponentially-weighted singlet yield is close to reaching its final value. The filled circles are the values of the yields for $\omega = 0$ in the limit $t \to \infty$, namely $\frac{1}{8}$ and $\frac{5}{8}$. The stars are the values of the yields for $\omega = 0$ for this given time parameter.}
   \label{31}
  \end{figure}

\begin{figure}[H]
  \centering
  \includegraphics[width=\textwidth]{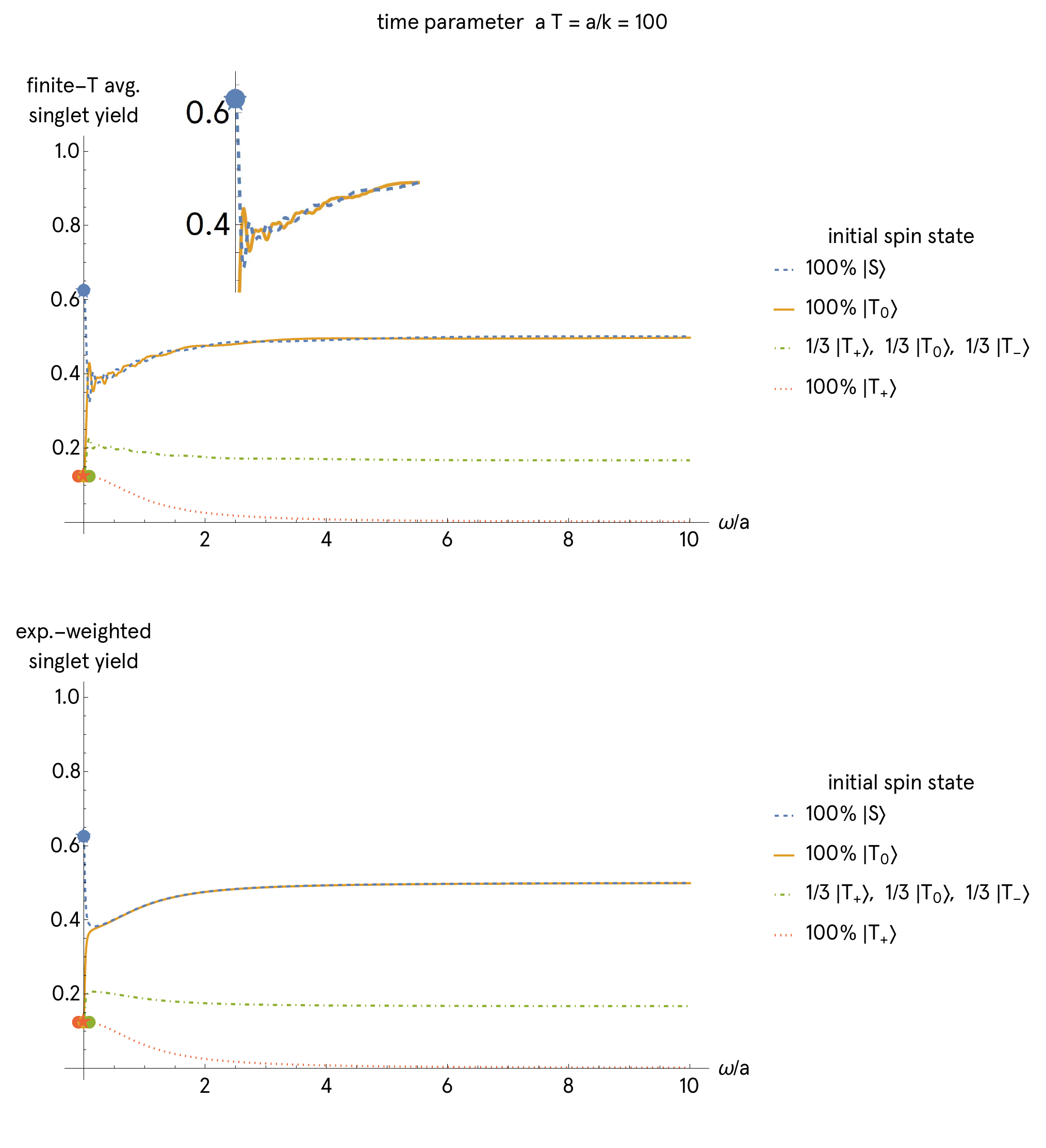}
  \captionsetup[figure]{aboveskip=4pt,belowskip=8pt}
  \captionof{figure}{\justifying \textbf{Singlet yields for time parameter \(aT = \frac{a}{k} = 100\).}
  The oscillatory character of the finite-$T$ average singlet yield is very fast and the oscillations almost wash out, and the low-field feature in the exponentially-weighted singlet yield is even closer to reaching its final value. The filled circles are the values of the yields for $\omega = 0$ in the limit $t \to \infty$, namely $\frac{1}{8}$ and $\frac{5}{8}$.  The stars are the values of the yields for $\omega = 0$ for this given time parameter. For the finite-$T$ average yield, we additionally show a zoomed-in window over $0 \le \frac{\omega}{a} \le 2$ to make the rapid oscillations in the curves for the initial $\ket{S}$ and $\ket{T_0}$ states clearly visible.}
   \label{100}
  \end{figure}

\noindent In Figs.~\ref{1} through~\ref{100} we plot the two considered singlet yields, as a function of the \linebreak dimensionless parameter $\frac{\omega}{a}$ for five representative values of the 
dimensionless time \linebreak parameter \(aT = \frac{a}{k}\)
chosen to be logarithmically spaced between \(1\) and \(100\); the time parameter corresponds to the duration of the observation window for the first singlet yield, and to the radical lifetime for the second. The embedded video of Fig.~\ref{P5} extends the same visualization to a much denser sampling of the time parameter. These plots make explicit how the singlet yield curves evolve as the maximal averaging time is increased. For short time parameters, the sharp time cutoff produces pronounced sinc-like oscillatory structure, with visible ripples, as a function of \(\omega\) for the finite-$T$ average singlet yield, and a Lorentzian-smoothed character for the exponentially-weighted singlet yield. Both curves retain nonmonotonic behavior and a broadened low-field feature. As the averaging window increases, these oscillatory features are progressively washed out for the first singlet yield, while the low-field structure becomes narrower in \(\omega\)-space for both yields. By \(aT=\frac{k}{a} =100\), the curves are already very close to the \(t\to\infty\), long-time average results of Fig.~\ref{P1}, except that the singlet yields still smooth the zero-field region and therefore replace the strict discontinuity of the infinite-time limit by a narrow but continuous crossover.

\vfill
\newpage

\begin{table}[H]
    \centering
\begin{tabular}{|c|c|}
\hline
initial state &  singlet yield $\langle\textcolor{InstituteBlue}{\underline{\left|\ket{S(t,\omega)}\right|^2}}\rangle_T$  \\ \hline
100\% $\ket{T_{+}}$                    &  $\frac{a^2}{8\Omega^2}
-
\frac{a^2}{8\Omega^3T}\sin\!\left(\Omega T\right)$  \\ \hline
100\% $\ket{T_{0}}$                    &  $\frac{3a^2+4\omega^2}{8\Omega^2}
+
\frac{a^2}{8\Omega^3T}\sin\!\left(\Omega T\right)
-
\frac12\,\mathcal K(T)$\\ \hline
$\frac{1}{3} \ket{T_{+}}$, \ $\frac{1}{3} \ket{T_{0}}$, \ $\frac{1}{3} \ket{T_{-}}$                                  &  $\frac{5a^2+4\omega^2}{24\Omega^2}
-
\frac{a^2}{24\Omega^3T}\sin\!\left(\Omega T\right)
-
\frac16\,\mathcal K(T)$ \\ \hline
100\% $\ket{S}$                                                                   & $\frac{3a^2+4\omega^2}{8\Omega^2}
+
\frac{a^2}{8\Omega^3T}\sin\!\left(\Omega T\right)
+
\frac12\,\mathcal K(T) $  \\ \hline
$\frac{1}{4} \ket{T_{+}}, \ \frac{1}{4} \ket{T_{0}}, \ \frac{1}{4} \ket{T_{-}}, \ \frac{1}{4} \ket{S}$ & $ \frac{1}{4} $  \\ \hline
$\alpha \ket{T_{+}}, \ \beta  \ket{T_{0}}, \ \gamma  \ket{T_{-}}, \ (1-\alpha-\beta-\gamma) \ket{S}$ & $\frac{(1+2\Sigma)a^2+4\Sigma\omega^2}{8\Omega^2}
+
\frac{(2\Sigma-1)a^2}{8\Omega^3T}
\sin\!\left(\Omega T\right)
+
\frac{\Delta}{2}\,\mathcal K(T) $ \\ \hline
\end{tabular}
   \captionsetup[table]{aboveskip=4pt,belowskip=8pt}
  \captionof{table}{\justifying \textbf{Finite-$T$ average singlet yield, expressed for different initial
spin initializations. } All results are self-consistent, and reduce to the long-time averages of Table~\ref{Table2} for $T \to \infty$. Closed expressions can be found in the accompanying SymPy code. 
The initial spin states appear in the curves only via two parameters, $\Sigma$ and $\Delta$, which measure the total initial weight in the $\{ \ket{S}, \ket{T_0}\}$ sector, and the initial difference between the populations in these states. All singlet yields are smoothly continuous for $\omega$ going from $0$ to $0^{+}$.}
    \label{tab:mfe_T}
\end{table}

\begin{table}[H]
\centering
\begin{tabular}{|c|c|}
\hline
initial state & singlet yield $\Phi_k(\omega)$ \\
\hline
$100\%\,|T_+\rangle$
&
$
\frac{a^2}{8(k^2+\Omega^2)}
$
\\ \hline

$100\%\,|T_0\rangle$
&
$
\frac{1}{2}-\frac{a^2}{8(k^2+\Omega^2)}-\frac12\,K(\omega)
$
\\ \hline

$\frac{1}{3}|T_+\rangle,\ \frac{1}{3}|T_0\rangle,\ \frac{1}{3}|T_-\rangle$
&
$
\frac{1}{6}+\frac{a^2}{24(k^2+\Omega^2)}-\frac{1}{6}\,K(\omega)
$
\\ \hline

$100\%\,|S\rangle$
&
$
\frac{1}{2}-\frac{a^2}{8(k^2+\Omega^2)}+\frac{1}{2}\,K(\omega)
$
\\ \hline

$\frac{1}{4}|T_+\rangle,\ \frac{1}{4}|T_0\rangle,\ \frac{1}{4}|T_-\rangle,\ \frac{1}{4}|S\rangle$
&
$
\frac{1}{4}
$
\\ \hline

$\alpha|T_+\rangle,\ \beta|T_0\rangle,\ \gamma|T_-\rangle,\ (1-\alpha-\beta-\gamma)|S\rangle$
&
$
\frac{\Sigma}{2}
-
\frac{(2\Sigma-1)a^2}{8(k^2+\Omega^2)}
+
\frac{\Delta}{2}\,K(\omega)
$
\\
\hline
\end{tabular}
 \captionof{table}{\justifying \textbf{Exponentially-weighted singlet yield, expressed for different initial
spin initializations. } All results are self-consistent, and reduce to the long-time averages of Table~\ref{Table2} for $k \to 0$. Closed expressions can be found in the accompanying SymPy code. The initial spin states appear in the curves only via two parameters, $\Sigma$ and $\Delta$, which measure the total initial weight in the $\{ \ket{S}, \ket{T_0}\}$ sector, and the initial difference between the populations in these states. All singlet yields are smoothly continuous for $\omega$ going from $0$ to $0^{+}$. Unlike the finite-$T$ average singlet yield of Table~\ref{tab:mfe_T}, these expressions were obtained using a more standard definition of the singlet yield involving the recombination rate $k$.}
\label{tab:mfe_k}
\end{table}

\vfill
\newpage

\vfill
\newpage

\vfill
\newpage

\vfill
\newpage

\vfill
\newpage

\vfill
\newpage

\begin{figure}[H]
  \centering
  \captionsetup[figure]{aboveskip=4pt, belowskip=8pt}
  
\includemedia[
  activate=onclick,
  addresource=Figures/SingletYieldsXTime.mp4,
  flashvars={source=Figures/SingletYieldsXTime.mp4}
]{%
  \fbox{%
    \parbox[c][1.09\linewidth][c]{1\linewidth}{%
      \centering
      \normalsize video link:\\[1em]
      \normalsize\href{https://bit.ly/SingletYieldsXTime}{https://bit.ly/SingletYieldsXTime}
    }%
  }%
}{VPlayer.swf}
  
  \caption{\justifying \textbf{Video of the singlet yields for time parameters $aT = \frac{a}{k} = 0$ to $100$.}  The oscillatory features develop and are then almost washed out for the finite-$T$ average singlet yield. The exponentially-weighted singlet yield develops a smooth low-field feature early on.}
  \label{P5}
\end{figure}

\vfill
\newpage

\noindent We can also calculate the asymptotic behaviour of the yields in the limits \(\omega\to0\) and \(\omega\to\infty\); the results are displayed in Tables~\ref{tab:N1} and~\ref{tab:N2}. In the zero-field limit, the yield retains the singular interference contribution associated with the exact triplet degeneracy. As a result, the yield depends not only on \(\Sigma\), which controls the population-type terms, but also on \(\Delta\), which controls the interference term. 
In the high-field limit, however, the expressions simplify to an effective \(\ket S\leftrightarrow\ket{T_0}\) form: the background is set by \(\Sigma/2\), while the remaining \(\Delta\)-dependence appears only as a finite-time or finite-lifetime correction that disappears in the \(T\to\infty\) or \(k\to0\) limits. In this sense, the \(\omega\to0\) entries make explicit that the zero-field anomaly originates in the interference term that survives only at exact triplet degeneracy. At high field, the initial \(\ket S\leftrightarrow\ket{T_0}\) imbalance does not contribute to the strict long-time yield. Its effect appears only through oscillatory terms that survive when the observation time or the lifetime is finite, and it disappears in the limits \(T\to\infty\) or \(k\to0\).
\\

\vfill
\newpage

\begin{table}[H]
\hspace*{-0.5cm}
\begin{tabular}{|c|c|c|}
\hline
initial state 
& $\lim_{\omega\to0}$  $\langle \textcolor{InstituteBlue}{\underline{\left|\ket{S(t,\omega)}\right|^2}}\rangle_T$
& $\lim_{\omega\to\infty}$ $\langle \textcolor{InstituteBlue}{\underline{\left|\ket{S(t,\omega)}\right|^2}}\rangle_T$
\\ \hline

$100\%\,\ket{T_{+}}$
&
$\frac{1}{8}-\frac{1}{8}\text{sinc}(aT)$
&
$0$
\\ \hline

$100\%\,\ket{T_{0}}$
&
$\frac{1}{8}-\frac{1}{8}\text{sinc}(aT)$
&
$\frac{1}{2}-\frac{1}{2}\text{sinc}(\frac{aT}{2})$
\\ \hline

$\frac{1}{3}\ket{T_{+}},\ \dfrac{1}{3}\ket{T_{0}},\ \dfrac{1}{3}\ket{T_{-}}$
&
$\frac{1}{8}-\frac{1}{8}\text{sinc}(aT)$
&
$\frac{1}{6}-\frac{1}{6}\text{sinc}(\frac{aT}{2})$
\\ \hline
$100\%\,\ket{S}$
&
$\frac{5}{8}+\frac{3}{8}\text{sinc}(aT)$
&
$\frac{1}{2}+\frac{1}{2}\text{sinc}(\frac{aT}{2})$
\\ \hline

$\frac{1}{4}\ket{T_{+}},\ \dfrac{1}{4}\ket{T_{0}},\ \dfrac{1}{4}\ket{T_{-}},\ \dfrac{1}{4}\ket{S}$
&
$\frac{1}{4}$
&
$\frac{1}{4}$
\\ \hline

$\alpha\ket{T_{+}},\ \beta\ket{T_{0}},\ \gamma\ket{T_{-}},\ (1-\alpha-\beta-\gamma)\ket{S}$
&
$\frac{1+2\Sigma+2\Delta}{8}
+\left(\frac{2\Sigma+2\Delta-1}{8}\right)\text{sinc}(aT)$
&
$\frac{\Sigma}{2}
+\frac{\Delta}{2}\text{sinc}\left(\frac{aT}{2}\right)$
\\ \hline
\end{tabular}
 \captionof{table}{\justifying \textbf{Finite-$T$ average singlet yield at its low- and high-magnetic field limits, expressed for different initial
spin initializations. } All results are self-consistent, and reduce to the long-time averages of Table~\ref{Table2} for $T \to \infty$. Closed expressions can be found in the accompanying SymPy code. 
The initial spin states appear in the curves only via two parameters, $\Sigma$ and $\Delta$, which measure the total initial weight in the $\{ \ket{S}, \ket{T_0}\}$ sector, and the initial difference between the populations in these states.}
\label{tab:N1}
\end{table}


\begin{table}[H]
\centering
\begin{tabular}{|c|c|c|}
\hline
initial state 
& $\lim_{\omega\to0}$  $\Phi_k(\omega)$
& $\lim_{\omega\to\infty}$  $\Phi_k(\omega)$
\\ \hline

$100\%\,\ket{T_{+}}$
&
$\frac{a^2}{8(k^2+a^2)}$
&
$0$
\\ \hline

$100\%\,\ket{T_{0}}$
&
$\frac{a^2}{8(k^2+a^2)}$
&
$\frac{a^2}{2(4k^2+a^2)}$
\\ \hline

$\frac{1}{3}\ket{T_{+}},\ \dfrac{1}{3}\ket{T_{0}},\ \dfrac{1}{3}\ket{T_{-}}$
&
$\frac{a^2}{8(k^2+a^2)}$
&
$\frac{a^2}{6(4k^2+a^2)}$
\\ \hline

$100\%\,\ket{S}$
&
$\frac{8k^2+5a^2}{8(k^2+a^2)}$
&
$1-\frac{a^2}{2(4k^2+a^2)}$
\\ \hline

$\frac{1}{4}\ket{T_{+}},\ \frac{1}{4}\ket{T_{0}},\ \frac{1}{4}\ket{T_{-}},\ \frac{1}{4}\ket{S}$
&
$\frac{1}{4}$
&
$\frac{1}{4}$
\\ \hline

$\alpha\ket{T_{+}},\ \beta\ket{T_{0}},\ \gamma\ket{T_{-}},\ (1-\alpha-\beta-\gamma)\ket{S}$
&
$\frac{4(\Sigma+\Delta)k^2+(1+2\Sigma+2\Delta)a^2}{8(k^2+a^2)}$
&
$\frac{\Sigma}{2}+\frac{2\Delta k^2}{4k^2+a^2}$
\\ \hline
\end{tabular}
 \captionof{table}{\justifying \textbf{Exponentially-weighted singlet yield at its low- and high-magnetic field limits, expressed for different initial
spin initializations.} All results are self-consistent, and reduce to the long-time averages of Table~\ref{Table2} for $k \to 0$. Closed expressions can be found in the accompanying SymPy code. 
The initial spin states appear in the curves only via two parameters, $\Sigma$ and $\Delta$, which measure the total initial weight in the $\{ \ket{S}, \ket{T_0}\}$ sector, and the initial difference between the populations in these states.}
\label{tab:N2}
\end{table}

\vfill 
\newpage

\noindent In Supplementary Section~\ref{shape}, we examine in detail the full oscillatory $\omega$-dependence of the yields; we can demonstrate, for example, that the field location
of any nonzero extremum of these magnetic field effect curves scales linearly with the strength of the hyperfine interaction, $a$ (Supplementary Subsections~\ref{scaling} and~\ref{kmodel}). We also scrutinize the limit $\omega \to 0$  (Supplementary Subsections~\ref{osc},~\ref{lf},~\ref{halffield}, ~\ref{max}, and~\ref{kmodel}), which is relevant in the discussion of the so-called `low-field effect'.
\\

\noindent The low-field effect, in the usual chemical sense, refers to a near-zero-field feature of the yield curves. It is usually described in the literature as a local dip for singlet-born radical pairs and as a local peak for some triplet-born radical pairs~\cite{Hamilton1988HighResolution, Maeda2008ChemicalCompass, Evans2016SubMillitesla, Kattnig2016ChemicalAmplification,Ross2026MagLOV2}. A historical overview of how the low-field effect has been understood in the chemistry literature can be found in Supplementary Subsection~\ref{lfhist}.
\\

\noindent The present physics-inspired derivation and interpretation of the low-field effect differ from previous accounts in several important respects. First, although earlier treatments have been successful in describing the low-field effect, they do not fully account for the continuity between the zero-field and low-field regimes. By contrast, all of our closed-form populations, coherences and yields are continuous as \(\omega\) goes from \(0\) to \(0^{+}\). Second, the existing literature does not fully explore how the initial populations of \(\ket{S}\), \(\ket{T_{+}}\), \(\ket{T_{0}}\), and \(\ket{T_{-}}\) affect the magnitude and sign of the low-field effect; here, we address this question by considering a generic incoherent mixture of initial spin states. Finally, previous works have focused primarily on long-time averages ($t \to \infty$, which generates the zero-field singularity) and have not fully examined the frequency content of both the singlet population and the singlet yield for different observation windows $T$ or radical lifetimes $k^{-1}$, like we do here.
\\

\noindent We now summarize the results of our analysis of the $\omega \to 0$ limit. We start by deriving, for $\omega \to 0$, a characteristic low-field oscillation scale, which originates from the $\mathcal{K}(T)$ and $K(\omega)$ terms; this oscillation scale can be shown to be proportional to $\frac{1}{T}$ or, equivalently, proportional to $k$; the low-field feature is a natural outcome of coherent oscillations being sampled over a limited time $T$ or lifetime scale $k^{-1}$. Importantly, the feature can only exist if $\Delta \neq 0$ because that is the constant multiplying $\mathcal{K}(T)$ and $K(\omega)$ in the yield expressions.  This initial population imbalance is thus a necessary condition for a low-field correction term to be present, but a genuine dip or peak additionally requires this correction to be strong enough to cancel and reverse the slope of the remaining terms in the full yield curve. In the strict long-time limit $t\to\infty$, we can show that, if $\Delta > 0$ ($\Delta < 0$), the magnetic field effect curve indeed produces a local dip (peak) for $\omega > 0$. In the same limit, the visibility of the low-field structure is maximized, if it is a dip (peak), for initial spins in a pure $\ket{S}$ ($\ket{T_0}$) state. 
\\

\noindent We stress that the singular feature at zero field is best understood as a coherence effect rather than simply as enhanced spin mixing. Both the singlet population and the singlet yield contain three contributions: the dark-state population, the bright-state population, and the bright--dark interference term \(\mathrm{Re}\!\left[\rho_{BD}(t)\right]\). As shown in Eqs.~(\ref{27}) through~(\ref{new37}), this coherence behaves qualitatively differently at exactly \(\omega=0\) than at any nonzero field. At \(\omega=0\), \(\mathrm{Re}\!\left[\rho_{BD}(t)\right]\) contains a constant component, whereas for \(\omega>0\) it is purely oscillatory and its long-time average vanishes. This is precisely the contribution proportional to \(\Delta\), and this change in the bright--dark coherence underlies the singular zero-field behavior. 
\\


\noindent Finally, one may ask under which conditions such a nonmonotonic low-field feature can be observed experimentally. In practice, one never measures a singlet yield curve after averaging for an infinite time nor with arbitrarily fine field resolution. Instead, one can only average over a finite time \(T\), decoherence suppresses the coherent contribution on a timescale \(T_2^\star\), and field drift or inhomogeneity effectively convolves the signal with a distribution of \(\omega\). It is therefore natural to define an effective coherent observation window
\begin{equation}
T_{\mathrm{eff}}\sim \min(T,T_2^\star) \ .
\end{equation}

\noindent The low-field effect discussed above, which originates from coherent dynamics, is then observable only if its intrinsic field-space width is not washed out by these experimental limitations. Since, near \(\omega\to0\), the sinc or Lorentzian terms set a characteristic oscillation scale of approximately \(\frac{1}{T_{\mathrm{eff}}}\), this coherent low-field feature can only be resolved when the overall experimentally determined field broadening \(\Delta\omega_{\mathrm{exp}}\), at $\omega \to 0$, is comparable to, or smaller than, this scale. Longer averaging times or longer coherence times therefore make the feature intrinsically narrower in \(\omega\)-space; this sharpens the structure, but also makes it more demanding to resolve experimentally.
\vfill
\newpage

\vfill

\newpage
\begin{center}
\sectionfont{Take-home messages}
\end{center}
\begin{enumerate}
    \item The experimentally relevant observable in typical radical-pair measurements is the singlet yield, or magnetic field effect curve, obtained by integrating the singlet \linebreak population signal over time.

\item We give explicit expressions for two typical yield integrals; these should be understood as functions of
$\omega$, for a fixed time-like parameter. 

    \item These singlet yield curves vary with $\omega$ nonlinearly, exhibiting many different frequencies of oscillation and, potentially, bends, ripples and local extrema. They behave \linebreak qualitatively differently at exactly $\omega = 0$. 


    \item The `low-field effect' in the yield curve is best understood as a coherence effect rather than simply as enhanced spin mixing: it originates from the bright--dark coherence term that contributes to the singlet yield. A nonzero \(\Delta\) guarantees the presence of this coherence-driven low-field correction, but a genuine low-field dip or peak appears only if that term is strong enough to generate a turning point in the full yield curve. When such a feature exists, its width is set by the inverse of the relevant time-like parameter. In the strict limit $t \to \infty$, such a feature does exist for $\Delta \neq 0$; if $\Delta > 0$ ($\Delta < 0$), the magnetic field effect curve has a local dip (peak) for $\omega > 0$.
\end{enumerate}


\newpage

\clearpage
\section{Solution for good magnetometric states}
\label{sec:mag}

\noindent A further novel aspect of this work is the import of concepts and tools from technological quantum sensing into the analysis of radical-pair dynamics. In particular, by framing the model in terms of magnetic sensitivity, we recast the radical pair not only as a spin-chemical system but also as a quantum sensor, thereby making direct contact with a well-developed body of methods for quantifying field response, optimal initial state preparation, and the trade-off between coherent phase accumulation and signal averaging. To our knowledge, this perspective has not been systematically developed in the radical-pair context. It provides a complementary viewpoint to the traditional spin-chemistry analysis, and helps isolate which features of the dynamics are genuinely advantageous for magnetosensing, as opposed to merely producing visible field dependence.
\\

\noindent In quantum sensing, given a signal $\mathcal{S}(t, \omega)$, the shot-noise-limited sensitivity $\eta$ with respect to $\omega$ is defined as~\cite{Aiello2014}
\begin{equation}
    \eta \equiv \lim_{\omega \to 0}\frac{\Delta \mathcal{S}}{\left| \frac{\partial \mathcal{S}}{\partial \omega}  \right|} \sqrt{\tau} = 
     \lim_{\omega \to 0}\frac{\sqrt{\frac{\mathcal{S}(1-\mathcal{S})}{N}}}{\left| \frac{\partial \mathcal{S}}{\partial \omega}  \right|} \sqrt{N \cdot t}  
    = \lim_{\omega \to 0}\frac{\sqrt{\mathcal{S}(1-\mathcal{S})}}{\left| \frac{\partial \mathcal{S}}{\partial \omega}  \right|} \sqrt{t} \ ; 
    \label{eta}
\end{equation}

\noindent here, $\Delta \mathcal{S}$ is the signal's standard deviation, and $\tau$ the total experiment time. For $N$ \linebreak realizations, the total time is $\tau = N \cdot t$, where $t$ is the time of a single experiment. Moreover,  if each run produces a binary outcome (success/failure) with probability $\mathcal{S}$, then the outcome is a Bernoulli random variable with variance $\mathcal{S}(1-\mathcal{S})$; therefore the intrinsic (projection/\linebreak binomial) standard deviation of the measured signal for an average over $N$ independent runs is $\Delta\mathcal{S} =\sqrt{\frac{\mathcal{S}(1-\mathcal{S})}{N}}$ --- hence the final expression for $\eta$. The quantity $\omega$ being a frequency, $\eta$ can be understood as the sensor's minimum detectable change in frequency. The factor $\sqrt{t}$ can be understood as a penalty: the longer the measurement takes, the larger (i.e., the worse) the sensitivity. 
\\

\noindent Because the full eight-level singlet observable (Eq.~(\ref{gennew})) is an even function of the Larmor \linebreak frequency,
\textcolor{InstituteBlue}{\underline{$\left|\ket{S(t, \omega)}\right|^2$}} $=$ \textcolor{InstituteBlue}{\underline{$\left|\ket{S(t, -\omega)}\right|^2$}}, it cannot provide first-order sensitivity to signed field changes in the zero-field limit: 
\begin{equation}
\left.\frac{\partial \text{\textcolor{InstituteBlue}{\underline{$\left|\ket{S(t, \omega)}\right|^2$}}}}{\partial \omega}\right|_{\omega=0}=0 \ ,
\end{equation}
so the standard slope-based shot-noise sensitivity diverges when evaluated at $\omega=0$. The same is true for the singlet yields $\langle$\textcolor{InstituteBlue}{\underline{$\left|\ket{S(t, \omega)}\right|^2$}}$\rangle_T$ and $\Phi_{k}(\omega)$. Note that the addition of more terms to the Hamiltonian might induce the symmetry breaking that will permit a zero-field linear magnetosensing response (namely, $\mathcal{S(\omega) \neq \mathcal{S(-\omega)}}$) --- we are neglecting this possibility here.
\\

\noindent Even if, indeed, the derivative vanishes as $\omega \to 0$, this might not necessarily be bad news for biological magnetosensing: we know that the three above observables can indeed be field-sensitive; rather, it means that zero field is not an appropriate operating point for linear-response magnetometry with the considered scalar signals (\textcolor{InstituteBlue}{\underline{$\left|\ket{S(t, \omega)}\right|^2$}}, $\langle$\textcolor{InstituteBlue}{\underline{$\left|\ket{S(t, \omega)}\right|^2$}}$\rangle_T$, and $\Phi_{k}(\omega)$).
\\

\noindent A more natural interpretation is therefore that biological magnetosensing, if based on
observables such as those above, and if under a symmetric Hamiltonian that is close to this toy model's, operates around a finite ambient bias field rather than around $\hat{B} = 0$. In other words, spins under this toy-model Hamiltonian are naturally a finite-bias magnitude sensor, not a zero-bias sign-sensitive sensor. In particular, the geomagnetic field provides a plausible operating point: the biologically relevant signal would then be the local variation of the singlet population or yield around that nonzero background
field. In that case, the relevant sensitivity is not a function of the signal derivative evaluated at $\omega = 0$, but rather
\begin{equation}
\left.\frac{\partial \mathcal{S}}{\partial \omega}\right|_{\omega=\omega_{\rm E}} \ ,
\end{equation}
where $\omega_{\rm E}$ denotes the Larmor frequency associated with the Earth's magnetic
field of \linebreak $\sim 50~\mu T$. Moreover, we posit that it is not unlikely that each protein has an optimum operating point that depends on local magnetic field conditions, for example, inside a cell. 
\\

\noindent Equivalently, one may say that spins under the Hamiltonian considered here can sense magnetic field magnitude departures from a finite reference field, rather than the sign of an
infinitesimal perturbation around zero. Such a spin quantum sensor can still distinguish whether the field
becomes stronger or weaker relative to its operating point, even though it cannot distinguish
$+\Delta\omega$ from $-\Delta\omega$ in the strict zero-bias limit.
\\

\noindent Hence, a natural 
calibrated differential signal is~\cite{Aiello2014}
\begin{equation}
\mathcal{D}(\omega,t)\equiv \frac12\Big(\mathcal{S}(\omega,t)-\mathcal{S}(\omega_{\rm E},t)\Big) \ , 
\label{eq:D_inst_def}
\end{equation}
\noindent where the signal $\mathcal{S}$ is, for example, one of \textcolor{InstituteBlue}{\underline{$\left|\ket{S(t, \omega)}\right|^2$}}, $\langle$\textcolor{InstituteBlue}{\underline{$\left|\ket{S(t, \omega)}\right|^2$}}$\rangle_T$, or $\Phi_{k}(\omega)$. Note that $\mathcal{D}(\omega, t)$ vanishes at the operating point $\omega_{\rm E}$ and is linear in $(\omega-\omega_{\rm E})$ to first order:
\begin{equation}
\mathcal{D}(\omega,t)\approx \frac{1}{2}\left.\frac{\partial \mathcal{S}}{\partial \omega}\right|_{\omega=\omega_{\rm E}} (\omega - \omega_{\rm E}) \ .
\label{eq:D_inst_expand}
\end{equation}

\noindent Although the calibrated differential signal $\mathcal{D}(\omega,t)$ is convenient conceptually, it is only an affine transformation of the original signal about the operating point. Consequently, a factor $\frac{1}{2}$ cancels between the shot noise and the slope, so the sensitivity can be written directly in terms of the local signal statistics and local slope of $\mathcal{S}$ at $\omega=\omega_{\rm E}$. Indeed, if the reference value $\mathcal{S}(\omega_{\rm E},t)$ is treated as known, and if each run produces a
binary singlet/triplet outcome, then the shot-noise-limited sensitivity is
\begin{equation}
\eta_{\rm E}(t)=
\frac{\sqrt{\mathcal{S}(\omega_{\rm E},t)\left(1 - \mathcal{S}(\omega_{\rm E},t)\right)}}{\big| \left.\frac{\partial \mathcal{S}}{\partial \omega}\right|_{\omega=\omega_{\rm E}}\big|}\sqrt{t} \ .
\label{eq:eta_inst_bias}
\end{equation}

\noindent If, instead, the reference (i.e., `sensing Earth's magnetic field') is itself obtained from an independent measurement with the same
statistics rather than treated as perfectly known, an additional factor $\sqrt{2}$ multiplies Eq.~\eqref{eq:eta_inst_bias}. Which of these scenarios is most relevant for biological magnetosensing is unclear. A biological sensor need not compare two \linebreak simultaneously acquired signals; it may instead compare the instantaneous response against an internal or slowly varying baseline associated with the ambient geomagnetic field (for example, a characteristic balance of chemical products accumulated under geomagnetic conditions). From this perspective, Eq.~\eqref{eq:eta_inst_bias} provides the natural idealized benchmark, whereas the additional factor $\sqrt{2}$ should be viewed as a conservative penalty applicable when the reference must itself be estimated with similar noise.
\\

\noindent Here, we calculate explicit sensitivity expressions for the two singlet yields, which we argue in Supplementary Subsection~\ref{mod} might be more appropriate proxies for biologically-\linebreak relevant magnetosensing than the instantaneous singlet population. 
\\

\noindent We wish to determine which initial spin states are optimal for magnetometry; \linebreak equivalently, which initial spin states minimize the sensitivity. Incidentally, like most quantities previously derived, we will prove that all magnetometry-related quantities are only a function of $\Sigma$ and $\Delta$, not of $\zeta$: how the population outside the $\{\ket{S},\ket{T_{0}}\}$ organizes itself will be of no consequence to how the spins perform magnetosensing.
\\

\noindent However, in general, finding an optimal $(\Sigma,\Delta)$ will not be possible without having recourse to numerics: one noteworthy conclusion from this section is that, even for the very simple toy model we are using, the calculated sensitivities are complicated functions that strongly depend: on both $\Sigma$ and $\Delta$; on $\omega_E$ and how it compares to the hyperfine strength $a$; and on the averaging interval $T$ --- or, equivalently, on the singlet lifetime $k^{-1}$. 
\\

\noindent All the symbolic calculations for the magnetometry-relevant quantities can be followed along using the accompanying SymPy code \href{https://bit.ly/Magnetometry_SymPy}{\texttt{Magnetometry\_SymPy.py}}.
\\

\noindent The shot-noise-limited sensitivities for the yields are thus
\begin{align}
    \eta_{\rm E}(T) &= \frac{\sqrt{\langle \textcolor{InstituteBlue}{\underline{\left|\ket{S(t, \omega_{\rm E})}\right|^2}}\rangle_T\bigl(1-\langle \textcolor{InstituteBlue}{\underline{\left|\ket{S(t,  \omega_{\rm E})}\right|^2}}\rangle_T\bigr)}}{\left|(2\Sigma-1)A+\Delta B\right|}\,\sqrt{T} \ , \label{etaT}\\
        \eta_{\rm E}(k) &= \frac{\sqrt{\Phi_k(\omega_{\rm E})\bigl(1-\Phi_k(\omega_{\rm E})\bigr)}}{\left|(2\Sigma-1)A'+\Delta B'\right|}\,\sqrt{k^{-1}} \ , \label{etak}
\end{align}
where, for $\langle \textcolor{InstituteBlue}{\underline{\left|\ket{S(t, \omega_{\rm E})}\right|^2}}\rangle_T$,
\newpage
\begin{align}
      A&\equiv \frac{a^2\omega_{\rm E}}{8T\Omega_{\rm E}^5} \big( 2T\Omega_{\rm E}+T\Omega_{\rm E}\cos(\Omega_{\rm E} T)-3\sin(\Omega_{\rm E} T) \big) \ , \\
    B&\equiv \frac12 \mathcal{K}'(T,\omega_{\rm E}) \ ;
\end{align}
\noindent and where, for $\Phi_{k}(\omega)$,
\begin{align}
     A' &\equiv \frac{a^2\omega_{\rm E}}{4\bigl(k^2+\Omega_{\rm E}^2\bigr)^2} \ , \\
B' &\equiv \frac12 K'(\omega_{\rm E}) \ .
\end{align}

\noindent In the above, we have defined $\Omega_{\rm E}\equiv\sqrt{a^2+\omega_{\rm E}^2}$.
\\

\noindent It is noteworthy that the sensitivity does not depend on \(\alpha,\beta,\gamma\) separately: it depends only on $\Sigma$ and $\Delta$: a three-parameter optimization problem just collapsed into a two-parameter optimization problem. Note that the allowed region in \((\Sigma,\Delta)\) is
\begin{equation}
0\le \Sigma\le 1 \ , \qquad |\Delta|\le \Sigma \ .
\end{equation}

\noindent Given the symmetry between $\eta_{\rm E}(T)$ and $\eta_{\rm E}(k)$, we will discuss here the sensitivity derived for the finite-$T$ average yield; for $\eta_{\rm E}(k)$, results are very similar, and fully derived in \linebreak Supplementary Section~\ref{mag2}. 
\\

\noindent The sensitivity depends on the magnitudes of both its numerator and its denominator. 
\\

\noindent We show in Supplementary Subsection~\ref{numden} that the numerator is minimized for a pure initial \(\ket{S}\), a pure initial \(\ket{T_{0}}\), or any initial mixture of \(\ket{T_{\pm}}\) adding up to the full population weight. Other states such as \(50\%\ |T_0\rangle, 50\%\ |S\rangle\) cannot minimize the numerator except in accidental degenerate cases.
\\

\noindent In turn, the denominator splits into two distinct slope contributions: the \((2\Sigma-1)A\) term is the population (or bright-sector) contribution, whereas the \(\Delta B\) term is the coherence (or bright--dark interference) contribution. In Supplementary Section~\ref{maginsight}, we argue that the key interferometric resource for magnetometry is bright--dark coherence. In other words, a good magnetometric state is not `as dark as possible', but one that balances dark-state phase memory with bright-state readout. The denominator term thus confirms our intuition that the sensitivity is not controlled by one single mechanism, but rather by the competition or cooperation of two mechanisms.
\\

\noindent We show in Supplementary Subsection~\ref{numden} that the denominator is maximized when the initial spin state is either pure $\ket{S}$ or pure $\ket{T_{0}}$. Neither $\ket{S}$ nor $\ket{T_{0}}$ is always the optimal choice to maximize the denominator: the optimum can 
flip with $T$ or with $\omega_{\rm E}$. More explicitly, at fixed $\Sigma$, pure $\ket{S}$ gives the larger denominator when
$((2\Sigma-1)A)(\Sigma B)>0$, and pure $\ket{T_0}$ gives the larger denominator when $((2\Sigma-1)A)(\Sigma B)<0$. If $A=0$ or $B=0$, the magnitude of the slope is the same for an initial pure $\ket{S}$ or pure $\ket{T_0}$ state. If 
$B\sim 0$, however, the coherence channel becomes ineffective, and the distinction between $|S\rangle$ and $|T_0\rangle$ correspondingly weakens.
\\

\noindent Note the particular case of the maximally mixed initial spin state ($\alpha = \beta = \gamma = \frac{1}{4}$, \linebreak corresponding to $\Sigma = \frac{1}{2}$ and $\Delta = 0$), for which $\left|(2\Sigma-1)A+\Delta B\right| = 0$: the sensitivity diverges; we thus recover the previous result that some degree of spin initialization is required for magnetosensing under this model.
\\

\noindent Putting the numerator and denominator analyses together, we arrive at the following \linebreak conclusion. The denominator-only criterion often favors the
$\{\ket{S},\ket{T_0}\}$ edge for an initial state, because that is where the slope is largest. But the numerator can favor a very different initial state: 
a pure initial \(\ket{S}\), a pure initial \(\ket{T_{0}}\), or any initial mixture of \(\ket{T_{\pm}}\) adding up to the full population weight, even if its slope is not maximal. That is why the full sensitivity can sometimes favor initial states such as $\ket{T_{\pm}}$, even though the denominator-only argument favors the $\{\ket{S},\ket{T_0}\}$ sector. However, if no initial population is in $\ket{T_{\pm}}$, we can confidently say that pure initial $\ket{S}$ or pure initial $\ket{T_0}$ will minimize the sensitivity.
\\

\noindent In general, note that many different triples \((\alpha,\beta,\gamma)\) are exactly equivalent: if they have the same \(\Sigma\) and the same \(\Delta\), they result in the same signal and the same sensitivity. Whenever \(\Sigma<1\), there is a continuous family of degenerate states obtained by redistributing population between \(\ket{T_\pm}\). In turn, if \(\Sigma=1\), then \(\alpha=\gamma=0\), and the corresponding triple is unique.
\\


\noindent Finally, note that the presence of a low-field effect does not necessarily imply good \linebreak magnetometric performance. The low-field singularity originates from the survival of a stationary coherence contribution at exactly \(\omega=0\), whereas first-order magnetometry in this toy model requires an operating point at which the signal has a nonzero field derivative; for the signals considered here, that is not the case at \(\omega=0\). For example, a pure initial \(\ket{T_\pm}\) state does not generate the coherence-driven low-field feature, but this does not mean that such an initial state is necessarily unfavorable for magnetometry, since useful field sensitivity may still occur away from zero field.
\\

\noindent In conclusion, the sensitivity curve, even for this very simple toy model, is a nontrivial function of $\Sigma$, $\Delta$, the interrogation time $T$, and the ratio $\frac{\omega_{\rm E}}{a}$. Accordingly, once both the numerator and denominator of the sensitivity are taken into account, exact comparisons between candidate initial states are most reliably carried out numerically.
\\

\noindent On another note, one interesting difference between the two singlet yields studied \linebreak throughout is that the lifetime weighting acts as a low-pass filter. The finite-$T$ average uses a sharp cutoff in time, $T$, so its kernel carries sinc-like terms and can generate oscillatory features as a function of \(T\). For the second yield model, the exponential weighting is smooth. The parameter \(k\) suppresses late-time dynamics continuously rather than sharply. Thus, the lifetime-weighted sensitivity is smoother as a function of time, with less oscillatory structure.
\\

\begin{figure}[H]
  \centering
  \includegraphics[width=\textwidth]{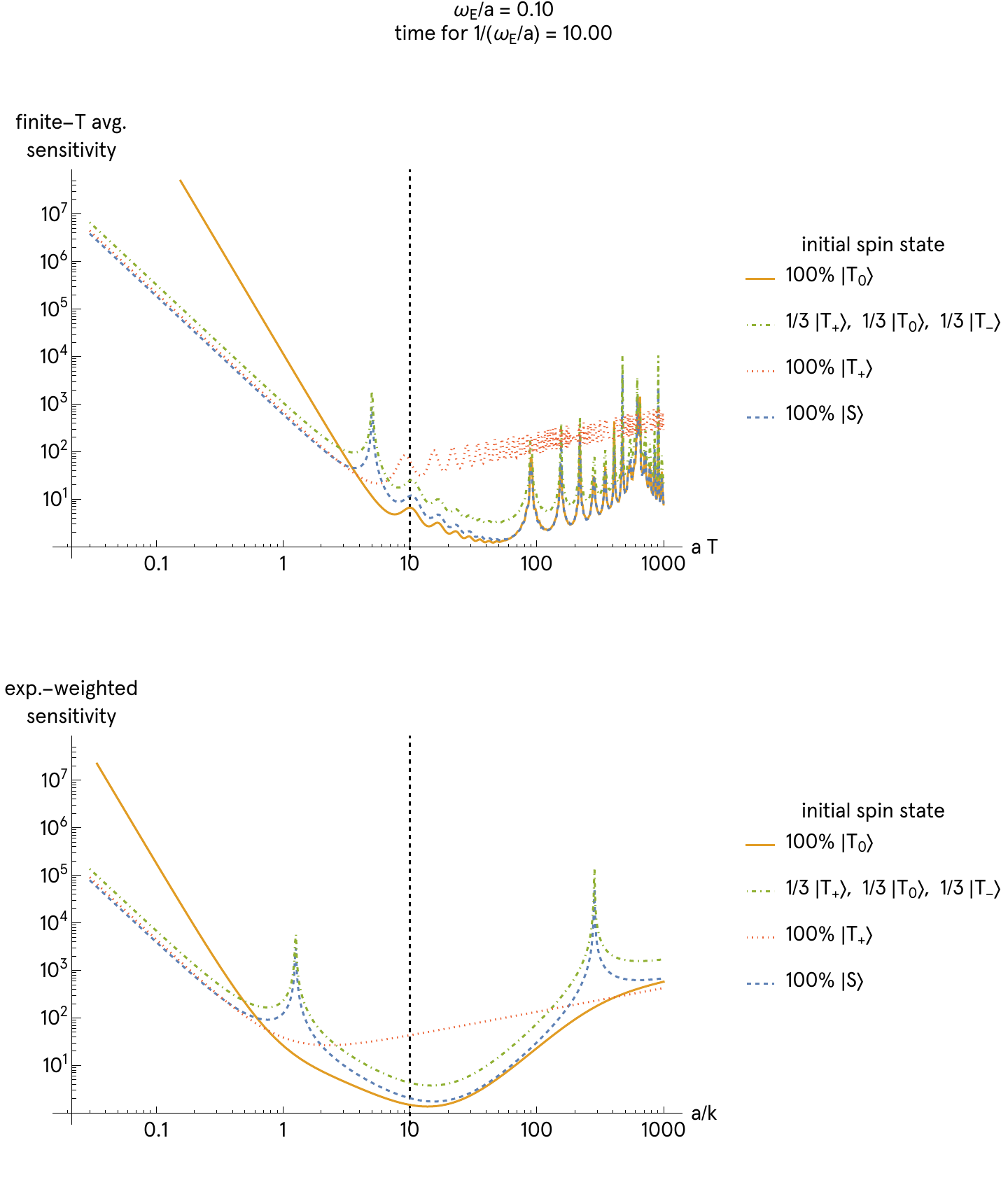}
  \captionsetup[figure]{aboveskip=4pt,belowskip=8pt}
  \captionof{figure}{\justifying \textbf{Sensitivities for the operating point \(\frac{\omega_{\rm E}}{a} = 0.1\).}
The vertical line marks the characteristic interrogation time associated with sensing the bias field $\omega_{\rm E}$.
The finite-$T$ average sensitivity is more oscillatory, while the exponentially-weighted sensitivity is smoother. $\ket{S}$ as an initial state almost always minimizes the sensitivity for shorter times, whereas $\ket{T_0}$ minimizes the sensitivity for longer times, in the time interval considered.}
   \label{eta01}
  \end{figure}

\begin{figure}[H]
  \centering
  \includegraphics[width=\textwidth]{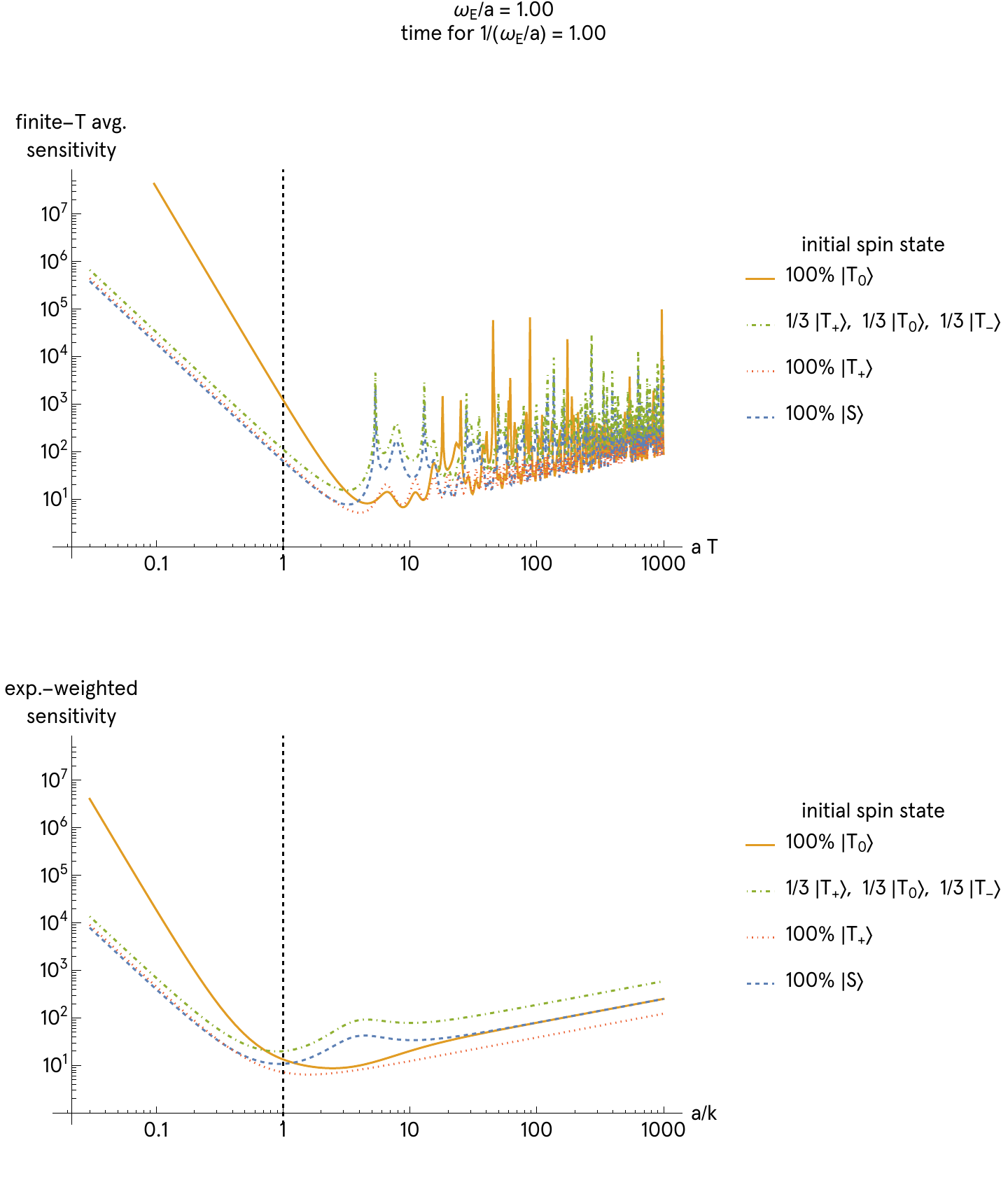}
  \captionsetup[figure]{aboveskip=4pt,belowskip=8pt}
  \captionof{figure}{\justifying \textbf{Sensitivities for the operating point \(\frac{\omega_{\rm E}}{a} = 1\).}
  The vertical line marks the characteristic interrogation time associated with sensing the bias field $\omega_{\rm E}$.
The finite-$T$ average sensitivity is more oscillatory, while the exponentially-weighted sensitivity is smoother. $\ket{S}$ and $\ket{T_{+}}$ as initial states minimize the sensitivity for shorter times in both curves, whereas $\ket{T_+}$ clearly minimizes the exponentially-weighted sensitivity for longer times, in the time interval considered. The oscillatory character of the finite-$T$ average sensitivity makes it hard for trends to emerge. Changes in the curve shapes and in the best performing initial state illustrate that optimal magnetometry depends nontrivially on both the time parameter and the operating point.}
   \label{eta1}
  \end{figure}

\begin{figure}[H]
  \centering
  \includegraphics[width=\textwidth]{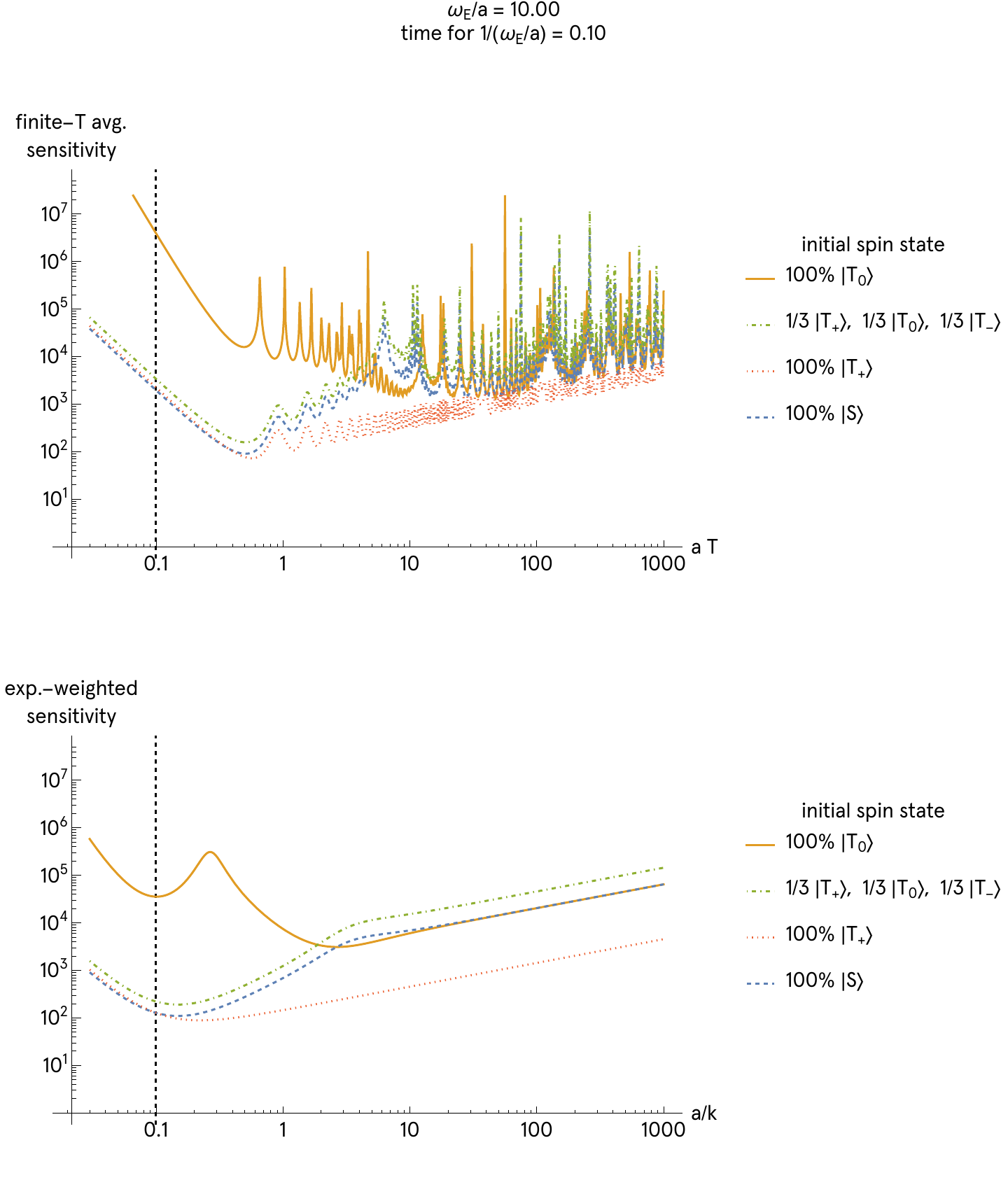}
  \captionsetup[figure]{aboveskip=4pt,belowskip=8pt}
  \captionof{figure}{\justifying \textbf{Sensitivities for the operating point \(\frac{\omega_{\rm E}}{a} = 10\).}
  The vertical line marks the characteristic interrogation time associated with sensing the bias field $\omega_{\rm E}$.
The finite-$T$ average sensitivity is more oscillatory, while the exponentially-weighted sensitivity is smoother. $\ket{S}$ and $\ket{T_{+}}$ as initial states minimize the sensitivity for shorter times in both curves, whereas $\ket{T_+}$ clearly minimizes the sensitivity for longer times, in the time interval considered.}
   \label{eta10}
  \end{figure}

\noindent Figs.~\ref{eta01} through \ref{eta10} show the shot noise-limited sensitivity of the two singlet yield observables, as a function of the time parameter $a T = \frac{a}{k}$, evaluated at three representative log-spaced operating points $\frac{\omega_{\rm E}}{a} = 0.1$, $1$,  and \(10\), with all quantities plotted in dimensionless form. The video embedded in Fig.~\ref{P6} extends this comparison across the range \(\frac{\omega_{\rm E}}{a} \in[0.1,10]\). The vertical line marks the characteristic timescale associated with sensing at the Earth-field operating point \(\omega_{\rm E}\), which is the natural finite-bias reference field considered in this section. Several general features are visible. The sensitivity for the finite-\(T\) average singlet yield indeed has strong oscillatory structure at long observation windows, whereas the sensitivity for the exponentially-weighted singlet yield is indeed much smoother. In addition, the figures make clear that both the overall value of the sensitivity and the initial state that minimizes it can change appreciably with the time parameter $a T = \frac{k}{a}$ and with the ratio \(\frac{\omega_{\rm E}}{a}\). 

\vfill
\newpage

\vfill
\newpage

\vfill
\newpage

\vfill
\newpage

\begin{figure}[H]
  \centering
  \captionsetup[figure]{aboveskip=4pt, belowskip=8pt}
  

\includemedia[
  activate=onclick,
  addresource=Figures/SensitivityXFrequencyRatio.mp4,
  flashvars={source=Figures/SensitivityXFrequencyRatio.mp4}
]{%
  \fbox{%
    \parbox[c][1.164\linewidth][c]{1\linewidth}{%
      \centering
      \normalsize video link:\\[1em]
      \normalsize\href{https://bit.ly/SensitivityXFrequencyRatio}{https://bit.ly/SensitivityXFrequencyRatio}
    }%
  }%
}{VPlayer.swf}

  \vspace*{-0.15cm}
  \caption{\justifying \textbf{Video of the sensitivities for operating points from $\frac{\omega_{\rm E}}{a} = 0.1$ to $10$.} 
The finite-$T$ average sensitivity is more oscillatory, while the exponentially-weighted sensitivity is smoother. Changes in the curve shapes and in the best performing initial state illustrate that optimal magnetometry depends nontrivially on both the time parameter and the operating point.}
  \label{P6}
\end{figure}

\vfill
\newpage

\noindent We can also investigate the time interval through which magnetometry works well. For small \(x=\Omega_{\rm E}T\),
\begin{equation}
    F(T,\omega_{\rm E})=2x+x\cos(x)-3\sin(x)
= \frac{x^5}{60}+O(x^7) \ ,
\end{equation}

\noindent so that
\begin{equation}
    A \sim \frac{a^2\omega_{\rm E}}{8T\Omega_{\rm E}^5}\cdot \frac{(\Omega_{\rm E}T)^5}{60}
= \frac{a^2\omega_{\rm E}T^4}{480} \ ;
\end{equation}

\noindent we can conclude that the slope contribution from the bright population sector is strongly suppressed at short times:
\begin{equation}
    A\propto T^4 \ .
\end{equation}

\noindent This result reflects the fact that the finite-$T$ average yield is almost magnetoinsensitive at short interrogation times, because there has not been enough time for oscillatory dynamics to build up a measurable field dependence. In addition, the coherence term \(B\) is also suppressed at short times for the same general reason: no appreciable bright--dark phase has accumulated yet. So one (trivially) expects $\eta_{\rm E}(T)\to \infty$ as $T\to 0$.
\\

\noindent It turns out that very long experimental or averaging times are not automatically good either: even if the slope term stabilizes, the sensitivity still carries an overall factor $\sqrt{T}$.
\\

\noindent Taken together, the two preceding observations indicate that using $\langle$\textcolor{InstituteBlue}{\underline{$\left|\ket{S(t, \omega)}\right|^2$}}$\rangle_T$ as a \linebreak  magnetometry signal typically requires a finite optimal time window. If $T$ is too short, there is no phase accumulation  and the slope is too small; if $T$ is too long, averaging washes out useful structure and the \(\sqrt{T}\) factor hurts the sensitivity.
\\

\vfill

\newpage
\begin{center}
\sectionfont{Take-home messages}
\end{center}

\begin{enumerate}
    \item Quantum sensing tools, such as magnetometric sensitivity, can and should be used to describe and understand radical-pair dynamics.
    \item For this lossless, symmetric toy model, magnetometry operates best at a finite bias, e.g., perhaps, the geomagnetic field. 
    \item With respect to the initial spin preparation, the derived magnetometric sensitivity is determined entirely by \(\Delta\) and \(\Sigma\).
    \item Using the bright–dark framework, the optimal magnetometric state is one that balances dark-state phase memory with bright-state readout, as captured by optimization over the two parameters $(\Delta, \Sigma)$.
    \item In general, sensitivities are complicated functions of $(\Delta, \Sigma)$, of the ratio between the bias field's Larmor frequency and the hyperfine strength, and of the averaging interval. While some insights can be obtained (for example: the equiprobable maximal mixture of initial spin states is indeed bad for magnetosensing; if no initial population is in the $\ket{T_{\pm}}$ manifold, optimal initial spin states are either pure $\ket{S}$ or pure $\ket{T_{0}}$), a comprehensive understanding requires numerics. 
\end{enumerate}
\clearpage
\section{Discussion}
\label{sec:dis}

\noindent Radical-pair magnetosensing is fundamentally an interferometric process. Our central message is that the sensing is carried by coherent electron spin superpositions, in much the same spirit as a Ramsey-style measurement. What the chemistry ultimately reads out is not the phase directly, but the time dependence that the phase imprints on the singlet population and on singlet-derived yields. 
\\

\noindent The analytical power of this work comes from choosing the simplest model that is still physically informative. The system consists of two spin-$\frac{1}{2}$ electrons, one effectively free and the other hyperfine-coupled to a single spin-$\frac{1}{2}$ nucleus, with no direct interaction between the two electrons. The hyperfine interaction is taken to be diagonal and isotropic, with strength $a$, and an external static field produces an electron Larmor frequency $\omega$. By working in the experimentally relevant singlet--triplet basis and allowing generic incoherent initial spin preparations of the form $\alpha \lvert T_{+}\rangle$, $\beta \lvert T_{0}\rangle$, $\gamma \lvert T_{-}\rangle$, and $(1-\alpha-\beta-\gamma)\lvert S\rangle$, we can solve the dynamics exactly and compare the zero-field, high-field, and long-time limits in closed form.
\\

\noindent That exact solution reveals a bright--dark decomposition that clarifies much of the familiar radical-pair phenomenology. In the appropriate basis, one sector is symmetry-protected and dark to spin mixing, while the other participates in oscillatory spin mixing. The singlet population signal, as well as the singlet yield signals, are therefore a sum of three qualitatively different pieces: a constant dark-sector contribution, an oscillatory bright-sector \linebreak contribution, and an oscillatory bright--dark coherence term. The oscillatory terms exhibit a rich frequency structure that, to our knowledge, had not previously been analyzed in detail. In more realistic anisotropic Hamiltonians, especially in the presence of loss, the dark sector will in general be only `approximately dark'.
\\

\noindent Within the bright--dark viewpoint, the particular behavior at exactly zero field (which becomes a singularity in the long-time limit) is best understood as phase-locking. The bright--dark coherence contains a stationary relative phase when $\omega = 0$, which stays sign-definite and survives long-time averaging; this indicates that the triplets are exactly degenerate, and their dynamics can interfere coherently. For any nonzero field, that contribution oscillates and averages away because the triplets are no longer degenerate. The so-called `low-field effect' in the singlet yield curves is therefore produced by the survival of a special coherence term at exactly zero field, being tied to coherent interference and indeed resulting in `more mixing' of a particular type described below. The sign and visibility of the low-field effect, similarly to multiple other dynamical features, are controlled by the initial preparation of the $\lvert S\rangle$ and $\lvert T_{0}\rangle$ manifold. The decisive quantity is the imbalance between the initial $\lvert S\rangle$ and $\lvert T_{0}\rangle$ populations. In the bright--dark language, that imbalance sets the strength of the relevant coherence term. If the two populations are equal, the frequencies that create the low-field feature are absent, so an initial $\lvert S\rangle\leftrightarrow\lvert T_{0}\rangle$ imbalance is necessary, though not by itself sufficient, for the effect to appear. In the strict $t \to \infty$ limit, the low-field effect does appear as a dip (peak) in the singlet yield curves when there is a larger (smaller) initial population in $\ket{S}$ than in $\ket{T_0}$, which is consistent with well-known experimental results. 
In addition, and in agreement with all the observations above, once a magnetic field becomes nonzero, one can identify a qualitatively distinct field-sensitive contribution in the imaginary part of the \(\ket{S}\leftrightarrow\ket{T_0}\) coherence, which controls population transfer; similar contributions are absent from the imaginary parts of the \(\ket{S}\leftrightarrow\ket{T_\pm}\) coherences. This should not be interpreted as the appearance of a new Hamiltonian pathway, since the \(\ket{S}\leftrightarrow\ket{T_0}\) and \(\ket{S}\leftrightarrow\ket{T_\pm}\) population transfer channels are all already dynamically active at zero field. This contribution, however, does enhance the mixing between \(\ket{S}\leftrightarrow\ket{T_0}\): this is the origin of the `pathway between $\ket{S}$ and $\ket{T_0}$ that opens', in the chemists' parlance, when a nonzero field is applied.
\\

\noindent The bright--dark interpretation also explains why states that are useful for magnetometry are not simply those that maximize dark-state weight: the most informative preparations are those that retain dark-sector phase memory while still coupling strongly enough to the bright sector to produce a readable signal. In addition, we suggest that, within this lossless model, magnetometry is more likely to occur around a finite bias field than exactly at zero field. The first-order sensitivities derived for the singlet-yield observables do not imply practical linear sensing at $\omega = 0$, because the signals considered here are even in $\omega$ and lack the needed nonzero slope. The discussion therefore points toward sensing about a nonzero working point, plausibly something like the geomagnetic field, while still allowing that each protein or local molecular environment could define its own effective bias. Additional Hamiltonian terms beyond the present toy model could in principle (and likely) break the symmetry and make a genuine zero-field linear response possible. Moreover, even in this stripped-down model, the question of which initial spin state is best for magnetometry does not have a universal answer. Performance depends on the bias field, the hyperfine strength, the averaging time or radical-pair lifetime, and the initial spin state. Still, the optimization problem simplifies dramatically: the sensitivity depends only on two combinations of the initial populations, which measure the total initial population in the $\{\lvert S\rangle,\lvert T_{0}\rangle\}$ sector and the imbalance within that sector. 
The magnetometry analysis also makes clear that either pure $\lvert S\rangle$ or pure $\lvert T_{0}\rangle$ will necessarily minimize the sensitivity in the special case when no initial population is within $\lvert T_{\pm}\rangle$, and why the maximally mixed state is poor for sensing. In fact, in this toy model, an equiprobable initial mixture of $\lvert S\rangle$, $\lvert T_{0}\rangle$, $\lvert T_{+}\rangle$, and $\lvert T_{-}\rangle$ produces no magnetic field-dependent modulation of the singlet population; some degree of state preparation is therefore required. This is an important conceptual point, because it connects radical-pair sensing to the broader language of bona fide quantum sensing, which requires initialization.
\\

\noindent A related insight is that the existence of the low-field effect and good magnetometric performance should not be conflated. The low-field singularity is caused by the survival of a stationary coherence term at exactly zero field, whereas useful first-order sensing requires operation at a bias point where the signal changes linearly with field. Those are different requirements. A configuration can therefore be poor for generating a dramatic low-field feature and yet still be compatible with good magnetometry; conversely, a conspicuous zero-field structure does not by itself guarantee good finite-bias sensitivity.
\\

\noindent Across all of these results, a single internal parameter organizes the scale of the problem: the hyperfine strength $a$. It defines the natural oscillation rate of the coherent dynamics, including the surviving frequencies in the limits of zero and very high fields; in this sense, $a$ sets the `internal clock' of the radical pair: appreciable coherent singlet–triplet population transfer happens for times longer than $\sim \frac{1}{a}$. The hyperfine strength also determines the overall magnetic field scale at which extrema in the yield curves appear (they appear as linear functions of $a$). Even the magnetosensitivity formulas simplify when written in dimensionless form, because the relevant control variables collapse naturally onto ratios such as $\frac{\omega}{a}$. This makes the toy model especially transparent: although the dynamics are rich, the internal scale that governs them is easy to identify.
\\

\noindent Again, at its core, this is a story of interference: most of the phenomena investigated here are simply different facets of the same coherent interference process. At the most basic level, destructive interference is what isolates the dark state and organizes the dynamics into bright and dark sectors. Once the singlet signal is reconstructed in the physical $\{\ket{S}, \ket{T_0}, \ket{T_{\pm}}\}$ basis, interference reappears through the bright--dark coherence term.
Coherence also clarifies why the zero-field point is singular: 
at exactly zero field the system cannot \linebreak energetically distinguish between the triplet substates, so amplitudes propagating through different triplet channels can interfere coherently indefinitely in this lossless model. The moment a field is turned on, however small, the triplet substates acquire distinct energies, their relative phases begin to drift, and the interference ceases to be stationary. 
For the same reason, coherence also enters the explanation of the singlet yield curves: it shapes them, and the bright--dark coherence term is precisely the contribution responsible for the low-field effect. Finally, the same interference physics underlies the magnetometric trade-off between retaining dark-sector phase memory and maintaining enough bright-\linebreak sector participation for readout.
\\

\noindent For all its simplicity, the model remains valuable precisely because it isolates the coherent mechanisms so cleanly. By neglecting recombination, decoherence, and exchange, dipolar, and anisotropic hyperfine interactions, it creates an idealized dark sector that is perfectly decoupled and a zero-field discontinuity that is correspondingly sharp. Real radical-pair systems will soften both of those features: dark states become only approximate, singularities broaden into finite-width structures, and additional interactions may either suppress or enhance low-field signatures depending on how they reshape the signals. Even so, the toy model provides a rare analytical benchmark and a clear conceptual starting point for more realistic treatments, including full open quantum system dynamics with distinct singlet and triplet rates, and explicitly directional magnetic field control.
\newpage
\clearpage
\section{Conclusion}
\label{sec:con}

\noindent We developed a physicist-friendly primer on the simplest radical-pair model. The \linebreak Hamiltonian was defined explicitly and solved analytically, and the resulting spin dynamics, singlet signal and yields, and magnetic field sensitivity were analyzed in closed form. \linebreak Decoherence and spin-selective recombination were neglected in order to isolate the basic coherent mechanisms at play. Even within this minimal setting, the model already reveals a rich structure: the analogy with Hamiltonians similar to those found in electromagnetically-\linebreak induced transparency physics; the role of the ensuing bright--dark coherence; features of experimentally relevant curves; and magnetometric optimization criteria.
\\

\noindent A particular advantage of the present treatment is that all central quantities can be written analytically and explored using both the accompanying SymPy code and the embedded Mathematica-generated interactive plots. Although exact expressions are sometimes too cumbersome to be maximally intuitive on their own, they become much more transparent when combined with limiting cases, such as $\omega \to 0$, $\omega \to \infty$, short and long interrogation times, and fast or slow lifetime weighting. In this sense, the toy model provides both a solvable benchmark and a conceptual guide for more realistic radical-pair descriptions.
\\

\noindent The natural next step is to move beyond the closed Hamiltonian picture and treat the full open quantum system dynamics with distinct singlet and triplet recombination rates, \(k_S\) and \(k_T\). 
\\

\noindent A further extension will be to include explicit magnetic field control through transverse field components \(\omega_x\) and \(\omega_y\), so that the field is no longer described only by its magnitude. This should make it possible to study directional control, anisotropic response, and more realistic magnetometric protocols. 
\\

\noindent The hope is that the present primer serves as a clean analytical starting point from which these more realistic calculations can be systematically developed.
\\

\noindent \textit{No pathways were harmed, opened, or closed during these calculations.}

\clearpage
\section{Author contributions}

\noindent Following the CRediT taxonomy \cite{Credit}, the authors confirm the following contributions:
\begin{enumerate}[leftmargin=0pt, label=\arabic*), itemsep=0pt]
\item Clarice D.\ Aiello: conceptualization; methodology; investigation; data curation; validation; formal analysis; software; writing (original draft); writing (review and editing); visualization; and funding acquisition.
\item Brian L.\ Ross: conceptualization; validation; writing (original draft); and writing (review and editing). 
\item Alessandro Lodesani: conceptualization; validation; software; and funding acquisition.
\item Morgan L.\ Sosa: validation; and software.
\end{enumerate}

\clearpage
\section{Acknowledgements}

\noindent We wholeheartedly thank the whole team at the Quantum Biology Institute for financial, operational, moral, and scientific support. 
\\

\noindent We acknowledge the use of ChatGPT for improving the readability of the text and assisting with debugging Mathematica code. The first author used Mathematica to derive all the expressions and to \href{https://bit.ly/MathematicaPlots}{plot} the results. All SymPy code was entirely written by ChatGPT, and subsequently verified by the first and fourth authors.
\\

\noindent This project was funded by a grant from the \href{http://www.quantumbiology.xyz}{Quantum Biology DAO}.
\\

\noindent The \href{http://quantumbiology.org}{Quantum Biology Institute} is a California nonprofit (501(c)(3)) focused research \linebreak organization~\cite{FRO} that performs basic research underpinning the quantum biology field in an open-science fashion. It is part of the \href{http://www.quantumbiology.eco}{Quantum Biology Ecosystem}.

\vfill 

\begin{textblock*}{3cm}(\dimexpr(\paperwidth-3cm)/2,25cm) 
    \includegraphics[width=3cm]{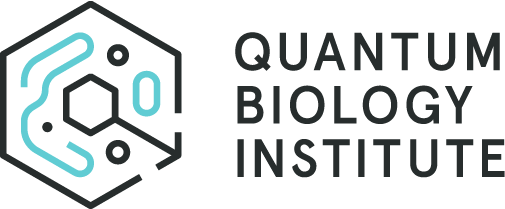} 
\end{textblock*}

\newpage

\renewcommand{\bibsection}{
  \centering\textbf{\large\sectionfont Bibliography} 
}
\addcontentsline{toc}{section}{Bibliography}  
\bibliography{bibliography.bib} 

\newpage

\providecommand{\contentsname}{\large\sectionfont Contents \vspace*{-0.097cm}}
\renewcommand{\contentsname}{\large\sectionfont Contents \vspace*{-0.097cm}}

\addtocontents{toc}{\protect\thispagestyle{tocstyle}}

\makeatletter
\renewcommand{\@cfttocfinish}{\relax}
\makeatother

\tableofcontents
\newpage

\newpage 
\vspace*{\fill} 
\begin{center} 
  \textbf{\sectionfont \Large Supplementary Information} 
\end{center}
\thispagestyle{empty} 
\vspace*{\fill} 
\newpage
\setcounter{section}{0}  
\renewcommand{\thesection}{\sectionfont SI\arabic{section}}
\renewcommand{\thesubsection}{\sectionfont SI\arabic{section}.\arabic{subsection}}
\setcounter{figure}{0}  
\renewcommand{\thefigure}{\sectionfont SI\arabic{figure}}  
\renewcommand{\theequation}{SI \arabic{equation}}  
\setcounter{equation}{0}  
\setcounter{table}{0}  
\renewcommand{\thetable}{\sectionfont SI\arabic{table}}  
\renewcommand{\figurename}{Suppl.~Fig.}
\clearpage
\section{Notation table}
\label{not}

\begin{table}[H]
\centering
\small
\begin{tabularx}{0.98\textwidth}{@{}lX@{}}
\toprule
symbol & definition \\
\midrule

\multicolumn{2}{@{}l}{\textit{field and dynamical parameters}} \\[2pt]
$\omega$ & electron Larmor frequency due to a constant external magnetic field,
\(\omega \equiv \frac{\mu g B}{\hbar}\) \\

$a$ & isotropic hyperfine coupling strength, assumed here $a > 0$ \\

$\Omega$ & effective oscillation frequency,
\(\Omega \equiv \sqrt{a^2+\omega^2}\) \\

$\theta$ & mixing angle,
\(\theta \equiv \frac{\Omega t}{2}\) \\[4pt]

\phantom{a} & \phantom{a} \\

\multicolumn{2}{@{}l}{\textit{state amplitudes and density matrix}} \\[2pt]
$\ket{X(t)}$ & component of the state \(\ket{\psi(t)}\) along \(\ket{X}\),
$|X\rangle\langle X|\psi(t)\rangle$ \\

$\bigl|\ket{X(t)}\bigr|^2$ & population in state \(\ket{X}\) at time \(t\),
$\langle\psi(t) | X \rangle \langle X | \psi(t)\rangle$ \\

$\rho_{XY}(t)$ & density matrix element between \(\ket{X}\) and \(\ket{Y}\),
$\langle X | \rho(t) | Y \rangle$ \\[4pt]

\phantom{a} & \phantom{a} \\

\multicolumn{2}{@{}l}{\textit{initial state populations}} \\[2pt]
$\alpha$ & initial population in \(\ket{T_+}\) \\

$\beta$ & initial population in \(\ket{T_0}\) \\

$\gamma$ & initial population in \(\ket{T_-}\) \\

$(1-\alpha-\beta-\gamma)$ & initial population in \(\ket{S}\) \\

$\Sigma$ & total initial population in the \(\{\ket{S},\ket{T_0}\}\) sector,
\(\Sigma \equiv (1-\alpha-\beta-\gamma)+\beta\) \\

$\Delta$ & initial population imbalance in the \(\{\ket{S},\ket{T_0}\}\) sector,
\(\Delta \equiv (1-\alpha-\beta-\gamma)-\beta\) \\

$\zeta$ & polarized triplet imbalance,
\(\zeta \equiv \alpha-\gamma\) \\[4pt]

\phantom{a} & \phantom{a} \\

\multicolumn{2}{@{}l}{\textit{yield definitions}} \\[2pt]
$T$ & observation window used in the finite-\(T\) average singlet yield \\

$\langle \textcolor{InstituteBlue}{\underline{\left|\ket{S(t,\omega)}\right|^2}}\rangle_T$ 
& finite-\(T\) average singlet yield \\

$k$ & radical-pair recombination rate used in the exponentially-weighted singlet yield \\

$\Phi_k(\omega)$ & exponentially-weighted singlet yield \\[4pt]

\phantom{a} & \phantom{a} \\

\multicolumn{2}{@{}l}{\textit{magnetometry notation}} \\[2pt]
$\eta$ & shot noise-limited magnetometry sensitivity \\

$\omega_{\rm E}$ & electron Larmor frequency due to the geomagnetic field \\

$\Omega_{\rm E}$ & effective oscillation frequency under the geomagnetic field,
\(\Omega_{\rm E} \equiv \sqrt{a^2+\omega_{\rm E}^2}\) \\

\bottomrule
\end{tabularx}
\end{table}

\renewcommand{\theequation}{SI~\arabic{equation}}
\renewcommand{\theHequation}{SI.\arabic{equation}}
\setcounter{equation}{0}

\clearpage
\section{Hamiltonian definition: extended discussion}
\label{sisec:hamiltonian}

\noindent We reconsider the Hamiltonian generating the electron spin dynamics of interest, Eq.~(\ref{eq:H}). 
\\

\noindent We let the magnetic field be along all directions in the lab frame, $\hat{B} = (B_x, B_y, B_z)$; and we let the hyperfine tensor be expressed in its most generic form, which is symmetric ($A_{jk} = A_{kj}$):
\begin{equation}
    \hat{A} = \begin{pmatrix}
A_{xx} & A_{xy} & A_{xz} \\
A_{xy} & A_{yy} & A_{yz} \\
A_{xz} & A_{yz} & A_{zz}
\end{pmatrix} \ ;
\end{equation}
\noindent antisymmetric components of the hyperfine tensor generally arise only when relativistic spin-orbit effects are included. Fermi-contact terms are obviously symmetric, since they are proportional to
$\mathbb{1}_{3}$, where $\mathbb{1}_{3}$ is the $3\times3$ identity matrix, and dipole-dipole interaction terms yield symmetric, traceless contributions. 
\\

\noindent The Hamiltonian now reads:
\begin{equation}
    \mathcal{H} =
\omega_x (\hat{S}_{x,1} + \hat{S}_{x,2}) + \omega_y (\hat{S}_{y,1} + \hat{S}_{y,2}) + \omega_z (\hat{S}_{z,1} + \hat{S}_{z,2}) + \hat{S}_2\cdot \hat{A} \cdot \hat{I} \ ,
\label{H}
\end{equation}
\noindent where the Larmor frequencies of the electrons are given by $\omega_q \equiv \frac{\mu\,|g|\,B_{q}}{\hbar}$, for $q = \{x, y, z\}$.
\\

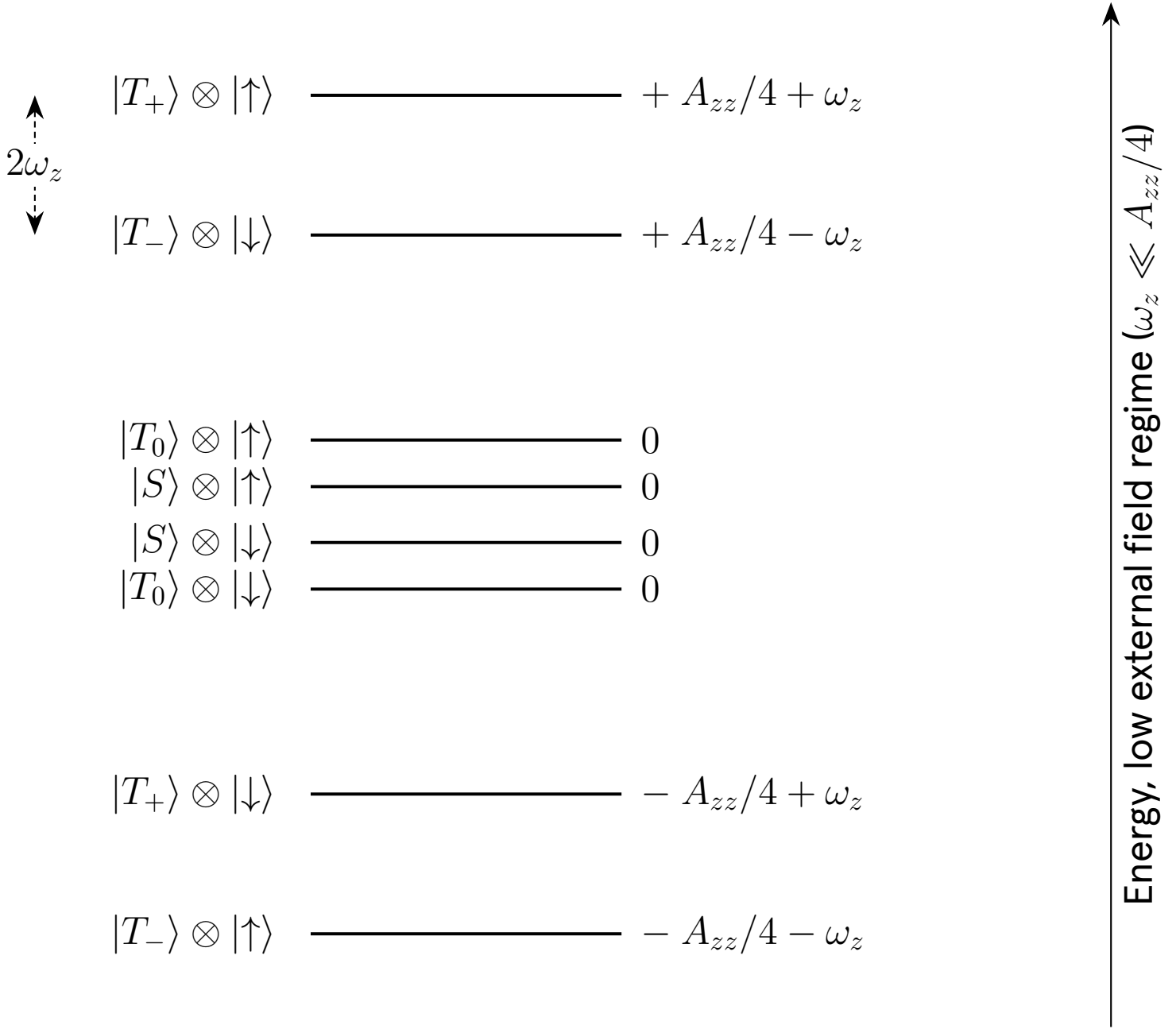
\begin{figure}[H]
  \centering

  \pgfmathsetmacro{\Scale}{1.3}
  \pgfmathsetlengthmacro{\LevelLW}{\Scale*1.5pt}

  \makebox[\textwidth][c]{%
  \begin{tikzpicture}[
      scale=\Scale,
      y=1.2cm,
      every node/.style={transform shape},
      level/.style={draw, line width=1.5pt, color=black},
      energy/.style={font=\large, color=black},
      state/.style={font=\large, color=black},
  ]

  \def\levelwidth{4}
  \def\labeloffset{3.2}



  \coordinate (E2) at (0, 4.5);
  \draw[level] (-\levelwidth/2, 4.5) -- (\levelwidth/2, 4.5);
  \node[state, left] at (-\levelwidth/2-0.3, 4.5) {$\ket{T_+} \otimes \ket{\uparrow}$};
  \node[energy, right] at (\levelwidth/2+0.1, 4.5) {$+\ A_{zz}/4 + \omega_z$};

  \coordinate (E7) at (0, 3.0); 
  \draw[level] (-\levelwidth/2, 3.0) -- (\levelwidth/2, 3.0);
  \node[state, left] at (-\levelwidth/2-0.3, 3.0) {$\ket{T_-} \otimes \ket{\downarrow}$};
  \node[energy, right] at (\levelwidth/2+0.1, 3.0) {$+\ A_{zz}/4 - \omega_z $};


\coordinate (E4) at (0, 0.8);
\draw[level] (-\levelwidth/2, 0.8) -- (\levelwidth/2, 0.8);
\node[state, left]  at (-\levelwidth/2-0.3, 0.8) {$\ket{T_0} \otimes \ket{\uparrow}$};
\node[energy, right] at (\levelwidth/2+0.1, 0.8) {$0$};

\coordinate (E1) at (0, 0.3);
\draw[level] (-\levelwidth/2, 0.3) -- (\levelwidth/2, 0.3);
\node[state, left]  at (-\levelwidth/2-0.3, 0.3) {$\ket{S} \otimes \ket{\uparrow}$};
\node[energy, right] at (\levelwidth/2+0.1, 0.3) {$0$};

\coordinate (E5) at (0, -0.3);
\draw[level] (-\levelwidth/2, -0.3) -- (\levelwidth/2, -0.3);
\node[state, left]  at (-\levelwidth/2-0.3, -0.3) {$\ket{S} \otimes \ket{\downarrow}$};
\node[energy, right] at (\levelwidth/2+0.1, -0.3) {$0$};

\coordinate (E8) at (0, -0.8);
\draw[level] (-\levelwidth/2, -0.8) -- (\levelwidth/2, -0.8);
\node[state, left]  at (-\levelwidth/2-0.3, -0.8) {$\ket{T_0} \otimes \ket{\downarrow}$};
\node[energy, right] at (\levelwidth/2+0.1, -0.8) {$0$};

  \coordinate (E6) at (0, -3.0);
  \draw[level] (-\levelwidth/2, -3.0) -- (\levelwidth/2, -3.0);
  \node[state, left] at (-\levelwidth/2-0.3, -3.0) {$\ket{T_+} \otimes \ket{\downarrow}$};
  \node[energy, right] at (\levelwidth/2+0.1, -3.0) {$-\ A_{zz}/4 + \omega_z $};

  \coordinate (E3) at (0, -4.5);
  \draw[level] (-\levelwidth/2, -4.5) -- (\levelwidth/2, -4.5);
  \node[state, left] at (-\levelwidth/2-0.3, -4.5) {$\ket{T_-} \otimes \ket{\uparrow}$};
  \node[energy, right] at (\levelwidth/2+0.1, -4.5) {$ -\ A_{zz}/4 - \omega_z$};

  \draw[{Stealth[length=4mm]}-{Stealth[length=4mm]}, thick, black, densely dashed]
    (-\levelwidth/2-1.8-0.75-1, 4.5) -- (-\levelwidth/2-1.8-0.75-1, 3.0);
  \node[font=\large, black, fill=white, inner sep=2pt]
      at (-\levelwidth/2-1.8-0.75-1, 3.75) {$2\omega_z$};

  \draw[-{Stealth[length=4mm]}, thick, black]
      (\levelwidth/2 + 6 - 1+0.9 + 0.4, -5.5) -- (\levelwidth/2 + 6 -1+0.9 +0.4, 5.5);
  \node[font=\large, rotate=90, black]
      at (\levelwidth/2 + 6.8 -1+0.9, 0) {Energy, low external field regime ($\omega_z \ll A_{zz}/4$)};

  \end{tikzpicture}%
  }

  \captionsetup[figure]{aboveskip=4pt,belowskip=8pt}
  \captionof{figure}{\justifying \textbf{Energy level diagram (diagonal terms of the Hamiltonian $\mathcal{H}$ of Eq.~\eqref{H}) in the low external magnetic field regime ($\omega_z \ll A_{zz}/4$; typically, e.g., geomagnetic conditions).}
  Quantum states are labeled in the basis $\ket{\text{electron spins}} \otimes \ket{\text{nuclear spin}}$.  The quantization axis is along $z$. The exact Zeeman splitting between the $\ket{T_{-}}$ and $\ket{T_{+}}$ manifolds is $2 \omega_z$; this term does not dominate the energy structure and is a perturbation on top of the hyperfine coupling $A_{zz}$. 
  The hyperfine coupling $A_{zz}$ dominates the energy structure, creating a clear separation between nuclear spin manifolds. Transverse magnetic fields (along $x, y$) induce transitions between quantum states and do not appear here. (The electron energies could be time-modulated by $\omega_z(t) = \omega_z^{\text{DC}} + \omega_z^{\text{AC}} \cos(2\pi f_z t)$.) }
   \label{SIFig:energy_low}
  \end{figure}

\noindent Solving for the energies in the above basis yields the equivalent diagrams of Suppl.~Figs.~\ref{SIFig:energy_low} and~\ref{SIFig:energy_high}, depicting regimes in which the externally applied magnetic field is, respectively, much smaller (larger) than typical hyperfine interaction strengths along a quantization axis, say along $z$. These figures differ from those in the main text in that the subindex $_z$ is kept in $\omega_z$; and in that the term $A_{zz}$ has not yet been replaced by the main text value's $a$.
\\

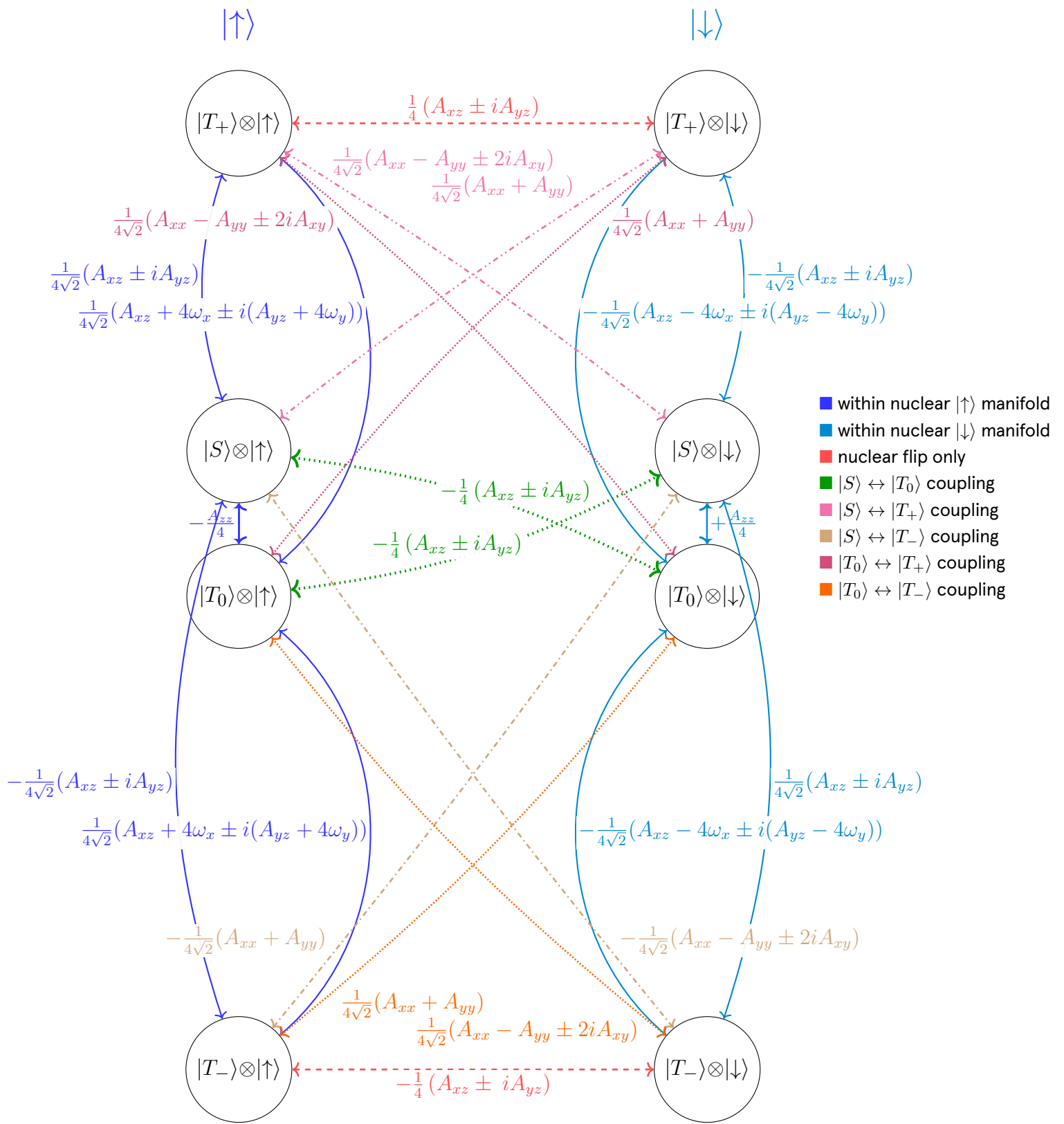
\begin{figure}[H]
\vspace*{-2cm}
\centering

\hspace*{-2cm}\begin{tikzpicture}[
    state/.style={draw, circle, minimum size=2cm, font=\Large\bfseries},
    arrow/.style={-{Stealth[length=2.5mm]}, thick},
    label/.style={fill=white, inner sep=1pt, opacity=0.95, text opacity=1},
] 

\def\colsep{9}
\def\rowsep{2.8}

\node[state] (S_up)  at (0,  1*\rowsep) {$\scriptstyle\ket{S} \otimes \ket{\uparrow}$};
\node[state] (Tp_up) at (0,  3.25*\rowsep) {$\scriptstyle\ket{T_+} \otimes \ket{\uparrow}$};
\node[state] (Tm_up) at (0, -3.25*\rowsep) {$\scriptstyle\ket{T_-} \otimes \ket{\uparrow}$};
\node[state] (T0_up) at (0,  0*\rowsep) {$\scriptstyle\ket{T_0} \otimes \ket{\uparrow}$};

\node[state] (S_down)  at (\colsep,  1*\rowsep) {$\scriptstyle\ket{S} \otimes \ket{\downarrow}$};
\node[state] (Tp_down) at (\colsep,  3.25*\rowsep) {$\scriptstyle\ket{T_+} \otimes \ket{\downarrow}$};
\node[state] (Tm_down) at (\colsep, -3.25*\rowsep) {$\scriptstyle\ket{T_-} \otimes \ket{\downarrow}$};
\node[state] (T0_down) at (\colsep,  0*\rowsep) {$\scriptstyle\ket{T_0} \otimes \ket{\downarrow}$};


\draw[arrow, <->, blue!80, line width=1.2pt]
  (S_up) -- (T0_up)
  node[label, midway, left] {$-\frac{A_{zz}}{4}$};

\draw[arrow, <->, blue!80, bend left=18]
  (S_up) to node[label, pos=0.55, left]
  {$\frac{1}{4\sqrt{2}}(A_{xz}\pm iA_{yz})$} (Tp_up);

\draw[arrow, <->, blue!80, bend right=18]
  (S_up) to node[label, pos=0.55, left]
  {$-\frac{1}{4\sqrt{2}}(A_{xz}\pm iA_{yz})$} (Tm_up);

\draw[arrow, <->, blue!80, bend left=50]
  (Tp_up) to node[label, pos=0.40, left]
  {$\frac{1}{4\sqrt{2}}(A_{xz}+4\omega_x \pm i(A_{yz}+4 \omega_y))$} (T0_up);

\draw[arrow, <->, blue!80, bend right=50]
  (Tm_up) to node[label, pos=0.50, left]
  {$\frac{1}{4\sqrt{2}}(A_{xz}+4\omega_x \pm i(A_{yz} + 4 \omega_y))$} (T0_up);


\draw[arrow, <->, cyan!70!blue, line width=1.2pt]
  (S_down) -- (T0_down)
  node[label, midway, right] {$+\frac{A_{zz}}{4}$};

\draw[arrow, <->, cyan!70!blue, bend right=18]
  (S_down) to node[label, pos=0.55, right]
  {$-\frac{1}{4\sqrt{2}}(A_{xz}\pm iA_{yz})$} (Tp_down);

\draw[arrow, <->, cyan!70!blue, bend left=18]
  (S_down) to node[label, pos=0.55, right]
  {$\frac{1}{4\sqrt{2}}(A_{xz}\pm iA_{yz})$} (Tm_down);

\draw[arrow, <->, cyan!70!blue, bend right=50]
  (Tp_down) to node[label, pos=0.40, right]
  {$-\frac{1}{4\sqrt{2}}(A_{xz}-4\omega_x \pm i(A_{yz}-4\omega_y))$} (T0_down);

\draw[arrow, <->, cyan!70!blue, bend left=50]
  (Tm_down) to node[label, pos=0.50, right]
  {$-\frac{1}{4\sqrt{2}}(A_{xz}-4\omega_x \pm i (A_{yz}-4 \omega_y))$} (T0_down);




\draw[arrow, <->, red!70, dashed, line width=1pt]
  (Tp_up) -- (Tp_down)
  node[label, midway, above] {$\frac{1}{4}\left(A_{xz} \pm iA_{yz}\right)$};

\draw[arrow, <->, red!70, dashed, line width=1pt]
  (Tm_up) -- (Tm_down)
  node[label, midway, below] {$-\frac{1}{4}\left(A_{xz} \pm\ iA_{yz}\right)$};


\draw[arrow, <->, green!60!black, dotted, line width=1.5pt]
  (S_up) to[bend left=10]
  node[label, pos=0.60, above] {$-\frac{1}{4}\left(A_{xz} \pm iA_{yz}\right)$} (T0_down);

\draw[arrow, <->, green!60!black, dotted, line width=1.5pt]
  (T0_up) to[bend right=10]
  node[label, pos=0.40, above] {$-\frac{1}{4}\left(A_{xz} \pm iA_{yz}\right)$} (S_down);

\draw[arrow, <->, purple!70, densely dotted, line width=1pt]
  (Tp_up) to[bend left=6]
  node[label, pos=0.12, below left] {$\frac{1}{4\sqrt{2}}(A_{xx}-A_{yy}\pm2 iA_{xy})$} (T0_down);

\draw[arrow, <->, purple!70, densely dotted, line width=1pt]
  (T0_up) to[bend left=6]
  node[label, pos=0.88, below right] {$\frac{1}{4\sqrt{2}}(A_{xx}+A_{yy})$} (Tp_down);

\draw[arrow, <->, brown!70, dash dot, line width=1pt]
  (Tm_up) to[bend right=2]
  node[label, pos=0.12, above left] {$-\frac{1}{4\sqrt{2}}(A_{xx}+A_{yy})$} (S_down);

\draw[arrow, <->, brown!70, dash dot, line width=1pt]
  (Tm_down) to[bend left=2]
  node[label, pos=0.12, above right] {$-\frac{1}{4\sqrt{2}}(A_{xx}-A_{yy}\pm2 iA_{xy})$} (S_up);

\draw[arrow, <->, orange!80!red, densely dotted, line width=1pt]
  (Tm_up) to[bend right=6]
  node[label, pos=0.12, below right] {$\frac{1}{4\sqrt{2}}(A_{xx}+A_{yy})$} (T0_down);

\draw[arrow, <->, orange!80!red, densely dotted, line width=1pt]
  (T0_up) to[bend right=6]
  node[label, pos=0.95, below left] {$\frac{1}{4\sqrt{2}}(A_{xx}-A_{yy}\pm2 iA_{xy})$} (Tm_down);

\draw[arrow, <->, magenta!70, dash dot dot, line width=1pt]
  (Tp_up) to[bend left=3]
  node[label, pos=0.10, above right] {$\frac{1}{4\sqrt{2}}(A_{xx}-A_{yy}\pm2 iA_{xy})$} (S_down);

\draw[arrow, <->, magenta!70, dash dot dot, line width=1pt]
  (Tp_down) to[bend right=3]
  node[label, pos=0.20, above left] {$\frac{1}{4\sqrt{2}}(A_{xx}+A_{yy})$} (S_up);

\node[font=\Large, blue!80] at (0, 11) {$\ket{\uparrow}$};
\node[font=\Large, cyan!70!blue] at (\colsep, 11) {$\ket{\downarrow}$};

\node[anchor=north west, align=left, font=\footnotesize] at (11, 4) {   
  \textcolor{blue!80}{$\blacksquare$} within nuclear $\ket{\uparrow}$ manifold \\
  \textcolor{cyan!70!blue}{$\blacksquare$} within nuclear $\ket{\downarrow}$ manifold\\
  \textcolor{red!70}{$\blacksquare$} nuclear flip only\\
  \textcolor{green!60!black}{$\blacksquare$} $\ket{S} \leftrightarrow \ket{T_{0}}$ coupling\\
  \textcolor{magenta!70}{$\blacksquare$} $\ket{S} \leftrightarrow \ket{T_{+}}$ coupling\\
  \textcolor{brown!70}{$\blacksquare$} $\ket{S} \leftrightarrow \ket{T_{-}}$ coupling\\
  \textcolor{purple!70}{$\blacksquare$} $\ket{T_{0}} \leftrightarrow \ket{T_{+}}$ coupling\\
  \textcolor{orange!80!red}{$\blacksquare$} $\ket{T_{0}} \leftrightarrow \ket{T_{-}}$ coupling\\
};

\end{tikzpicture}

 \captionsetup[figure]{aboveskip=4pt,belowskip=8pt}
  \captionof{figure}{\justifying \textbf{Couplings diagram (off-diagonal terms of the Hamiltonian $\mathcal{H}$ of Eq.~\eqref{H}) for a generic (symmetric) hyperfine tensor.}
  Quantum states are labeled in the basis $\ket{\text{electron spins}} \otimes \ket{\text{nuclear spin}}$.  The quantization axis is along $z$. 
   Magnetic fields along the quantization direction $z$ change energy levels and do not appear here.}
   \label{Fig:couplings_full}
\end{figure}

\noindent In turn, solving for the couplings in the above basis results in the somewhat unilluminating diagram of Suppl.~Fig.~\ref{Fig:couplings_full}. Defining the following constants, which are the absolute values of complex matrix elements, where the choice of sign is determined by the specific coupling being considered:
\begin{align}
    C_1 &\equiv \left| \pm \frac{1}{4} (A_{xz} \pm i A_{yz}) \right| \ ; \\
    C_2 &\equiv \left| \pm \frac{1}{4\sqrt{2}} (A_{xx} + A_{yy}) \right| \ ; \\ 
    C_3 &\equiv \left| \pm \frac{1}{4\sqrt{2}} (A_{xx} - A_{yy} \pm 2 i A_{xy}) \right | \ ; \\
    B^{+} &\equiv \left| \pm \frac{1}{4\sqrt{2}} [A_{xz} + 4
    \omega_x \pm i (A_{yz} + 4\omega_y)] \right| \ ; \\
    B^{-} &\equiv \left| \pm \frac{1}{4\sqrt{2}} [A_{xz} - 4\omega_x \pm i (A_{yz} - 4\omega_y)] \right| \ ,
\end{align}
\noindent we plot the slightly less cumbersome couplings diagram of Supplementary Fig.~\ref{Fig:couplings_cst}. $C_1$, $C_2$ and $C_3$ are uniquely a function of the hyperfine tensor, whereas $B^{\pm}$ depend both on the hyperfine tensor and on the transverse magnetic field (via the terms on $\omega_{x,y}$). Note that, for simplicity, in this figure, the only couplings that have retained their sign are the $\ket{S} \leftrightarrow \ket{T_{0}}$ couplings $\pm \frac{A_{zz}}{4}$.
\\

\noindent Arguably, the simplest hyperfine tensor that reproduces key experimental features of the radical-pair model  is the diagonal tensor for which $A_{xx} = A_{yy} = A_{zz} \equiv a$. In the main text, we assume $a > 0$; it is our understanding that, if $a < 0$, none of the qualitative conclusions of our work changes. Changing the sign of 
$a$ changes phase and coherence conventions but leaves the population and yields observables studied here invariant.
\\

\noindent Using $a$, the much simplified resulting couplings diagram is depicted in the main text in Fig.~\ref{fig:couplings}. There are smarter choices of parameters that simplify calculations, for example $A_{xx} = A_{yy}
= \frac{a\sqrt{2}}{2}$ and $A_{zz} = a$. However, here we want to demonstrate that an isotropic hyperfine tensor in the lab frame is enough to reproduce most well-known features of the radical pair mechanism; the exception is magnetic field directionality sensitivity, which will not arise with an isotropic hyperfine sensor for the following reason.  Because the tensor has the form \(\hat A=a \cdot \mathbb{1}_{3}\), the hyperfine interaction is simply \(a \cdot \hat{S}_2\!\cdot\!\hat{I}\), which is rotationally invariant: it does not single out any preferred molecular axis. If, in addition, the Zeeman coupling is isotropic (same scalar g-factor for the electrons) and there are no other anisotropic interactions (for example, dipolar interactions), then changing only the direction of the applied magnetic field amounts to a global rotation of the coordinate system. Such a rotation changes the basis labels but not the energies or any basis-independent observable, so the dynamics will only depend on the scalar field strength \(|\hat{B}|\), i.e., on \(\omega=\mu g|\hat{B}|\), and not on the orientation of \(\hat{B}\). In other words, an isotropic hyperfine tensor can still produce a magnetic field effect with respect to the magnitude of the field, because \(|\hat{B}|\) competes with the hyperfine scale \(a\), but it cannot by itself encode compass-like directional information: directional magnetosensitivity requires a fixed anisotropy in the molecular frame (for example, an anisotropic hyperfine tensor or an anisotropic g-tensor) so that rotating \(\hat{B}\) relative to the molecule changes the Hamiltonian in a physically meaningful way.
\\


\vfill

\newpage
\vspace*{-2cm}

\vspace*{-2cm}
\begin{figure}[H]
  \centering

  \pgfmathsetmacro{\Scale}{1.3} 
  \pgfmathsetlengthmacro{\LevelLW}{\Scale*1.5pt} 

  \makebox[\textwidth][c]{%
  \begin{tikzpicture}[
      scale=\Scale,
      y=1.2cm, 
      every node/.style={transform shape},
      level/.style={draw, line width=1.5pt, color=black},
      energy/.style={font=\large, color=black},
      state/.style={font=\large, color=black},
  ]

  \def\levelwidth{4}
  \def\labeloffset{3.2}

  \coordinate (E1) at (0, 0.3);
  \draw[level] (-\levelwidth/2, 0.3) -- (\levelwidth/2, 0.3);
  \node[state, left] at (-\levelwidth/2-0.3, 0.3) {$\ket{S} \otimes \ket{\uparrow}$};
  \node[energy, right] at (\levelwidth/2+0.1, 0.3) {$0$};

  \coordinate (E2) at (0, 4.5);
  \draw[level] (-\levelwidth/2, 4.5) -- (\levelwidth/2, 4.5);
  \node[state, left] at (-\levelwidth/2-0.3, 4.5) {$\ket{T_+} \otimes \ket{\uparrow}$};
  \node[energy, right] at (\levelwidth/2+0.1, 4.5) {$+\ \omega_z + A_{zz}/4$};

  \coordinate (E3) at (0, -4.5);
  \draw[level] (-\levelwidth/2, -4.5) -- (\levelwidth/2, -4.5);
  \node[state, left] at (-\levelwidth/2-0.3, -4.5) {$\ket{T_-} \otimes \ket{\uparrow}$};
  \node[energy, right] at (\levelwidth/2+0.1, -4.5) {$-\ \omega_z - A_{zz}/4$};

  \coordinate (E4) at (0, 0.8);
  \draw[level] (-\levelwidth/2, 0.8) -- (\levelwidth/2, 0.8);
  \node[state, left] at (-\levelwidth/2-0.3, 0.8) {$\ket{T_0} \otimes \ket{\uparrow}$};
  \node[energy, right] at (\levelwidth/2+0.1, 0.8) {$0$};

  \coordinate (E5) at (0, -0.3);
  \draw[level] (-\levelwidth/2, -0.3) -- (\levelwidth/2, -0.3);
  \node[state, left] at (-\levelwidth/2-0.3, -0.3) {$\ket{S} \otimes \ket{\downarrow}$};
  \node[energy, right] at (\levelwidth/2+0.1, -0.3) {$0$};

  \coordinate (E6) at (0, 4.0);
  \draw[level] (-\levelwidth/2, 4.0) -- (\levelwidth/2, 4.0);
  \node[state, left] at (-\levelwidth/2-0.3, 4.0) {$\ket{T_+} \otimes \ket{\downarrow}$};
  \node[energy, right] at (\levelwidth/2+0.1, 4.0) {$+\ \omega_z - A_{zz}/4$};

  \coordinate (E7) at (0, -4.0);
  \draw[level] (-\levelwidth/2, -4.0) -- (\levelwidth/2, -4.0);
  \node[state, left] at (-\levelwidth/2-0.3, -4.0) {$\ket{T_-} \otimes \ket{\downarrow}$};
  \node[energy, right] at (\levelwidth/2+0.1, -4.0) {$-\ \omega_z + A_{zz}/4$};

  \coordinate (E8) at (0, -0.8);
  \draw[level] (-\levelwidth/2, -0.8) -- (\levelwidth/2, -0.8);
  \node[state, left] at (-\levelwidth/2-0.3, -0.8) {$\ket{T_0} \otimes \ket{\downarrow}$};
  \node[energy, right] at (\levelwidth/2+0.1, -0.8) {$0$};

  \draw[{Stealth[length=4mm]}-{Stealth[length=4mm]}, thick, black, densely dashed]
    (-\levelwidth/2-1.8-0.75-1, 4.5) -- (-\levelwidth/2-1.8-0.75-1, -4.0);
  \node[font=\large, black, fill=white, inner sep=2pt]
      at (-\levelwidth/2-1.8-0.75-1, 0.25) {$2\omega_z$};

  \draw[-{Stealth[length=4mm]}, thick, black]
      (\levelwidth/2 + 6 - 1+0.9 + 0.4, -5.5) -- (\levelwidth/2 + 6 -1+0.9 + 0.4, 5.5);
  \node[font=\large, rotate=90, black]
      at (\levelwidth/2 + 6.8 -1+0.9, 0) {Energy, high external field regime ($\omega_z \gg A_{zz}/4$)};

  \end{tikzpicture}%
  }
  \captionsetup[figure]{aboveskip=4pt,belowskip=8pt}
  \captionof{figure}{\justifying \textbf{Energy level diagram (diagonal terms of the Hamiltonian $\mathcal{H}$ of Eq.~\eqref{H}) in the high external magnetic field regime ($\omega_z \gg A_{zz}/4$).}
  Quantum states are labeled in the basis $\ket{\text{electron spins}} \otimes \ket{\text{nuclear spin}}$.  The quantization axis is along $z$. The exact Zeeman splitting between the $\ket{T_{-}}$ and $\ket{T_{+}}$ manifolds is $2 \omega_z$; this term dominates the energy structure. Transverse magnetic fields (along $x, y$) induce transitions between quantum states and do not appear here. (The electron energies could be time-modulated by $\omega_z(t) = \omega_z^{\text{DC}} + \omega_z^{\text{AC}} \cos(2\pi f_z t)$.)}
   \label{SIFig:energy_high}
  
  \end{figure}

  \newpage

  \newpage
  \begin{figure}[H]
\vspace*{-2cm}
\centering

\hspace*{-1cm}\begin{tikzpicture}[
    state/.style={draw, circle, minimum size=2cm, font=\Large\bfseries},
    arrow/.style={-{Stealth[length=2.5mm]}, thick},
    label/.style={fill=white, inner sep=1pt, opacity=0.95, text opacity=1},
] 

\def\colsep{9}
\def\rowsep{2.8}

\node[state] (S_up)  at (0,  1*\rowsep) {$\scriptstyle\ket{S} \otimes \ket{\uparrow}$};
\node[state] (Tp_up) at (0,  3.25*\rowsep) {$\scriptstyle\ket{T_+} \otimes \ket{\uparrow}$};
\node[state] (Tm_up) at (0, -3.25*\rowsep) {$\scriptstyle\ket{T_-} \otimes \ket{\uparrow}$};
\node[state] (T0_up) at (0,  0*\rowsep) {$\scriptstyle\ket{T_0} \otimes \ket{\uparrow}$};

\node[state] (S_down)  at (\colsep,  1*\rowsep) {$\scriptstyle\ket{S} \otimes \ket{\downarrow}$};
\node[state] (Tp_down) at (\colsep,  3.25*\rowsep) {$\scriptstyle\ket{T_+} \otimes \ket{\downarrow}$};
\node[state] (Tm_down) at (\colsep, -3.25*\rowsep) {$\scriptstyle\ket{T_-} \otimes \ket{\downarrow}$};
\node[state] (T0_down) at (\colsep,  0*\rowsep) {$\scriptstyle\ket{T_0} \otimes \ket{\downarrow}$};


\draw[arrow, <->, blue!80, line width=1.2pt]
  (S_up) -- (T0_up)
  node[label, midway, left] {$-\frac{A_{zz}}{4}$};

\draw[arrow, <->, blue!80, bend left=18]
  (S_up) to node[label, pos=0.55, left]
  {$\frac{1}{\sqrt{2}}C_{1}$} (Tp_up);

\draw[arrow, <->, blue!80, bend right=18]
  (S_up) to node[label, pos=0.55, left]
  {$\frac{1}{\sqrt{2}}C_{1}$} (Tm_up);

\draw[arrow, <->, blue!80, bend left=50]
  (Tp_up) to node[label, pos=0.40, left]
  {$B^{+}$} (T0_up);

\draw[arrow, <->, blue!80, bend right=50]
  (Tm_up) to node[label, pos=0.50, left]
  {$B^{+}$} (T0_up);


\draw[arrow, <->, cyan!70!blue, line width=1.2pt]
  (S_down) -- (T0_down)
  node[label, midway, right] {$+\frac{A_{zz}}{4}$};

\draw[arrow, <->, cyan!70!blue, bend right=18]
  (S_down) to node[label, pos=0.55, right]
  {$\frac{1}{\sqrt{2}}C_{1}$} (Tp_down);

\draw[arrow, <->, cyan!70!blue, bend left=18]
  (S_down) to node[label, pos=0.55, right]
  {$\frac{1}{\sqrt{2}}C_{1}$} (Tm_down);

\draw[arrow, <->, cyan!70!blue, bend right=50]
  (Tp_down) to node[label, pos=0.40, right]
  {$B^{-}$} (T0_down);

\draw[arrow, <->, cyan!70!blue, bend left=50]
  (Tm_down) to node[label, pos=0.50, right]
  {$B^{-}$} (T0_down);




\draw[arrow, <->, red!70, dashed, line width=1pt]
  (Tp_up) -- (Tp_down)
  node[label, midway, above] {$C_{1}$};

\draw[arrow, <->, red!70, dashed, line width=1pt]
  (Tm_up) -- (Tm_down)
  node[label, midway, below] {$C_{1}$};


\draw[arrow, <->, green!60!black, dotted, line width=1.5pt]
  (S_up) to[bend left=10]
  node[label, pos=0.60, above] {$C_{1}$} (T0_down);

\draw[arrow, <->, green!60!black, dotted, line width=1.5pt]
  (T0_up) to[bend right=10]
  node[label, pos=0.40, above] {$C_{1}$} (S_down);

\draw[arrow, <->, purple!70, densely dotted, line width=1pt]
  (Tp_up) to[bend left=6]
  node[label, pos=0.12, below left] {$C_{3}$} (T0_down);

\draw[arrow, <->, purple!70, densely dotted, line width=1pt]
  (T0_up) to[bend left=6]
  node[label, pos=0.88, below right] {$C_{2}$} (Tp_down);

\draw[arrow, <->, brown!70, dash dot, line width=1pt]
  (Tm_up) to[bend right=2]
  node[label, pos=0.12, above left] {$C_{2}$} (S_down);

\draw[arrow, <->, brown!70, dash dot, line width=1pt]
  (Tm_down) to[bend left=2]
  node[label, pos=0.12, above right] {$C_{3}$} (S_up);

\draw[arrow, <->, orange!80!red, densely dotted, line width=1pt]
  (Tm_up) to[bend right=6]
  node[label, pos=0.12, below right] {$C_{2}$} (T0_down);

\draw[arrow, <->, orange!80!red, densely dotted, line width=1pt]
  (T0_up) to[bend right=6]
  node[label, pos=0.95, below left] {$C_{3}$} (Tm_down);

\draw[arrow, <->, magenta!70, dash dot dot, line width=1pt]
  (Tp_up) to[bend left=3]
  node[label, pos=0.10, above right] {$C_{3}$} (S_down);

\draw[arrow, <->, magenta!70, dash dot dot, line width=1pt]
  (Tp_down) to[bend right=3]
  node[label, pos=0.20, above left] {$C_{2}$} (S_up);

\node[font=\Large, blue!80] at (0, 11) {$\ket{\uparrow}$};
\node[font=\Large, cyan!70!blue] at (\colsep, 11) {$\ket{\downarrow}$};

\node[anchor=north west, align=left, font=\footnotesize] at (11, 4) {   
  \textcolor{blue!80}{$\blacksquare$} within nuclear $\ket{\uparrow}$ manifold \\
  \textcolor{cyan!70!blue}{$\blacksquare$} within nuclear $\ket{\downarrow}$ manifold\\
  \textcolor{red!70}{$\blacksquare$} nuclear flip only\\
  \textcolor{green!60!black}{$\blacksquare$} $\ket{S} \leftrightarrow \ket{T_{0}}$ coupling\\
  \textcolor{magenta!70}{$\blacksquare$} $\ket{S} \leftrightarrow \ket{T_{+}}$ coupling\\
  \textcolor{brown!70}{$\blacksquare$} $\ket{S} \leftrightarrow \ket{T_{-}}$ coupling\\
  \textcolor{purple!70}{$\blacksquare$} $\ket{T_{0}} \leftrightarrow \ket{T_{+}}$ coupling\\
  \textcolor{orange!80!red}{$\blacksquare$} $\ket{T_{0}} \leftrightarrow \ket{T_{-}}$ coupling\\
};

\end{tikzpicture}

 \captionsetup[figure]{aboveskip=4pt,belowskip=8pt}
  \captionof{figure}{\justifying \textbf{Couplings diagram (off-diagonal terms of the Hamiltonian $\mathcal{H}$ of Eq.~\eqref{H}) for a generic (symmetric) hyperfine tensor, with defined constants for increased readability.}
  Quantum states are labeled in the basis $\ket{\text{electron spins}} \otimes \ket{\text{nuclear spin}}$.  Note that, in the text, the constants $C_1$, $C_2$, $C_3$, $B^{+}$, and $B^{-}$ were defined as absolute values for simplicity; thus, in this figure, the only couplings that have retained their sign are the $\ket{S} \leftrightarrow \ket{T_{0}}$ couplings $\pm \frac{A_{zz}}{4}$.}
   \label{Fig:couplings_cst}
\end{figure}
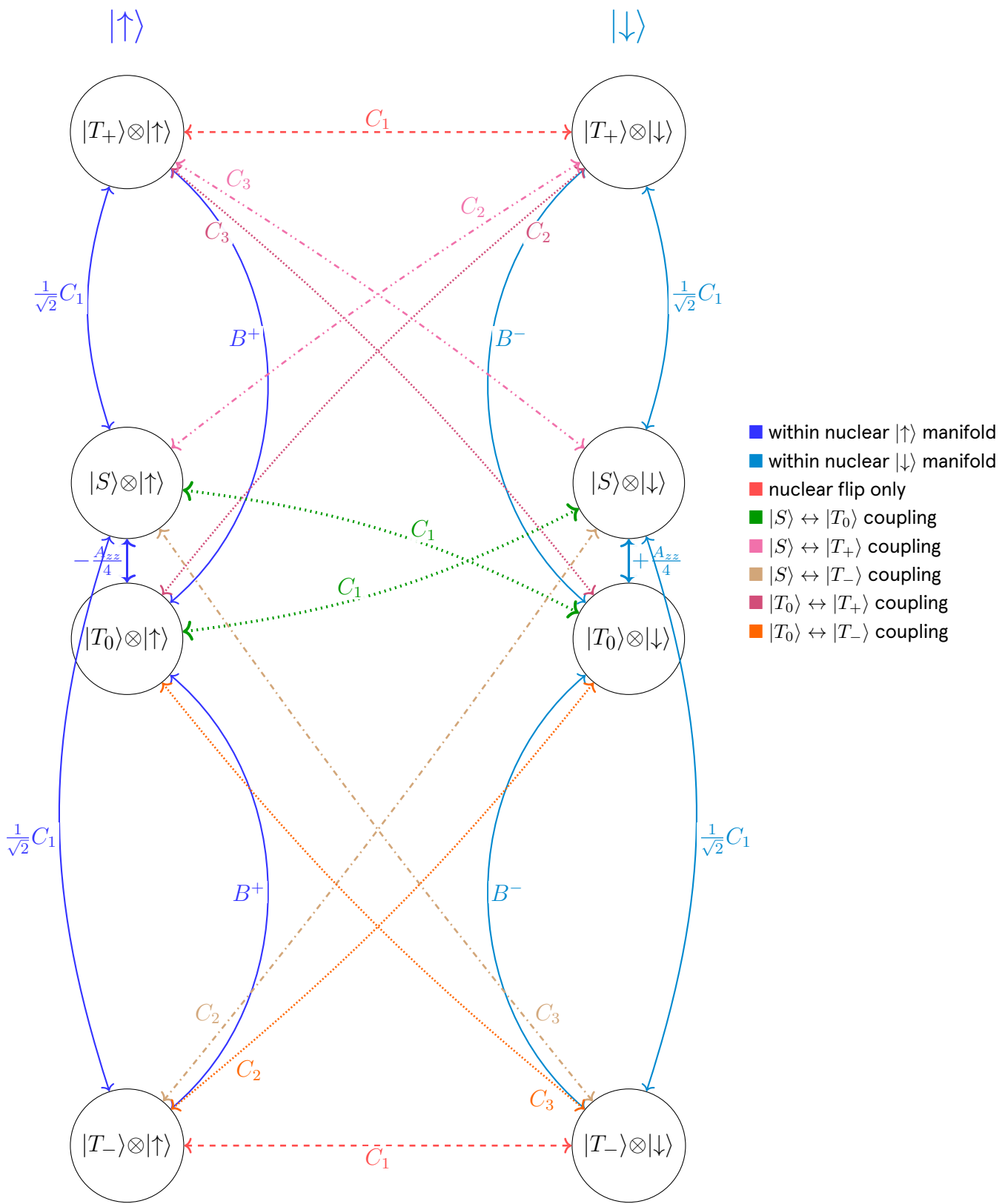

\section{Solution for the singlet population and density matrix: extended discussion}

\subsection{Signals for the full eight-level dynamics can be derived using the dynamics of only one interacting three-state subspace}
\label{explanation}

\noindent Here, we make explicit how the dynamics of the second subspace is related to the first subspace's, which was solved in the main text.
\\

\noindent Using the basis $\left\{
\ket{S}\otimes\ket{\downarrow},
\ket{T_0}\otimes\ket{\downarrow},
\ket{T_-}\otimes\ket{\uparrow}
\right\}$, the Hamiltonian of the second subspace is 
\begin{equation}
\mathcal{H}_{5}=
\begin{pmatrix}
0 & \dfrac{a}{4} & -\dfrac{a}{2\sqrt2}\\[4pt]
\dfrac{a}{4} & 0 & \dfrac{a}{2\sqrt2}\\[4pt]
-\dfrac{a}{2\sqrt2} & \dfrac{a}{2\sqrt2} & -\dfrac{a}{4}-\omega
\end{pmatrix} \ .
\label{eq:Hdown1}
\end{equation}

\noindent One may also make the $\ket{S} \leftrightarrow \ket{T_{0}}$ symmetry manifest in the same way as for the first subspace. Defining
\begin{equation}
\ket{D'}\equiv \frac{\ket{T_0}+\ket{S}}{\sqrt2} \ ,
\qquad
\ket{B'}\equiv \frac{\ket{T_0}-\ket{S}}{\sqrt2} \ ,
\label{eq:ABminus}
\end{equation}
so that
\[
\ket{S}=\frac{\ket{D'}-\ket{B'}}{\sqrt2} \ ,
\qquad
\ket{T_0}=\frac{\ket{D'}+\ket{B'}}{\sqrt2}\ ,
\]
then the Hamiltonian of Suppl.~Eq.~\eqref{eq:Hdown1} becomes, in the basis $\{\ket{D'},\ket{B'},\ket{T_{-}}\}$,
\begin{equation}
\mathcal{H}_{6}=
\begin{pmatrix}
\dfrac{a}{4} & 0 & 0\\[4pt]
0 & -\dfrac{a}{4} & \dfrac{a}{2}\\[4pt]
0 & \dfrac{a}{2} & -\dfrac{a}{4}-\omega
\end{pmatrix} \ 
= \ 
\mathcal{H}_{2}\Big|_{\ket{T_+}\to\ket{T_-},\,\omega\to-\omega} \ .
\label{eq:Hdown2}
\end{equation}

\noindent Suppl.~Eq.~\eqref{eq:Hdown2} makes explicit that the second subspace has exactly the same structure as the first: $\ket{D'}$ is again a dark state, while the nontrivial dynamics takes place in the two-level subspace spanned by $\{\ket{B'},\ket{T_-}\}$. Therefore, once the singlet population dynamics has been solved for the first subspace, the corresponding result for the second subspace is obtained by the replacements
\begin{equation}
\ket{T_+} \to \ket{T_-} \ ,
\qquad
\omega \to -\omega \ ,
\label{sym}
\end{equation}
together with the appropriate nuclear spin flip.
\\

\noindent To understand Eq.~(\ref{signalform}), we note that the nuclear spin is by assumption incoherent and in a maximally mixed state $\frac{1}{2}\left(\ket{\uparrow}\bra{\uparrow} + \ket{\downarrow}\bra{\downarrow}  \right)$; there are no interference terms between the two interacting subspaces, because they live in orthogonal nuclear-spin sectors and because the initial nuclear state carries no coherence between $\ket{\uparrow}$ and $\ket{\downarrow}$. Hence, the singlet population signal is obtained by summing subspace probabilities, not amplitudes. 
\\

\noindent The full eight-level signal is thus given by Eq.~(\ref{signalform}), after identifying the second subspace \linebreak probabilities with the first subspace ones under the symmetries found in Suppl.~Eq.~(\ref{sym}).
\vfill
\newpage

\noindent In Supplementary Subsections~\ref{c1} through~\ref{c5}, we explicitly derive the singlet population expressions for particular choices of initial spin populations. Limits of these expressions for $\omega \to 0$, $\omega \to \infty$, and $t \to \infty$ have already been presented in Tables~\ref{Table1} and~\ref{Table2}. 

\subsection{Case 1: $\ket{T_{+}}$ as initial spin state} 
\label{c1}

\noindent In this case, $\ket{D(0)} = 0$, $\ket{B(0)} = 0$, and $\ket{T_{+}(0)} = 1$. Thus,
\begin{equation}
\begin{pmatrix}
  B(t) \\
  T_{+}(t)
\end{pmatrix} = e^{+ i t\frac{a}{4}} \ \cdot \  e^{- i t\mathcal{H}_{4}} \ \cdot \ \begin{pmatrix}
  0 \\
  1
\end{pmatrix} \ . 
\end{equation}

\noindent Explicitly, this yields for $\ket{B(t)}$:
\begin{equation}
\ket{B(t)} = e^{+ i t \frac{a}{4}} \cdot e^{- i t \frac{\omega}{2}} \cdot \left(-i\ \frac{a}{\Omega} \sin(\theta)\right) \ .
\end{equation}

\noindent Also, $\ket{D(t)} = e^{-i t\frac{a}{4}}\ket{D(0)} = 0$. Using $\ket{S(t)} = \frac{\ket{B(t)} - \ket{D(t)}}{\sqrt{2}} = \frac{\ket{B(t)}}{\sqrt{2}}$, we derive the signal:
\begin{equation}
    \left|\ket{S(t)}\right|^2_{100\% \ket{T_{+}} \ @ \ t=0} = \left|\frac{1}{\sqrt{2}} \cdot e^{+ i t\frac{a}{4}} \cdot e^{- i t \frac{\omega}{2}} \cdot \left(-i\ \frac{a}{\Omega} \sin(\theta)\right)\right|^2 = \frac{a^2}{2\Omega^2}\sin(\theta)^2 \ . 
\end{equation}

\noindent For an initial electron state $100\%\ket{T_+}$ (fractions $\alpha = 1$, $\beta = \gamma = 0$), the true eight-level signal is reduced by a factor of $\frac{1}{2}$ relative to the 3-state subspace result, because only the $\ket{T_+}\otimes\ket{\downarrow}$ branch participates in the interacting dynamics, whereas $\ket{T_+}\otimes\ket{\uparrow}$ is decoupled. We then have:
\begin{equation}
   \textcolor{InstituteBlue}{\underline{ \left|\ket{S(t)}\right|^2_{100\% \ket{T_{+}} \ @ \ t=0}}} = \frac{a^2}{4\Omega^2} \sin(\theta)^2 \ . 
   \label{tpsi}
\end{equation}

\noindent In the absence of a magnetic field ($\omega \rightarrow 0$), 
\begin{equation}
   \textcolor{InstituteBlue}{\underline{ \left|\ket{S(t)}\right|^2_{100\% \ket{T_{+}} \ @ \ t=0}}}  = \frac{1}{4} \sin(\frac{a t}{2})^2 \ .
\end{equation}

\noindent For very large magnetic fields ($\omega \rightarrow \infty$), the Zeeman splitting between $\ket{T_{0}} \leftrightarrow \ket{T_{\pm}}$ is very large; it is thus expected that, if the spin population starts in $\ket{T_{+}}$, it remains trapped there and never makes it to the singlet state. Indeed, 
\vfill
\newpage
\begin{equation}
    \lim_{\omega \to \infty}	\textcolor{InstituteBlue}{\underline{\left|\ket{S(t)}\right|^2_{100\% \ket{T_{+}} \ @ \ t=0}}}  = 0 \ . 
\end{equation}

\noindent Since we are assuming no loss takes place, the spin population coherently oscillates into and out of the singlet state indefinitely. However, \textcolor{InstituteBlue}{\underline{$\left|\ket{S(t)}\right|^2_{100\% \ket{T_{+}} \ @ \ t=0}$}} can be artificially time-averaged; the assumption being that we perform an average for $t \rightarrow \infty$. This yields:
\begin{equation}
  \langle  \textcolor{InstituteBlue}{\underline{\left|\ket{S(t)}\right|^2_{100\% \ket{T_{+}} \ @ \ t=0}}} \rangle_{t\to\infty}
   = \frac{a^2}{8\Omega^2}  \ . 
\end{equation}

\noindent For very large magnetic fields, also as expected, $\lim_{\omega \to \infty}  \langle  \textcolor{InstituteBlue}{\underline{\left|\ket{S(t)}\right|^2_{100\% \ket{T_{+}} \ @ \ t=0}}} \rangle_{t\to\infty}  = 0$.

\vfill
\newpage
\subsection{Case 2: $\ket{T_{0}}$ as initial spin state} 
\label{c2}

\noindent In this case, $\ket{D(0)} = \ket{B(0)} = \frac{1}{\sqrt{2}}$ and $\ket{T_{+}(0)} = 0$. Thus,
\begin{equation}
\begin{pmatrix}
  B(t) \\
  T_{+}(t)
\end{pmatrix} = e^{+ i t\frac{a}{4}} \ \cdot \  e^{- i t\mathcal{H}_{4}} \ \cdot \ \begin{pmatrix}
  \frac{1}{\sqrt{2}} \\
  0
\end{pmatrix} \ .
\end{equation}

\noindent Explicitly, this yields for $\ket{B(t)}$:
\begin{equation}
\ket{B(t)} = \frac{1}{\sqrt{2}} \cdot e^{+ i t\frac{a}{4}} \ \cdot \  e^{- i t\frac{\omega}{2}} \cdot \left( \cos(\theta) + i \frac{\omega}{\Omega}\sin(\theta) \right) \ ,
\end{equation}
\noindent and for $\ket{D(t)}$:
\begin{equation}
    \ket{D(t)} = e^{-i t\frac{a}{4}} \cdot \frac{1}{\sqrt{2}} \ . 
\end{equation}

\noindent Hence,
\begin{align}
    \ket{S(t)}  &= \frac{1}{2} \left[    e^{+ i t\frac{a}{4}} \ \cdot \  e^{- i t\frac{\omega}{2}} \cdot \left( \cos(\theta) + i \frac{\omega}{\Omega}\sin(\theta) \right) -  e^{-i t\frac{a}{4}}     \right] \nonumber \\
    &=  \frac{1}{2} \cdot e^{-i t\frac{a}{4}} \cdot \left[-1+ e^{i t\frac{(a-\omega)}{2}} \left(
\cos(\theta) +i \frac{\omega}{\Omega} \sin(\theta)\right) \right]  \ , 
\end{align}
whose absolute squared value can be analytically calculated to be
\begin{align}
 &\left|\ket{S(t)}\right|^2_{100\% \ket{T_{0}} \ @ \ t=0} = \nonumber \\
&\frac{1}{4} \left[1+\cos(\theta)^2-2 \cos(\theta) \cos\left(\frac{(a-\omega)t}{2}\right) 
+\frac{\omega \sin(\theta)}{\Omega^2} \left(\omega  \sin(\theta)+2 \Omega  \sin\left(\frac{(a-\omega )t}{2}\right)\right)\right] \ .
\end{align}

\noindent For an initial electron state $100\%\ket{T_0}$ (fractions $\beta = 1$, $\alpha = \gamma = 0$), the true eight-level signal is obtained by averaging the contributions from the two interacting nuclear spin branches, $\ket{T_0}\otimes\ket{\uparrow}$ and $\ket{T_0}\otimes\ket{\downarrow}$. Unlike the case of Supplementary Subsection~\ref{c1}, neither branch is decoupled, so there is no overall suppression by a factor of $\frac{1}{2}$; instead, the full signal is the average of the two 3-state subspace signals related by $\omega\to-\omega$:
\begin{equation}
   \textcolor{InstituteBlue}{\underline{ \left|\ket{S(t)}\right|^2_{100\% \ket{T_{0}} \ @ \ t=0}}} = \frac{1}{2} \bigg(  \left|\ket{S(t, \omega)}\right|^2_{100\% \ket{T_{0}} \ @ \ t=0} +  \left|\ket{S(t, -\omega)}\right|^2_{100\% \ket{T_{0}} \ @ \ t=0}  \bigg)  \ . 
\end{equation}

\noindent This yields:
\begin{align}
\textcolor{InstituteBlue}{\underline{ \left|\ket{S(t)}\right|^2_{100\% \ket{T_{0}} \ @ \ t=0}}} = 
\frac{1}{2}
-\frac{a^{2}}{4\Omega^{2}}\sin(\theta)^2
-\frac{1}{2}\cos\left(\frac{at}{2}\right)
\left[
\cos(\theta)\cos\left(\frac{\omega t}{2}\right)
+\frac{\omega}{\Omega}\sin(\theta)\sin\left(\frac{\omega t}{2}\right)
\right] \ . 
\label{t0si}
\end{align}

\noindent In the absence of a magnetic field ($\omega \rightarrow 0$), 
\begin{equation}
    \textcolor{InstituteBlue}{\underline{\left|\ket{S(t)}\right|^2_{100\% \ket{T_{0}} \ @ \ t=0}}}  = \frac{1}{4}\sin\left(\frac{at}{2}\right)^2 \ ,
    \label{34}
\end{equation}
\noindent and, for very large magnetic fields ($\omega \rightarrow \infty$), 
\begin{equation}
    \lim_{\omega \to \infty}	\textcolor{InstituteBlue}{\underline{\left|\ket{S(t)}\right|^2_{100\% \ket{T_{0}} \ @ \ t=0}}}  = \sin\left(\frac{a t}{4}\right)^2 \ . 
\end{equation}

\noindent The signal can be artificially time-averaged to yield:
\begin{equation}
  \langle  \textcolor{InstituteBlue}{\underline{\left|\ket{S(t)}\right|^2_{100\% \ket{T_{0}} \ @ \ t=0}}} \rangle_{t\to\infty} = \frac{3a^{2}+4\omega^{2}}{8\Omega^{2}} \ 
\end{equation}
\noindent if $\omega \neq 0$, and 
\begin{equation}
  \langle  \textcolor{InstituteBlue}{\underline{\left|\ket{S(t)}\right|^2_{100\% \ket{T_{0}} \ @ \ t=0}}} \rangle_{t\to\infty} = \frac{1}{8} \ 
\end{equation}
\noindent if $\omega = 0$. Note that the long-time averaged signal is discontinuous between $\omega = 0$ and $\omega = 0^{+}$. This is not the case for the full signal of Suppl.~Eq.~(\ref{t0si}), which reduces smoothly to Suppl.~Eq.~(\ref{34}). The discontinuity in the long-time average signal arises because the limits $\omega \to 0$ and $t \to \infty$ do not commute.
\\

\noindent We can once more check a simple limit in the absence of decoherence. For $\omega \to \infty$, Zeeman splittings are large; it is thus expected that, if the spin population starts in $\ket{T_{0}}$, it remains coherently oscillating between $\ket{S}$ and $\ket{T_{0}}$. Indeed, $\lim_{\omega \to \infty}  \langle  \textcolor{InstituteBlue}{\underline{\left|\ket{S(t)}\right|^2_{100\% \ket{T_{0}} \ @ \ t=0}}} \rangle_{t\to\infty} = \frac{1}{2}$.

\vfill
\newpage
\subsection{Case 3: equiprobable incoherent mixture of $\ket{T_{+}}$, $\ket{T_{0}}$, and $\ket{T_{-}}$ as initial spin state} 
\label{c3}

\noindent A more realistic spin starting condition is an equiprobable (classical) mixture of the triplet states $\ket{T_{+}}$, $\ket{T_{0}}$, and $\ket{T_{-}}$; in this case ($\alpha = \beta = \gamma = \frac{1}{3}$), we have simply:

\begin{equation}
\begin{split}
\textcolor{InstituteBlue}{%
\underline{\left|\ket{S(t)}\right|^2_{\substack{
\frac{1}{3}\ket{T_{+}},\;
\frac{1}{3}\ket{T_{0}},\;
\frac{1}{3}\ket{T_{-}}\ @\ t=0
}}}}
={}& \phantom{\ + \ }
\frac{1}{6}
\left|\ket{S(t,\omega)}\right|^2_{100\% \ket{T_{+}} \ @ \ t=0}
+
\frac{1}{6}
\left|\ket{S(t,-\omega)}\right|^2_{100\% \ket{T_{+}} \ @ \ t=0}
\\
&+
\frac{1}{6}\Big(
\left|\ket{S(t,\omega)}\right|^2_{100\% \ket{T_{0}} \ @ \ t=0}
+
\left|\ket{S(t,-\omega)}\right|^2_{100\% \ket{T_{0}} \ @ \ t=0}
\Big)\ .
\end{split}
\end{equation}

\noindent The signal can still be analytically calculated with the help of symbolic computation software:

\begin{equation}
\textcolor{InstituteBlue}{%
\underline{\left|\ket{S(t)}\right|^2_{\substack{
\frac{1}{3}\ket{T_{+}},\;
\frac{1}{3}\ket{T_{0}},\;
\frac{1}{3}\ket{T_{-}}\ @\ t=0
}}}}
=
\frac{1}{6}
+\frac{a^{2}}{12\Omega^{2}}\sin(\theta)^2
-\frac{1}{6}\cos\!\left(\frac{at}{2}\right)
\left[
\cos(\theta) \cos\!\left(\frac{\omega t}{2}\right)
+\frac{\omega}{\Omega}\sin(\theta) \sin\!\left(\frac{\omega t}{2}\right)
\right]\ .
\label{eq}
\end{equation}

\noindent In the absence of a magnetic field ($\omega \rightarrow 0$), 
\begin{equation}
   \textcolor{InstituteBlue}{\underline{\left|\ket{S(t)}\right|^2_{
  \frac{1}{3} \ket{T_{+}},\;
  \frac{1}{3}  \ket{T_{0}},\;
  \frac{1}{3}  \ket{T_{-}}\  @\ t=0}   
}}   = 
   \frac{1}{4}\sin\left(\frac{at}{2}\right)^{2} \ ,
\end{equation}
\noindent and, for very large magnetic fields ($\omega \rightarrow \infty$), 
\begin{equation}
    \lim_{\omega \to \infty} \textcolor{InstituteBlue}{\underline{\left|\ket{S(t)}\right|^2_{
  \frac{1}{3} \ket{T_{+}},\;
  \frac{1}{3}  \ket{T_{0}},\;
  \frac{1}{3}  \ket{T_{-}}\  @\ t=0}   
}}	 = 
   \frac{1}{3}\sin\left(\frac{a t}{4}\right)^{2} \ .
\end{equation}

\noindent The signal can be artificially time-averaged to yield:
\begin{equation}
  \langle \textcolor{InstituteBlue}{\underline{\left|\ket{S(t)}\right|^2_{
  \frac{1}{3} \ket{T_{+}},\;
  \frac{1}{3}  \ket{T_{0}},\;
  \frac{1}{3}  \ket{T_{-}}\  @\ t=0}   
}} \rangle_{t\to\infty} = \frac{5a^{2}+4\omega^{2}}{24\Omega^{2}}  \ 
\end{equation}
\noindent if $\omega \neq 0$, and 
\begin{equation}
  \langle\textcolor{InstituteBlue}{\underline{\left|\ket{S(t)}\right|^2_{
  \frac{1}{3} \ket{T_{+}},\;
  \frac{1}{3}  \ket{T_{0}},\;
  \frac{1}{3}  \ket{T_{-}}\  @\ t=0}   
}} \rangle_{t\to\infty} = \frac{1}{8} \ 
\end{equation}
\noindent if $\omega = 0$. Note that the long-time averaged signal is discontinuous between $\omega = 0$ and $\omega = 0^{+}$.
\\

\noindent For large $\omega$, $\lim_{\omega \to \infty}  \langle  \textcolor{InstituteBlue}{\underline{\left|\ket{S(t)}\right|^2_{\frac{1}{3} \ket{T_{+}},\;
  \frac{1}{3}  \ket{T_{0}},\;
  \frac{1}{3}  \ket{T_{-}}\  @\ t=0}}} \rangle_{t\to\infty} = \frac{1}{6}$ as expected. 

\vfill
\newpage
\subsection{Case 4: $\ket{S}$ as initial spin state} 
\label{c4}

\noindent For an initial electron spin state $100\%\ket{S}$ ($\alpha = \beta = \gamma = 0$), a calculation similar to those above yields for the subspace signal:
\begin{align}
 &\left|\ket{S(t)}\right|^2_{100\% \ket{S} \ @ \ t=0} = \nonumber \\
 & 
\frac{1}{4} \left[1+\cos(\theta)^2 \textcolor{red}{+} 2 \cos(\theta) \cos\left(\frac{(a-\omega )t}{2}\right) +\frac{\omega  \sin(\theta)}{\Omega^2} \left(  \omega  \sin(\theta)     \textcolor{red}{-}2 \Omega \sin\left(
\frac{(a-\omega )t}{2}\right) \right)\right] \ . 
\end{align}

\noindent We highlighted in \textcolor{red}{red} the only two signs that make the above expression differ from that for $\left|\ket{S(t)}\right|^2_{100\% \ket{T_{0}} \ @ \ t=0}$; note the (expected) high symmetry between the two results. 
\\

\noindent The true eight-level signal is obtained by averaging the contributions from the two interacting nuclear-spin branches, $\ket{S}\otimes\ket{\uparrow}$ and $\ket{S}\otimes\ket{\downarrow}$. Since neither branch is decoupled, the full signal is the average of the two 3-state subspace signals related by $\omega\to-\omega$. We obtain:
\begin{align}
\textcolor{InstituteBlue}{\underline{\left|\ket{S(t)}\right|^2_{100\% \ket{S} \ @ \ t=0}}} =  
\frac{1}{2}
-\frac{a^{2}}{4\Omega^{2}}\sin(\theta)^2
\textcolor{dred}{+}\frac{1}{2}\cos\left(\frac{at}{2}\right)
\left[
\cos(\theta)\cos\left(\frac{\omega t}{2}\right)
+\frac{\omega}{\Omega}\sin(\theta)\sin\left(\frac{\omega t}{2}\right)
\right] \ . 
\label{s}
\end{align}

\noindent The above expression differs from $\textcolor{InstituteBlue}{\left|\ket{S(t)}\right|^2_{100\% \ket{T_{0}} \ @ \ t=0}}$ only in the sign highlighted in red.
\\

\noindent In the absence of a magnetic field ($\omega \rightarrow 0$), 
\begin{equation}
   \textcolor{InstituteBlue}{\underline{\left|\ket{S(t)}\right|^2_{100\% \ket{S} \ @ \ t=0}}}  = 1-\frac{3}{4}\sin\left(\frac{at}{2}\right)^{2} \ ,
\end{equation}
\noindent and, for very large magnetic fields ($\omega \rightarrow \infty$), 
\begin{equation}
    \lim_{\omega \to \infty}	\textcolor{InstituteBlue}{\underline{\left|\ket{S(t)}\right|^2_{100\% \ket{S} \ @ \ t=0}}}  = \cos\left(\frac{at}{4}\right)^{2} \ . 
\end{equation}

\noindent We can, as usual, artificially time-average the signal to yield:
\begin{equation}
  \langle  \textcolor{InstituteBlue}{\underline{\left|\ket{S(t)}\right|^2_{100\% \ket{S} \ @ \ t=0}}} \rangle_{t\to\infty} =  \frac{3a^{2}+4\omega^{2}}{8\Omega^{2}} \ 
\end{equation}
\noindent if $\omega \neq 0$, and 
\begin{equation}
  \langle  \textcolor{InstituteBlue}{\underline{\left|\ket{S(t)}\right|^2_{100\% \ket{S} \ @ \ t=0}}} \rangle_{t\to\infty} = \frac{5}{8} \ 
\end{equation}
\noindent if $\omega = 0$. Note that the long-time averaged signal is discontinuous between $\omega = 0$ and $\omega = 0^{+}$.
\\

\noindent For high fields, we indeed find $\lim_{\omega \to \infty}  \langle  \textcolor{InstituteBlue}{\underline{\left|\ket{S(t)}\right|^2_{100\% \ket{S} \ @ \ t=0}}} \rangle_{t\to\infty} = \frac{1}{2}$, reflecting the fact that, in the absence of decoherence, the singlet population continues to oscillate coherently between $\ket{S} \leftrightarrow \ket{T_{0}}$ indefinitely.


\vfill
\newpage
\subsection{Case 5: equiprobable incoherent mixture of $\ket{T_{+}}$, $\ket{T_{0}}$, $\ket{T_{-}}$, and $\ket{S}$ as initial spin state} 
\label{c5}

\noindent A maximally incoherent electron-spin starting condition is an equiprobable incoherent (classical) mixture of the four states $\ket{T_{+}}$, $\ket{T_{0}}$, $\ket{T_{-}}$, and $\ket{S}$; in this case, we have simply $\alpha=\beta=\gamma=\frac{1}{4}$. Using all the expressions derived above, we obtain:
\begin{equation}
\begin{split}
\textcolor{InstituteBlue}{%
\underline{\left|\ket{S(t)}\right|^2_{\substack{
\frac{1}{4}\ket{T_{+}},\;
\frac{1}{4}\ket{T_{0}},\;
\frac{1}{4}\ket{T_{-}},\\
\frac{1}{4}\ket{S}\ @\ t=0
}}}}
={}&\phantom{\ + \ }
\frac{1}{8}
\left|\ket{S(t,\omega)}\right|^2_{100\% \ket{T_{+}} \ @ \ t=0}
+
\frac{1}{8}
\left|\ket{S(t,-\omega)}\right|^2_{100\% \ket{T_{+}} \ @ \ t=0}
\\
&+
\frac{1}{8}\Big(
\left|\ket{S(t,\omega)}\right|^2_{100\% \ket{T_{0}} \ @ \ t=0}
+
\left|\ket{S(t,-\omega)}\right|^2_{100\% \ket{T_{0}} \ @ \ t=0}
\Big)
\\
&+
\frac{1}{8}\Big(
\left|\ket{S(t,\omega)}\right|^2_{100\% \ket{S} \ @ \ t=0}
+
\left|\ket{S(t,-\omega)}\right|^2_{100\% \ket{S} \ @ \ t=0}
\Big)
\\
={}&\phantom{\ + \ }
\frac{1}{4}\ .
\end{split}
\end{equation}

\noindent This is to say, when all electron spin states are initially equipopulated, the singlet population is time-independent: there is no magnetic-field effect and no coherent signal modulation. Hence, quantum sensing in proteins necessarily relies on some degree of spin initialization, at least for this simplified toy model, in the sense that the initial electron spin state cannot be a maximally incoherent mixture of all possible states.

\vfill
\newpage
\subsection{Derivation of the meaning of the real and imaginary parts of the coherence}
\label{cohexpl}

\noindent The relation between coherence and population transfer can be seen explicitly in a generic two-level problem. Consider a Hamiltonian
\begin{equation}
\mathcal{H}=
\begin{pmatrix}
E_1 & J\\
J & E_2
\end{pmatrix} \ ,
\end{equation}
where \(J\) is the off-diagonal coupling. In the considered radical-pair toy model, \(J\) plays the role of the singlet--triplet mixing term in the Hamiltonian, analogous to an external `drive'. In turn, the magnetic field contributes instead through the energy difference \((E_1-E_2)\). Writing the density matrix in the same basis, the evolution equation \(\dot\rho=-i[H,\rho]\) gives
\begin{align}
\dot\rho_{11}=-2J \ \mathrm{Im}(\rho_{12}) \ , \\
\dot\rho_{22}=+2J \ \mathrm{Im}(\rho_{12}) \ ,
\end{align}
so the imaginary part of the coherence is directly proportional to the instantaneous \linebreak population current between the two states. On the other hand, the evolution of the coherences follows
\begin{equation}
\dot\rho_{12}=-i(E_1-E_2)\rho_{12}-iJ(\rho_{22}-\rho_{11}) \ .
\end{equation}

\noindent The evolution equation for \(\rho_{12}\) shows two things: the energy splitting \((E_1-E_2)\) causes the coherence to precess, rotating its real and imaginary parts into one another, while the off-diagonal coupling \(J\) converts population imbalance into coherence. Since the population current is proportional to \(\mathrm{Im}(\rho_{12})\), the real part \(\mathrm{Re}(\rho_{12})\) does not directly represent transfer, but for \((E_1-E_2) \neq 0\) it can be converted into \(\mathrm{Im}(\rho_{12})\) and thereby influence the transfer dynamics.
\\


\vfill
\newpage
\subsection{Particular case in which population transfer is quenched at zero field, but allowed for nonzero fields: not a new pathway}
\label{particular}

\noindent Under particular circumstances, a small nonzero field can produce a nonzero imaginary part in a coherence channel whose imaginary part vanishes exactly at zero field. This, however, does not signify the appearance of a new Hamiltonian coupling. Rather, the zero-field result can arise from a cancellation between contributions that is specific to the initial state preparation. We illustrate this with an explicit example.
\\

\noindent Take the $\ket{S} \leftrightarrow \ket{T_{+}}$ coherence in the first subspace, given by Eq.~(\ref{rhoST+}). Its imaginary part is
\begin{equation}
\Im (\rho_{ST_+}(t))
=
\frac{a}{4\sqrt{2}\,\Omega}\sin\theta
\left[
\Delta \cos\!\left(\frac{(a-\omega)t}{2}\right)
+
(2\Sigma-1-\zeta)\cos\theta
\right] \ .
\end{equation}
In the zero-field limit this becomes
\begin{equation}
\Im (\rho_{ST_+}(t)) 
=
\frac{1}{8\sqrt{2}}\,
(2\Sigma+\Delta-1-\zeta)\sin(at) \ .
\end{equation}
Hence, for specially chosen initial populations satisfying
\begin{equation}
(2\Sigma+\Delta-1-\zeta)=0 \ ,
\end{equation}
the imaginary part vanishes identically at \(\omega=0\), even though the coherence channel itself is allowed by the Hamiltonian. For example, choosing
$\Sigma=\frac{3}{8}$, $\Delta=\frac{1}{4}$, and $\zeta=0$ 
gives
\begin{equation}
\Im (\rho_{ST_+}(t)) = 0 \ .
\end{equation}
For any small nonzero field, however,
\begin{equation}
\Im (\rho_{ST_+}(t))
=
\frac{a}{16\sqrt{2}\,\Omega}\sin\theta
\left[
\cos\!\left(\frac{(a-\omega)t}{2}\right)-\cos\theta
\right] \ ,
\end{equation}
\noindent which is generically nonzero for \(t\neq 0\). 

\noindent Thus, a weak field removes a zero-field cancellation between contributions whose exact balance depends on the initial state preparation. The resulting nonzero imaginary part should therefore not be interpreted as the opening of a new Hamiltonian pathway, but \linebreak as the lifting of a preparation-dependent cancellation already present in the zero-field case.


\vfill
\newpage
\subsection{Our results are consistent with a previously published result, I/III}
\label{wdcon}

\noindent Our present results are consistent with a previous result from the literature~\cite{woodward2023} in which density matrix elements were calculated for a similar toy model that differs from ours just on their assumption that, when any population hits the singlet state, it is instantaneously removed. The initial population considered was an equiprobable mixture of all triplet states, $\alpha = \beta = \gamma = \frac{1}{3}$ (i.e., $\frac{1}{6}$ of the total population in each of the six triplet states, considering the full eight-level dynamics). 
\\

\noindent Without explicitly using this language, what was essentially  calculated was the density matrix for all states that do not participate in the spin-mixing dynamics: the isolated $\ket{T_{\mp}}$ states, in addition to the remaining triplet components that do not couple to the singlet state, $\ket{\chi} \equiv c_0 \ket{T_{0}} + c_{\pm}\ket{T_{\pm}}$ and $\ket{\chi'} \equiv c_0' \ket{T_{0}} + c_{\mp}'\ket{T_{\mp}}$. For the first (respectively, second) subspace, we can use $\mathcal{H}_1$ ($\mathcal{H}_5$) and determine $\ket{\chi}$ ($\ket{\chi'}$) via $\langle S | \mathcal{H}_1 | \chi \rangle = 0$ ($\langle S | \mathcal{H}_5 | \chi' \rangle = 0$):
\begin{align}
    \ket{\chi} = \sqrt{\frac{2}{3}} \ket{T_{0}} + \frac{1}{\sqrt{3}} \ket{T_{+}} \ ; \\
     \ket{\chi'} = \sqrt{\frac{2}{3}} \ket{T_{0}} + \frac{1}{\sqrt{3}} \ket{T_{-}} \ . 
\end{align}
\noindent Note that $\ket{\chi}$ ($\ket{\chi'}$) is the same for $\omega = 0$ and $\omega \neq 0$. However, $\ket{\chi}$ ($\ket{\chi'}$) is only an eigenstate of $\mathcal{H}_1$ ($\mathcal{H}_5$) for $\omega = 0$. The formulas above correspond to a normalized version of the triplet-only zero-field eigenstates calculated in~\cite{woodward2023} ($\ket{\chi}$ is state $\ket{10}$ in the paper; it also corresponds to the state that has $J = \frac{3}{2}$, $m_J = \frac{1}{2}$, and $S = 1$ in the paper's notation; $\ket{\chi'}$ is state $\ket{12}$, with $J = \frac{3}{2}$, $m_J = -\frac{1}{2}$, and $S = 1$).
\\

\noindent The expression for the surviving density matrix of that paper is obtained by:
\begin{equation}
    \rho_{\text{surviving}} = \underbrace{\frac{1}{6}\Big(|T_{+}\rangle\langle T_{+}|\Big) + \frac{1}{6}\Big( |\chi\rangle\langle \chi|   \Big)}_{\text{first subspace}} + \underbrace{\frac{1}{6}\Big(|T_{-}\rangle\langle T_{-}|\Big)  +  \frac{1}{6}\Big( |\chi'\rangle\langle \chi'|   \Big) }_{\text{second subspace}}\ .
\end{equation}

\noindent In particular, for example for the first subspace, the surviving projector component that has no singlet overlap is:
\vfill
\newpage
\begin{align}
    \frac{1}{6} \Big(|\chi\rangle\langle \chi|   \Big) &= \frac{1}{6} \Bigg[  
    \frac{2}{3}\Big(|T_{0}\rangle\langle T_{0}|\Big) 
    + \frac{1}{3}\Big(|T_{+}\rangle\langle T_{+}|\Big)   
    + \frac{\sqrt{2}}{3}\Big(|T_{+}\rangle\langle T_{0}|\Big) 
    + \frac{\sqrt{2}}{3}\Big(|T_{0}\rangle\langle T_{+}|\Big)\Bigg] \nonumber \\
    &= \frac{1}{9}\Big(|T_{0}\rangle\langle T_{0}|\Big)
    +  \frac{1}{18}\Big(|T_{+}\rangle\langle T_{+}|\Big)   
    +  \frac{1}{9\sqrt{2}}\Big(|T_{+}\rangle\langle T_{0}|\Big) + \frac{1}{9\sqrt{2}}\Big(|T_{0}\rangle\langle T_{+}|\Big) \ ;
\end{align}
\noindent the above coefficients are identical to those of the paper's Table I (c). The large-field limit calculated  in Table I (a) is also consistent with our results. 
\\

\noindent In other words, in the model considered, the reaction is assumed to instantaneously remove the singlet-coupled component, leaving only the triplet population orthogonal to the singlet, together with the population in the nonreactive triplet states. By contrast, in the closed model considered here, the singlet population is retained and evolves unitarily. The two approaches are therefore internally consistent, but they address different physical questions: our closed model describes how radical-pair spin states evolve when no population is lost, whereas Woodward’s model describes the radical-pair spin state that remains after all singlet-reactive population has been removed.

\clearpage
\section{\seclq Bright-dark\secrq \ interpretation of the Hamiltonian: extended discussion}



\subsection{The considered Hamiltonian is analogous to those found in the physics of electromagnetically-induced transparency}
\label{subsec:EIT}

\noindent This simple toy model of radical-pair dynamics has a Hamiltonian-level analogue of \linebreak electromagnetically-induced transparency~\cite{Fleischhauer2005EIT}, but with a different observable.
\\

\noindent In a $\Lambda$-type electromagnetically-induced transparency system, the dark state is the \linebreak superposition that has zero coupling to a lossy excited state. Population pumped into that state cannot absorb a probe (usually a laser), so the medium becomes `transparent' to that probe. In other words, typical optical electromagnetically-induced transparency experiments measure how much population avoids the bright channel: the dark state is dark to the driving field, and therefore avoids the lossy excited state. 
\\

\noindent In the radical-pair case, chemical experiments can measure how much population avoids the bright spin-mixing channel --- but the observables are not absorption, but rather the singlet population, or the integrated singlet yield (for the latter, see Section~\ref{sec:MFE}). More dark-state trapping means less participation in the oscillatory bright (or spin-coupled) sector, but both the instantaneous singlet population and the yield also depend on the strength of the bright--dark coherence term.
\vfill
\newpage

\subsection{Explicit derivation of signal contributions from the bright and the dark sectors}
\label{sigcontr}

\noindent Using dark and bright state bases, for example $\{\ket{D}, \ket{B}, \ket{T_{+}}\}$, we will show that the signal \linebreak $\textcolor{InstituteBlue}{\underline{\left|\ket{S(t)}\right|^2_{
  \alpha  \ket{T_{+}},\;
  \beta  \ket{T_{0}},\;
  \gamma  \ket{T_{-}},\;
  (1-\alpha-\beta-\gamma)\ket{S}\ @\ t=0
}}}$ can be decomposed into contributions from both the dark and bright state populations, and from the bright--dark coherence.
\\

\noindent We saw in Section~\ref{sec:analogy}, Eq.~(\ref{sigdb}), that the signal in one subspace is given by  \linebreak
$\left|\ket{S(t)}\right|^2 
= \frac12\left(\left|\ket{B(t)}\right|^2+\left|\ket{D(t)}\right|^2\right)- \text{Re}\left[\rho_{BD}(t)\right]$.
\\

\noindent For the subspace spanned by the basis $\{\ket{D}, \ket{B}, \ket{T_{+}}\}$, the $3\times3$ density matrix is, at $t = 0$:
\begin{align}
\rho(0)
&=
\frac{\alpha}{2} |T_{+} \rangle \langle T_{+}|
+\frac{\beta}{2} |T_{0} \rangle \langle T_{0}|
+\frac{(1-\alpha - \beta -\gamma)}{2}|S \rangle \langle S| \ , 
\end{align}

\noindent whereas, for the subspace spanned by the basis $\{\ket{D'}, \ket{B'}, \ket{T_{-}}\}$, it is:
\begin{align}
\rho(0)
&=
\frac{\gamma}{2}|T_{-} \rangle \langle T_{-}|
+\frac{\beta}{2}|T_{0} \rangle \langle T_{0}|
+\frac{(1-\alpha - \beta -\gamma)}{2}|S \rangle \langle S| \ .
\end{align}

\noindent Explicitly, 
\begin{equation}
\begin{split}
\frac{\beta}{2}\ket{T_{0}}\bra{T_{0}}
+\frac{(1-\alpha-\beta-\gamma)}{2}\ket{S}\bra{S}
={}&\phantom{ \ - \ }
\frac{1-\alpha-\gamma}{4}\Big(
\ket{B}\bra{B}+\ket{D}\bra{D}
\Big)
\\
&-
\frac{(1-\alpha-\beta-\gamma)-\beta}{4}\Big(
\ket{B}\bra{D}+\ket{D}\bra{B}
\Big)
\\
={}&\phantom{ \ - \ }
\frac{\Sigma}{4}\Big(
\ket{B}\bra{B}+\ket{D}\bra{D}
\Big)
-\frac{\Delta}{4}\Big(
\ket{B}\bra{D}+\ket{D}\bra{B}
\Big)\ .
\end{split}
\end{equation}

\noindent At $t = 0$, the populations and coherences then read:
\begin{align}
    \left|\ket{B(0)}\right|^2 &= \frac{1-\alpha-\gamma}{4} = \frac{\Sigma}{4} \ ,  \\
    \left|\ket{D(0)}\right|^2 &= \frac{1-\alpha-\gamma}{4} = \frac{\Sigma}{4} \ , \\
    \rho_{BD}(0)&=-\frac{((1-\alpha -\beta -\gamma) - \beta)}{4} = -\frac{\Delta}{4} \ .
\end{align}

\noindent Now, the dark state population does not vary in time because it is decoupled, so
\vfill
\newpage
\begin{equation}
\left|\ket{D(t)}\right|^2 = \left|\ket{D(0)}\right|^2 = \frac{1-\alpha-\gamma}{4} = \frac{\Sigma}{4} \ . 
\end{equation}

\noindent For the bright state population evolution, we need to use Eqs.~(\ref{17}) and (\ref{18}). If we start in $\ket{B}$, we end up in $\ket{B}$ at time $t$ with a probability
\begin{equation}
\left|\ket{B(t)}\right|^2_{100\% \ket{B} \ @ \ t=0} = 1 - \frac{a^2}{\Omega^2}\sin(\theta)^2 \ , 
\end{equation}

\noindent whereas, if we start in $\ket{T_{+}}$, we end up in $\ket{B}$ at time $t$ with a probability
\begin{equation}
\left|\ket{B(t)}\right|^2_{100\% \ket{T_{+}} \ @ \ t=0}  = \frac{a^2}{\Omega^2}\sin(\theta)^2 \ . 
\end{equation}

\noindent Hence, generally in the first subspace,
\begin{equation}
\left|\ket{B(t)}\right|^2 = \frac{1-\alpha-\gamma}{4} \left( 1 - \frac{a^2}{\Omega^2}\sin(\theta)^2  \right) + \frac{\alpha}{2} \frac{a^2}{\Omega^2}\sin(\theta)^2 = \frac{\Sigma}{4}
+\frac{(1-2\Sigma+\zeta)a^2}{4\Omega^2}\sin(\theta)^2 \ .
\end{equation}

\noindent By symmetry, for the second subspace, we have:
\begin{equation}
\left|\ket{B'(t)}\right|^2 = \frac{1-\alpha-\gamma}{4} \left( 1 - \frac{a^2}{\Omega^2}\sin(\theta)^2  \right) + \frac{\gamma}{2} \frac{a^2}{\Omega^2}\sin(\theta)^2 = \frac{\Sigma}{4}
+\frac{(1-2\Sigma-\zeta)a^2}{4\Omega^2}\sin(\theta)^2 \ .
\end{equation}

\noindent The evolution of the bright state populations for the eight-level system is obtained by averaging the two subspace expressions, because the two nuclear spin manifolds enter the full signal with equal statistical weight of $\frac{1}{2}$:
\begin{equation}
    \frac{1}{2}\left(\left|\ket{B(t)}\right|^2 + \left|\ket{B'(t)}\right|^2\right) = \frac{1-\alpha-\gamma}{4} - \frac{(1-2\alpha-2\gamma)a^2}{4\Omega^2}\sin(\theta)^2 =  \frac{\Sigma}{4}
    - \frac{(2\Sigma-1)a^2}{4\Omega^2}\sin(\theta)^2 \ .
\end{equation}

\noindent Performing similar calculations for the coherences (as previously done in Eq.~(\ref{cohsub})), in the first and second subspaces, we obtain, respectively:
\begin{align}
\rho_{BD}(t)=& -\frac{\Delta}{4}\,
e^{\,i(a-\omega)t/2}
\left(
\cos(\theta)+i\frac{\omega}{\Omega}\sin(\theta)
\right) \nonumber \\
=& -\frac{\Delta}{4}
\bigg[
\phantom{ \ i \ }\left(
\cos(\theta)\cos\!\left(\frac{(a-\omega)t}{2}\right)
-
\frac{\omega}{\Omega}\sin(\theta)\sin\!\left(\frac{(a-\omega)t}{2}\right)
\right) \nonumber \\
&\quad \quad +i\left(
\cos(\theta)\sin\!\left(\frac{(a-\omega)t}{2}\right)
+
\frac{\omega}{\Omega}\sin(\theta)\cos\!\left(\frac{(a-\omega)t}{2}\right)
\right)
\bigg]
      \ , \\
\rho_{B'D'}(t)=& -\frac{\Delta}{4}\,
e^{\,i(a+\omega)t/2}
\left(
\cos(\theta)-i\frac{\omega}{\Omega}\sin(\theta)
\right) \nonumber \\
=& -\frac{\Delta}{4}
\bigg[\phantom{ \ i \ }
\left(
\cos(\theta)\cos\!\left(\frac{(a+\omega)t}{2}\right)
+
\frac{\omega}{\Omega}\sin(\theta)\sin\!\left(\frac{(a+\omega)t}{2}\right)
\right) \nonumber \\
&\quad \quad +i\left(
\cos(\theta)\sin\!\left(\frac{(a+\omega)t}{2}\right)
-
\frac{\omega}{\Omega}\sin(\theta)\cos\!\left(\frac{(a+\omega)t}{2}\right)
\right)
\bigg]
      \ .
\end{align}

\noindent Taking the real parts and adding them gives
\begin{equation}
  -\frac{\Delta}{2}
\cos\!\left(\frac{at}{2}\right)
\left[
\cos(\theta)\,\cos\!\left(\frac{\omega t}{2}\right)
+
\frac{\omega}{\Omega}\sin(\theta)\,\sin\!\left(\frac{\omega t}{2}\right)
\right] \ .
\end{equation}

\noindent Since the singlet projector contains a term in minus the real part of the coherences, the bright--dark coherence contribution is
\begin{equation}
    \frac{\Delta}{2}
\cos\!\left(\frac{at}{2}\right)
\left[
\cos(\theta)\,\cos\!\left(\frac{\omega t}{2}\right)
+
\frac{\omega}{\Omega}\sin(\theta)\,\sin\!\left(\frac{\omega t}{2}\right)
\right] \ .
\end{equation}

\noindent Putting all together, 
\begin{equation}
\begin{split}
\textcolor{InstituteBlue}{%
\underline{\left|\ket{S(t)}\right|^2_{\substack{
\alpha \ket{T_{+}},\;
\beta \ket{T_{0}},\;
\gamma \ket{T_{-}},\\
(1-\alpha-\beta-\gamma)\ket{S}\ @\ t=0
}}}}
={}&
\underbrace{\frac{\Sigma}{4}}_{\text{dark-state population}}
\\
&\quad+
\underbrace{
\left[
\frac{\Sigma}{4}
-\frac{(2\Sigma-1)a^2}{4\Omega^2}\sin(\theta)^2
\right]
}_{\text{bright-state population}}
\\
&\quad+
\underbrace{
\frac{\Delta}{2}
\cos\!\left(\frac{at}{2}\right)
\left[
\cos(\theta) \cos\!\left(\frac{\omega t}{2}\right)
+\frac{\omega}{\Omega}\sin(\theta) \sin\!\left(\frac{\omega t}{2}\right)
\right]
}_{\text{bright--dark coherence}}
\ . 
\end{split}
\tag{18}
\end{equation}

\noindent After combining the first two terms, this becomes exactly Eq.~(\ref{gennew}).

\vfill
\newpage
\subsection{Our results are consistent with a previously published result, II/III}

\noindent The only other paper we have identified that appears to note the decomposition of the radical-pair Hamiltonian into dark and bright sectors is~\cite{Xu2016}.
\\

\noindent In that paper, the conclusion is that the dark-state population, rather than any coherence term, determines the magnetosensitive observable (namely, the singlet yield) because the relevant bright--dark coherence oscillates rapidly and is taken to average to zero on the reaction timescale. In our formulation, meanwhile, the signal contains an explicit bright--dark interference contribution and therefore depends directly on \(\mathrm{Re}[\rho_{BD}(t)]\), see Eq.~(\ref{full}) and Supplementary Subsection~\ref{sigcontr} above. Moreover, in the singlet yields studied in Section~\ref{sec:MFE}, this coherence contribution is not absent from the outset either: it is precisely the term responsible for the low-field structure and, at exactly \(\omega=0\), it generates the stationary contribution that remains in both the finite-time average and the lifetime-weighted yields. 
\\

\noindent The difference is therefore not that the two bright--dark decompositions are incompatible, but that the observables and averaging procedures emphasize different parts of the \linebreak dynamics. In particular, for any \(\omega\neq 0\), our treatment agrees with the work in~\cite{Xu2016} in the strict long-time limit given by $t \to \infty$, where the vanishing bright--dark coherence does not contribute to the asymptotic yield; the distinction arises because, away from that limit, the same coherence remains essential to the zero-field and near-zero-field behavior.
\\
\vfill
\newpage
\section{Solution for the singlet yield: extended discussion}
\label{shape}

\noindent Supplementary Subsections~\ref{osc}, ~\ref{lf}, ~\ref{halffield}, ~\ref{max},  ~\ref{scaling} concern derivations made for the finite-$T$ average yield,  $\langle \textcolor{InstituteBlue}{\underline{\left|\ket{S(t,\omega)}\right|^2}}\rangle_T$. Derivations for the lifetime-weighted yield $\Phi_k(\omega)$ are very similar and appear grouped in Supplementary Subsection~\ref{kmodel}.

\subsection{The singlet yield curves contain oscillatory structure as a function of $\omega$, including for $\omega \to 0$: deriving the possible existence of the \seclq low-field effect\secrq}
\label{osc}

\noindent The normalized integrated signals of Table~\ref{tab:mfe_T}, regarded as functions of $\omega$ at fixed $T$, \linebreak generally display a nontrivial oscillatory structure rather than simple monotonic behavior. Indeed, finite-time integration leaves behind several oscillatory contributions depending nonlinearly on $\omega$: besides the nonoscillatory background (i.e., the terms in $\frac{1}{\Omega^2}$), the resulting expressions still contain oscillatory corrections involving $\sin\left(\Omega T\right)$, as well as the four sinc terms entering through $\mathcal{K}(T)$. 
\\

\noindent \noindent To identify the characteristic oscillation scales, in $\omega$ space, of the singlet yield curves, we inspect the \(\omega\)-dependent oscillating arguments entering the finite-\(T\) average expressions of Table~\ref{tab:mfe_T}. These are:
\begin{equation}
\Phi_0(\omega)\equiv \Omega T \ ,
\qquad
\Phi_{1,2,3,4}(\omega)\equiv\frac{(a\pm\omega\pm\Omega)T}{2} \ .
\label{osccontr}
\end{equation}

\noindent Linearizing locally around a given field value \(\omega'\),
\begin{equation}
\Phi_j(\omega)\approx \Phi_j(\omega')+ \frac{d\Phi_j}{d\omega}\bigg|_{\omega'}(\omega-\omega') \ ,
\end{equation}
we obtain that each individual oscillatory term appearing in the finite-$T$ average expression has a locally sinusoidal dependence on $\omega$
with wavenumber
\begin{equation}
\kappa_j(\omega')=\bigg| \ \frac{d\Phi_j}{d\omega}\bigg|_{\omega'} \ \bigg| \ ,
\end{equation}

\noindent yielding a local oscillation scale of approximately
\vfill
\newpage
\begin{equation}
\Delta\omega_j(\omega)\sim \frac{2\pi}{\kappa_j(\omega)} \ .
\end{equation}
Explicitly,
\begin{equation}
\kappa_0(\omega)=T \frac{|\omega|}{\Omega} \ ,
\qquad
\kappa_{\pm}(\omega)=\frac{T}{2}\left(1\pm\frac{\omega}{\Omega}\right) \ ,
\end{equation}
so that
\begin{equation}
\Delta\omega_0(\omega)\sim \frac{2\pi \ \Omega}{T|\omega|} \ ,
\qquad
\Delta\omega_{\pm}(\omega)\sim \frac{4\pi}{T\left(1\pm\omega/\Omega\right)} \ .
\end{equation}

\noindent Hence the normalized magnetic field effect curves are governed by several distinct field-\linebreak dependent oscillation scales, which in turn explain the appearance of the nonmonotonicity and the local extrema as $\omega$ is varied.
\\

\noindent In particular, for $\omega \to 0$, 
\begin{equation}
\Delta\omega_0(\omega) \to \infty \ ,
\qquad
\Delta\omega_{\pm}(\omega) \to \frac{4\pi}{T} \ .
\end{equation}

\noindent Thus, the $\sin(\Omega T)$ terms vary only slowly with $\omega$ near zero field, whereas the sinc terms set a characteristic low-field oscillation scale proportional to $\frac{1}{T}$. Recall that the sinc terms arise from $\mathcal{K}(T)$, and that in the singlet yield expression $\mathcal{K}(T)$ appears multiplied by a coefficient proportional to the difference between the initial populations in $\ket{S}$ and $\ket{T_0}$, $\Delta$. Therefore, whenever the initial populations of $\ket{S}$ and $\ket{T_0}$ differ, this nontrivial low-field correction is present, and it generically allows for a low-field structure within a window in
$\omega$ of order $\frac{1}{T}$. However, the mere presence of this correction does not by itself guarantee a genuine dip or peak in the singlet yield curve: such a feature requires the full curve to develop a turning point at some low-field $\omega_\ast$, i.e., that the curve's first derivative with respect to $\omega$ vanish at  $\omega_\ast$, see Supplementary Subsection~\ref{lf}. In other words, the population imbalance condition ensures that the nontrivial low-field correction term is present, but a genuine dip or peak additionally requires this correction to be strong enough to cancel and reverse the slope of the remaining terms in the full singlet yield curve.
\\

\noindent The analysis above shows how a nonmonotonic feature can arise near zero field, and what sets its characteristic width: in the finite-$T$ average expressions, the sinc terms, which are nonlinear in $\omega$, introduce a local field scale proportional to $\frac{1}{T}$. Such a nonmonotonic feature in the singlet yield curve is commonly referred to as the `low-field effect'. Whether this feature is actually present for a given initial spin state, and whether it generates a dip or a peak, is determined by the sign of the first derivative with respect to $\omega$ of the full singlet yield curve, as detailed in Supplementary Subsection~\ref{lf}. 

\vfill
\newpage
\subsection{Deriving whether the low-field effect is a peak or a dip}
\label{lf}

\noindent In Supplementary Subsection~\ref{osc}, we derived why such a low-field feature might exist, \linebreak and in the main text we argued under which conditions it might be experimentally resolvable. In turn, here we assess whether we can unambiguously characterize the feature as creating a local dip or a local peak, as a function of the different spin initializations; the answer will depend on whether the singlet yield curve develops a low-field turning point. 
\\

\noindent For general $T$, the resulting equations for this analysis are transcendental and not very enlightening. In turn, the strict long-time limit $T\to\infty$ can be analyzed exactly. In particular, one can derive an explicit criterion for whether the magnetic field effect curve exhibits a zero-field minimum or a zero-field maximum. In this time limit, talking about zero-field minima or maxima is a slight abuse of language, as the magnetic field curve is discontinuous at $\omega = 0$; however, 
we will derive that the extremum will appear at exactly $\omega = 0$.
\\

\noindent For a generic incoherent initial state, $\alpha \ket{T_{+}}, \ \beta  \ket{T_{0}}, \ \gamma  \ket{T_{-}}, \ (1-\alpha-\beta-\gamma) \ket{S}$, we already established that, for $T \to \infty$ and $\omega > 0$, 
\begin{equation}
 F_{>0}(\omega) \equiv \lim_{T\to\infty}
\langle
\textcolor{InstituteBlue}{\underline{\left|\ket{S(t,\omega)}\right|^2}}
\rangle_T   = \frac{(3-2\alpha-2\gamma)a^{2}+4(1-\alpha-\gamma)\omega^{2}}{8\Omega^{2}} = \frac{(1+2\Sigma)a^{2}+4\Sigma\omega^{2}}{8\Omega^{2}} \ ,
\label{tinf}
\end{equation}
\noindent i.e., that we retrieve the results of Table~\ref{Table2}. 
\\

\noindent Differentiating Suppl.~Eq.~\eqref{tinf} gives
\begin{equation}
\frac{dF_{>0}}{d\omega}
=
\frac{(1-2\alpha-2\gamma)a^2\,\omega}{4(a^2+\omega^2)^2}  = \frac{(2\Sigma-1)a^2\,\omega}{4(a^2+\omega^2)^2} \ .
\label{Fpos_derivative}
\end{equation}
\noindent Hence, $F_{>0}(\omega)$ is globally monotone on $]0,\infty]$ (i.e., there is no extremum for $\omega > 0$); in particular:
\begin{align}
(2\Sigma-1) > 0  \ &\Rightarrow \ 
F_{>0}(\omega)\ \text{is strictly increasing on }]0,\infty] \ , \\
(2\Sigma-1) <0  \ &\Rightarrow \ 
F_{>0}(\omega)\ \text{is strictly decreasing on }]0,\infty] \ , \\
(2\Sigma-1) = 0 \ &\Rightarrow \ 
F_{>0}(\omega)\ \text{is constant on }]0,\infty] \ .
\end{align}
\vfill
\newpage
\noindent Therefore, in the strict long-time limit that we can easily solve for, the only possible \linebreak nonmonotonic feature is a singular feature at exactly \(\omega=0\).
\\

\noindent To determine whether such a feature is a maximum or a minimum of the singlet yield curve, one must compare the exact zero-field long-time value with the \(\omega\to 0^+\) limit of the \(\omega>0\) branch. Setting \(\omega=0\) in the exact finite-$T$ average expression first, and then taking \(T\to\infty\), gives
\begin{equation}
F(0)
= \frac{1}{8}+
\frac{(1-\alpha-\beta-\gamma)}{2} = \frac{1}{8}+
\frac{(\Sigma + \Delta)}{4}\ .
\label{F0_exact}
\end{equation}
By contrast, taking the \(\omega\to 0^+\) limit of Suppl.~Eq.~\eqref{tinf} gives
\begin{equation}
F(0^+) \equiv
\lim_{\omega\to 0^+}F_{>0}(\omega)
=
\frac{3-2\alpha-2\gamma}{8} = \frac{(1+2\Sigma)}{8} \ .
\label{F0plus}
\end{equation}
Subtracting Suppl.~Eqs.~\eqref{F0_exact} and \eqref{F0plus}, one finds
\begin{equation}
F(0)-F(0^+)
=
\frac{((1-\alpha-\beta-\gamma) - \beta)}{4} = \frac{\Delta}{4} \ ,
\label{jump_formula}
\end{equation}
\noindent that is, the discontinuity is proportional to the difference in the initial populations between states $\ket{S}$ and $\ket{T_{0}}$.
\\

\noindent The discontinuity $F(0)-F(0+)$ in Suppl.~Eq.~(\ref{jump_formula}) therefore arises because a stationary bright--dark interference term survives only at exactly $\omega=0$; for any $\omega>0$, however small, that term oscillates and is averaged away. In this sense, we have proposed that the low-field feature should be understood as the phase-locked limit of the coherence term, rather than simply `more mixing at low field'.
\\

\noindent In the strict long-time limit for this toy model, 
\begin{equation}
\Delta > 0 \ \  \text{(more $\ket{S}$ than $\ket{T_{0}}$ initially)} \  \Rightarrow \ 
F(0) > F(0^+) 
\end{equation}
and the curve has a zero-field maximum, whereas
\begin{equation}
\Delta < 0 \ \  \text{(more $\ket{T_{0}}$ than $\ket{S}$ initially)} \ \Rightarrow \
F(0)<F(0^+) 
\end{equation}
and the curve has a zero-field minimum. Finally,
\begin{equation}
\Delta = 0 \ \  \text{(equal initial $\ket{S}$ and $\ket{T_{0}}$ population)} \ \Rightarrow \ 
F(0)=F(0^+)
\end{equation}
and there is no singular zero-field anomaly.
\\

\noindent It is crucial to realize that a local zero-field maximum implies that the magnetic field effect curve decreases for $\omega = 0^{+}$, creating a dip for $\omega > 0$; and that a local zero-field minimum implies that the magnetic field effect curve increases for $\omega = 0^{+}$, creating a peak for $\omega > 0$. 
\\

\noindent We can summarize the results for the different initial spin polarizations in Suppl.~Table~\ref{Table30}. The key message is that, in this toy model, and in the strict $T \to \infty$ regime that we can easily analyze without having recourse to numerics, the following is true: when the initial $\ket{S}$ population exceeds (is less than) that of $\ket{T_{0}}$, a dip (peak) in the magnetic field curve will occur for low fields $\omega > 0$. 
\\

\noindent Finally, it is important to stress that this argument is only valid in the strict long-time limit $T \to \infty$. It proves the existence of a zero-field maximum or minimum relative to the \(\omega>0\) branch, but does not by itself imply the existence of other finite-field extrema for finite $T$. Rather, for large but finite $T$, the singular zero-field anomaly is smoothed into a narrow low-field feature whose width shrinks as $T$ increases (on a scale $\propto \frac{1}{T}$, see Supplementary Subsection~\ref{osc}), and any other extrema that might exist for $\omega > 0$ are not available in closed form in general and must typically be determined numerically.

\vfill
\newpage

\begin{table}[H]
    \centering
\begin{tabular}{|c|c|}
\hline
initial state &  $T\to\infty$ limit of low-field feature at $\omega = 0$  \\ \hline
100\% $\ket{T_{+}}$                    &  $\nexists$  \\ \hline
100\% $\ket{T_{0}}$                    &  is at minimum $\Rightarrow$ peak for   $\omega > 0$ \\ \hline
$\frac{1}{3} \ket{T_{+}}$, \ $\frac{1}{3} \ket{T_{0}}$, \ $\frac{1}{3} \ket{T_{-}}$              &  is at minimum $\Rightarrow$ peak for   $\omega > 0$ \\ \hline 
100\% $\ket{S}$  & is at maximum $\Rightarrow$ dip for   $\omega > 0$   \\ \hline
$\frac{1}{4} \ket{T_{+}}, \ \frac{1}{4} \ket{T_{0}}, \ \frac{1}{4} \ket{T_{-}}, \ \frac{1}{4} \ket{S}$ &  $\nexists$ \\ \hline
$\alpha \ket{T_{+}}, \ \beta  \ket{T_{0}}, \ \gamma  \ket{T_{-}}, \ (1-\alpha-\beta-\gamma) \ket{S}$ &   depends on sign of $\Delta$      \\ \hline
\end{tabular}
   \captionsetup[table]{aboveskip=4pt,belowskip=8pt}
  \captionof{table}{\justifying \textbf{Character of low-field feature at $\omega = 0$, in the $T \to \infty$ limit, determining if a dip or peak occurs for $\omega > 0$, expressed for different initial spin initializations.} All results are self-consistent. In this $T \to \infty$ limit that is solvable without numerics, determining whether the low-field feature is a dip or a peak depends on the imbalance of the initial populations between states $\ket{S}$ and $\ket{T_{0}}$: more $\ket{S}$ ($\Delta > 0$) means there is a maximum at $\omega = 0$, and hence a dip in the singlet yield curve for $\omega > 0$; and more $\ket{T_{0}}$ ($\Delta < 0$) means there is a minimum at $\omega = 0$, and hence a peak in the singlet yield curve for $\omega > 0$.}
    \label{Table30}
\end{table}

\vfill
\newpage
\subsection{Deriving the `half-field parameter'}
\label{halffield}

\noindent Using the results of the previous subsection, we can estimate the parameter commonly referred to in the literature as $B_{\frac{1}{2}}$. This quantity, with corresponding Larmor frequency $\omega_{\frac{1}{2}}$, is defined as the characteristic `half-field' parameter governing the magnetic field dependence of the singlet yield. It typically encodes information about spin relaxation, radical recombination kinetics, electron--electron couplings, and hyperfine interactions. In the toy model considered here, however, $B_{\frac{1}{2}}$ and $\omega_{\frac{1}{2}}$ depend only on the hyperfine interaction. We now derive $\omega_{\frac{1}{2}}$.
\\

\noindent A natural definition of the half-saturation field for the singlet yield curve is the field
at which the $\omega>0$ branch has reached half of its total change between its
$\omega = 0$ value and its high-field plateau:
\begin{equation}
F_{>0}(\omega_{\frac{1}{2}})
\equiv
\frac{F(0)+F(\infty)}{2} \ .
\end{equation}

\noindent For finite $T$, the equation above is generally transcendental; however, in the strict long-time average, a simple expression for $\omega_{\frac{1}{2}}$ can be obtained. We need, however, to modify the above definition slightly to
\begin{equation}
F_{>0}(\omega_{\frac{1}{2}})
\equiv
\frac{F(0^+)+F(\infty)}{2} 
\end{equation}
\noindent because, for $t \to \infty$, $\omega = 0$ is a singularity and thus $F(0)$ is a singular point, not the endpoint of the smooth finite-field expression for the yield.
\\

\noindent In the strict long-time average, we saw previously that
\begin{equation}
 F_{>0}(\omega) \equiv \lim_{T\to\infty}
\langle
\textcolor{InstituteBlue}{\underline{\left|\ket{S(t,\omega)}\right|^2}}
\rangle_T   = \frac{(1+2\Sigma)a^{2}+4\Sigma\omega^{2}}{8\Omega^{2}} \ ,
\tag{SI 79}
\end{equation}
\noindent with
\begin{align}
F(0^+)=&\frac{1+2\Sigma}{8} \ , \tag{SI 85}\\
F(\infty)=&\frac{\Sigma}{2} \ .
\end{align}
\noindent Therefore,
\vfill
\newpage
\begin{align}
\frac{(1+2\Sigma)a^2+4\Sigma\omega_{\frac{1}{2}}^2}{8(a^2+\omega_{\frac{1}{2}}^2)}
&=
\frac{1}{2}
\left(
\frac{1+2\Sigma}{8}
+
\frac{\Sigma}{2}
\right) \nonumber \\
&=
\frac{1+6\Sigma}{16} \ .
\end{align}

\noindent Multiplying by $16(a^2+\omega_{\frac{1}{2}}^2)$ gives
\begin{equation}
2(1+2\Sigma)a^2+8\Sigma\omega_{\frac{1}{2}}^2
=
(1+6\Sigma)(a^2+\omega_{\frac{1}{2}}^2) \ .
\end{equation}
\noindent Rearranging the equation we obtain
\begin{equation}
(1-2\Sigma)\omega_{\frac{1}{2}}^2
=
(1-2\Sigma)a^2 \ .
\end{equation}
\noindent Hence, provided $\Sigma\neq \tfrac12$,
\begin{equation}
\omega_{\frac{1}{2}}=a \ . 
\end{equation}

\noindent Thus, within this toy model, the half-saturation field is exactly set by the hyperfine scale $a$, which is consistent with the literature~\cite{Wong2023MagneticFieldCryptochromes}.
\\



\noindent When \(\Sigma=\tfrac12\), the nonzero-field branch becomes
\begin{equation}
F_{>0}(\omega)= \frac{a^2 + \omega^2}{4\Omega^2}= \frac14 \ ,
\end{equation}
so the yield is completely flat for all nonzero fields. In other words, there is no meaningful \(B_{\frac{1}{2}}\) to define from the \(\omega>0\) branch. 

\newpage

\subsection{Deriving the initial spin states that maximize the visibility of the low-field feature}
\label{max}

\noindent We derive two metrics showing that the maximum visibility of the low-field feature occurs when the initial spin population is pure in $\ket{S}$ or in $\ket{T_{0}}$.
\\

\noindent The first metric is related to the jump between $\omega = 0$ and $\omega = 0^{+}$, which we calculated in Suppl.~Eq.~(\ref{jump_formula}); its maximum absolute value is $\frac{1}{4}$, which is achieved either for $\beta = 1$ (100\% $\ket{T_{0}}$) or for $\alpha = \beta = \gamma = 0$ (100\% $\ket{S}$).
\\

\noindent The second metric is related to the jump between the zero-field value and the high-field plateau value, which can be calculated to be:
\begin{equation}
F(\infty)-F(0) =
\frac{\beta}{2} -\frac{1}{8} = \frac{(2\Sigma - 2\Delta - 1)}{8} \ ;
\label{jump_formula2}
\end{equation}
\noindent Its maximum value is for $\beta = 1$ (100\% $\ket{T_{0}}$), $F(\infty)-F(0) = \frac{3}{8}$. Conversely, its largest magnitude when $F(\infty)-F(0) < 0$ is $\frac{1}{8}$, which is obtained whenever \(\beta=0\), i.e., for any spin initialization with no population in \(\ket{T_0}\). Thus, under the second metric alone, the strongest downward zero-to-high-field contrast is not unique: any state with \(\beta=0\) maximizes it. However, if one also wants this downward contrast to
correspond to a genuine low-field dip rather than merely a plateau offset, then the natural optimal choice is an initial pure $\ket{S}$ state since it also maximizes the first metric. 
\\

\noindent More generally, any population placed in \(\ket{T_+}\) or \(\ket{T_-}\) reduces the amount of population available in the \(\{\ket{S},\ket{T_0}\}\) sector and therefore weakens the genuine low-field structure. 

\vfill
\newpage
\subsection{Deriving a field scale for the extrema of the singlet yield curve}
\label{scaling}

\noindent A natural question is whether the global minimum or maximum of the singlet yield curve occurs at a field strength set by the hyperfine coupling $a$, as heuristically claimed in the radical-pair literature. In the present toy model, without numerics, we cannot demonstrate whether an extremum occurs at $\omega \sim a$; however, we can demonstrate that the field location of any nonzero extremum scales linearly with $a$.
\\

\noindent Defining the dimensionless variables:
\begin{equation}
x \equiv \frac{\omega}{a} \ , \qquad \tau \equiv aT \ ,
\end{equation}
we note that all singlet yield curves in Table~\ref{tab:mfe_T} can be written in the form
\begin{equation}
\langle \textcolor{InstituteBlue}{\underline{\left|\ket{S(t,\omega)}\right|^2}} \rangle_T
=
f(x,\tau) \ ,
\end{equation}
for some dimensionless function $f$. Therefore,
\begin{equation}
\frac{d}{d\omega}
\langle \textcolor{InstituteBlue}{\underline{\left|\ket{S(t,\omega)}\right|^2}} \rangle_T
=
\frac{1}{a}\,\frac{\partial f}{\partial x} \ ,
\end{equation}
and any interior extremum $x_\ast$  satisfies
\begin{equation}
\frac{\partial f}{\partial x}\bigg|_{(x_\ast,\tau)}=0 \ .
\end{equation}
Hence, the field value of any extremum can always be written as
\begin{equation}
\omega_\ast = a\,x_\ast(\tau)=a\,x_\ast(aT) \ \sim \ a \ .
\end{equation}
Thus, in this toy model, the hyperfine coupling does set the overall field scale of the extrema, although their precise locations depend on the dimensionless parameter $aT$ and must generally be calculated numerically.
\vfill
\newpage

\subsection{Results for the exponentially-weighted singlet yield are similar}
\label{kmodel}

\noindent Using
\begin{equation}
k\int_0^\infty e^{-kt}\sin^2\!\left(\frac{\Omega t}{2}\right)dt
=
\frac{\Omega^2}{2(k^2+\Omega^2)} \ ,
\label{eq:weighted_sin2}
\end{equation}
together with
\begin{equation}
k\int_0^\infty e^{-kt}\cos\!\left(\frac{\lambda t}{2}\right)dt
=
\Lambda_k(\lambda) \ ,
\qquad \text{with} \ 
\Lambda_k(\lambda)\equiv \frac{4k^2}{4k^2+\lambda^2} \ ,
\label{eq:weighted_cos}
\end{equation}
and the decomposition
\begin{equation}
\begin{split}
&\phantom{ \ + \ }\cos\!\left(\frac{at}{2}\right)
\left[
\cos\!\left(\frac{\Omega t}{2}\right)\cos\!\left(\frac{\omega t}{2}\right)
+
\frac{\omega}{\Omega}\sin\!\left(\frac{\Omega t}{2}\right)\sin\!\left(\frac{\omega t}{2}\right)
\right]
\\
={}&\phantom{ \ + \ }
\frac14\left(1+\frac{\omega}{\Omega}\right)
\left[
\cos\!\left(\frac{(a+\omega-\Omega)t}{2}\right)
+
\cos\!\left(\frac{(a-\omega+\Omega)t}{2}\right)
\right]
\\
&+
\frac14\left(1-\frac{\omega}{\Omega}\right)
\left[
\cos\!\left(\frac{(a-\omega-\Omega)t}{2}\right)
+
\cos\!\left(\frac{(a+\omega+\Omega)t}{2}\right)
\right]\ ,
\end{split}
\label{eq:weighted_trig_decomp}
\end{equation}
\noindent one derives the entries of Table~\ref{tab:mfe_k}.
\\

\noindent The role played in Supplementary Subsection~\ref{osc} by the sinc-based boxcar term $\mathcal{K}(T)$ is here played by the
Lorentzian-like function $K(\omega)$. For $k\ll a$ and $|\omega|\ll a$, two of the four
arguments entering Eq.~\eqref{eq:Kk_def} are close to zero, while the other two remain of
order $a$, so that
\begin{equation}
K(\omega)
=
\frac12\,\frac{4k^2}{4k^2+\omega^2}
+
O\!\left(\frac{k^2}{a^2},\frac{\omega}{|a|}\right) \ .
\label{eq:Kk_lowfield}
\end{equation}
Thus, for the usual exponentially-weighted singlet yield, the low-field structure is broadened
over a field scale of order $\Delta\omega \sim k$. Compare this with the finite-$T$ average case, for which the corresponding field scale is of order $\frac{1}{T}$; as it will be explicitly derived below, we will recover the same $\omega >0$ branch and the same jump as for that case --- what differs is how that singular jump gets smoothed at finite observation scale (i.e., the broadening mechanism is different).
\\

\noindent Other features are also the same as in the previous model for the singlet yield curve. Indeed, for any fixed $\omega\neq 0$,
\vfill
\newpage
\begin{equation}
\lim_{k\to 0}\Lambda_k(\lambda)=0 \ ,
\end{equation}
and therefore
\begin{equation}
\lim_{k\to 0}K(\omega)=0 \ .
\label{eq:Kk_nonzero_field_limit}
\end{equation}
\noindent At exactly zero field, however,
\begin{align}
K(0)
=
\frac12\left(1+\frac{k^2}{k^2+a^2}\right) \ , 
\label{eq:Kk_zero_field}
\end{align}
\noindent which reduces to $\frac{1}{2}$ for $k \to 0$. Hence, just as for the finite-$T$ average, the limits $k\to 0$ and $\omega\to 0$ do not commute.
\\

\noindent For fixed $\omega>0$, Suppl.~Eq.~\eqref{eq:Kk_nonzero_field_limit} immediately gives
\begin{equation}
F_{>0}(\omega)
\equiv
\lim_{k\to 0}\Phi_k(\omega)
=
\frac{(3-2\alpha-2\gamma)a^2+4(1-\alpha-\gamma)\omega^2}{8(a^2+\omega^2)} = \frac{(1+2\Sigma)a^2+4\Sigma\omega^2}{8(a^2+\omega^2)} \ ;
\label{eq:Fgt0_weighted}
\end{equation}
This is exactly the same $\omega>0$ branch as in Supplementary Section~\ref{lf}. Differentiating,
\begin{equation}
\frac{dF_{>0}}{d\omega}
=
\frac{(1-2\alpha-2\gamma)a^2\omega}{4(a^2+\omega^2)^2} = \frac{(2\Sigma-1)a^2\omega}{4(a^2+\omega^2)^2}\ .
\label{eq:Fgt0_weighted_derivative}
\end{equation}
Hence $F_{>0}(\omega)$ is globally monotone on $]0,\infty]$, exactly as in the previous case.
\\

\noindent At $\omega=0$, the weighted yield is
\begin{equation}
\Phi_k(0)
=
\frac{(5-4\alpha-4\beta-4\gamma)a^2+8(1-\alpha-\beta-\gamma)k^2}{8(a^2+k^2)} = \frac{(1+2\Sigma+2\Delta)a^2+4(\Sigma+\Delta)k^2}{8(a^2+k^2)} \ ,
\label{eq:Phi0_finitek}
\end{equation}
so that
\begin{equation}
F(0)
\equiv
\lim_{k\to 0}\Phi_k(0)
=
\frac{1}{8} + \frac{(1-\alpha-\beta-\gamma)}{2} = \frac{(1+2\Sigma+2\Delta)}{8}\ .
\label{eq:F0_weighted}
\end{equation}
In addition,
\begin{equation}
F(0^+)
\equiv
\lim_{\omega\to 0^+}F_{>0}(\omega)
=
\frac{3-2\alpha-2\gamma}{8} = \frac{1+2\Sigma}{8} \ .
\label{eq:F0plus_weighted}
\end{equation}
Subtracting,
\begin{equation}
F(0)-F(0^+)
=
\frac{((1-\alpha-\beta-\gamma)-\beta)}{4} = \frac{\Delta}{4} \ ,
\label{eq:weighted_anomaly_difference}
\end{equation}
which is again proportional to the difference between the initial populations in $\ket{S}$ and $\ket{T_{0}}$.
\\

\noindent Therefore, in the strict $k\to 0$ limit, the usual exponentially-weighted singlet yield leads to
exactly the same criteria as the finite-$T$ average:
\begin{align}
\Delta > 0 \ \  \text{(more $\ket{S}$ than $\ket{T_{0}}$ initially)} \  &\Rightarrow \ 
F(0) > F(0^+) 
 \ , 
\\
\Delta < 0 \ \  \text{(more $\ket{T_{0}}$ than $\ket{S}$ initially)} \ &\Rightarrow \
F(0)<F(0^+) 
 \ , 
\\
\Delta = 0 \ \  \text{(equal initial $\ket{S}$ and $\ket{T_{0}}$ population)} \ &\Rightarrow \ 
F(0)=F(0^+)
 \ , 
\end{align}

\noindent which yield precisely the same classification as before, respectively: zero-field maximum, hence dip for $\omega>0$; zero-field minimum, hence peak for $\omega>0$; and no singular zero-field anomaly.
\\

\noindent The exponentially-weighted yield also obeys the same scaling logic for the extrema as that found in Supplementary Section~\ref{scaling}.
\\

\noindent Introducing the dimensionless variables:
\begin{equation}
x\equiv \frac{\omega}{a} \ ,
\qquad
\kappa\equiv \frac{k}{a} \ ,
\label{eq:xkappa_def}
\end{equation}
all entries of Table~\ref{tab:mfe_k} can be written as
\begin{equation}
\Phi_k(\omega)=g(x,\kappa)
\label{eq:g_x_kappa}
\end{equation}
for some dimensionless function $g$. Therefore,
\begin{equation}
\frac{d}{d\omega}\Phi_k(\omega)
=
\frac{1}{a}\,\frac{\partial g}{\partial x} \ ,
\label{eq:g_derivative_scaling}
\end{equation}
and any interior extremum $x_*$ satisfies
\begin{equation}
\left.\frac{\partial g}{\partial x}\right|_{(x_*,\kappa)}=0 \ .
\end{equation}
Hence the field location of any nonzero extremum can be written as
\begin{equation}
\omega_*=a\,x_*(\kappa)=a\,x_*\!\left(\frac{k}{a}\right) \ ,
\label{eq:weighted_extremum_scaling}
\end{equation}
so the hyperfine coupling again sets the overall field scale, while the precise location depends
on the dimensionless ratio $k/a$. In particular, in the singular limit $k\to 0$, this reduces to the same $a$-scaling statement
as for the long-time $T \to \infty$ average.
\\

\noindent In summary, the finite-$k$ exponentially-weighted singlet yield is not identical to the finite-$T$ average, because the sinc structure is replaced by a Lorentzian broadening. However,
in the strict limits $k\to 0$ and $T\to\infty$, both observables yield the same $\omega>0$ branch, the same singular zero-field anomaly, the same dip/peak criterion, and the same
field-scaling law for extrema.

\vfill
\newpage

\subsection{Chemistry-based interpretations of the low-field effect}
\label{lfhist}

\noindent Previous works have often described the low-field effect in terms of symmetry breaking: lifting the zero-field degeneracy changes which singlet--triplet transfer processes contribute most strongly to the dynamics. In this picture, the low-field effect reflects a balance between two competing consequences of an external magnetic field: on the one hand, lifting the exact triplet degeneracy destroys the special zero-field interference structure and singles out a leading field-sensitive \(\ket{S}\leftrightarrow\ket{T_{0}}\) population transfer term, in addition to terms that were already present at zero field; on the other hand, Zeeman splitting suppresses mixing at larger fields by separating \(\ket{T_{+}}\) and \(\ket{T_{-}}\) in energy from \(\ket{S}\) and \(\ket{T_{0}}\). The peak or dip in the singlet-yield curve then corresponds to the field strength at which these competing effects are optimized. In the chemistry literature, this phenomenon has often been described as one in which low fields `open' additional singlet--triplet transitions that are absent at zero field.
\\


\noindent Brocklehurst and McLauchlan~\cite{brocklehurst1996freeradical} argued that, at zero field, only transitions that conserve total angular momentum are allowed. When a magnetic field is applied, however, only the component of the angular momentum along the field direction is conserved, allowing transitions that would be forbidden at zero field. For two electrons with a single nuclear spin-$\frac{1}{2}$ coupled to one electron, the relevant quantum numbers in this framework are \(J\) and \(m_J\), where: \(J = S_e + I_n\), with \(S_e\) the total electron spin (0 for singlet and 1 for triplet) and \(I_n\) the nuclear spin; and \(m_J\) the projection of \(J\) onto the quantization axis. Triplet states can have \(J = \frac{3}{2}\) or \(\frac{1}{2}\), whereas singlet states can only have \(J = \frac{1}{2}\). In the absence of an external field, because both \(J\) and \(m_J\) are conserved, transitions from the \(J = \frac{3}{2}\) component of the triplet manifold to the singlet state (\(J = \frac{1}{2}\)) are forbidden. Once a field is applied, however, this symmetry is broken and only \(m_J\) is conserved, thereby enabling some mixing between \(J = \frac{3}{2}\) triplet states and singlet states.
\\

\noindent Timmel et al.~\cite{timmel1998weakfields} expanded on this work and showed, through an eigenbasis decomposition of the density matrix, that an external field breaks the degeneracy of some eigenstates, making coherences between them time-dependent and thereby enhancing singlet--triplet mixing at low fields. Using this approach, a closed-form expression for the singlet yield of singlet-born radical pairs was derived, and the size of the low-field effect was shown to depend on the nuclear spins coupled to each electron. We show in Supplementary Subsection~\ref{timres} that our present results are consistent with those in~\cite{timmel1998weakfields}.
\\

\noindent There has been discussion in the literature about which states undergo the strongest \linebreak alteration in population transfer when a field is applied. Brocklehurst and McLauchlan argued, based on a vectorial picture of the spin dynamics, that this symmetry breaking enhances \(\ket{S} \leftrightarrow \ket{T_{\pm}}\) transitions. However, Lewis et al.~\cite{lewis2018lowfield} showed that the transitions that `open up' at low field are actually \(\ket{S} \leftrightarrow \ket{T_{0}}\) transitions, not \(\ket{S} \leftrightarrow \ket{T_{\pm}}\) transitions. Their approach was based on calculations of the spin dynamics in the Hamiltonian eigenbasis and on the projections of these eigenstates onto the singlet and triplet subspaces. Woodward~\cite{woodward2023} takes a different approach, showing that at zero field one can identify a triplet-only eigenstate that cannot mix with the singlet; in the presence of an applied field this state ceases to be an eigenstate, resulting in enhanced \(\ket{S} \leftrightarrow \ket{T_{0}}\) mixing. 
Woodward focused on triplet-born states and calculated product yields for the case in which singlets decay rapidly into products, while triplets remain coherent indefinitely. We showed in Supplementary Subsection~\ref{wdcon} that our present results are consistent with those in~\cite{woodward2023}. We agree with the results of~\cite{lewis2018lowfield} and of~\cite{woodward2023} in that, at nonzero low fields, population transfer between \(\ket{S} \leftrightarrow \ket{T_{0}}\), and not between $\ket{S} \leftrightarrow \ket{T_{\pm}}$, is enhanced: there is a completely new term $\propto \frac{\omega}{\Omega} \sim \omega + O(\omega^3)$ in $\text{Im}(\rho_{ST_0}(t))$ that is absent at exactly $\omega = 0$, see Eq.~(\ref{25}).
\\

\vfill
\newpage
\subsection{Our results are consistent with a previously published result, III/III}
\label{timres}


\noindent The fourth line of Table~\ref{tab:mfe_k}, describing the exponentially weighted singlet yield for an initially pure $\ket{S}$ state, is in fact algebraically identical to Eq.~(12) in a previously published work~\cite{timmel1998weakfields}, up to notation. 
\\

\noindent In particular, our kernel $\Lambda_k(\lambda)=\frac{4k^2}{4k^2+\lambda^2}$ is related to their $f(x)=\frac{k^2}{k^2+x^2}$ by $\Lambda_k(\lambda)=f(\lambda/2)$, so the two expressions coincide term by term after this change of variables. 
\\

\noindent Accordingly, our result also reproduces the limits quoted there: $\Phi_k(\omega)=\frac{5}{8}$ for $\omega=0$ with $a \gg k$, and $\Phi_k(\omega)=\frac{3}{8}$ for $a \gg \omega \gg k$.

\vfill
\newpage

\section{Solution for good magnetometric states: extended discussion}

\subsection{Suitable magnetometry observables must reflect biological accumulation of chemical products}
\label{mod}

\noindent The instantaneous signal \textcolor{InstituteBlue}{\underline{$\left|\ket{S(t, \omega)}\right|^2$}} provides a very clean, idealized metrological figure of merit. Its advantages are: that it corresponds to a well-defined time-resolved readout; that
it has a straightforward binomial noise model; and that one may optimize over the interrogation
time $t$ to exploit coherent transient slopes. Its disadvantage is that it assumes access to a
well-defined readout time, which is unrealistic for an asynchronized or continuously
replenished ensemble of radical pairs in a biological context.
\\

\noindent In contrast, both singlet yields considered here, 
$\langle$\textcolor{InstituteBlue}{\underline{$\left|\ket{S(t, \omega)}\right|^2$}}$\rangle_T$ and $\Phi_{k}(\omega)$, 
are more \linebreak coarse-grained observables. Their main advantage is robustness: they are less sensitive to timing jitter and do not require a sharply defined readout instant. Their disadvantages are: that the averaging generally suppresses oscillatory slope and therefore tends to reduce sensitivity; 
and that the simple Bernoulli shot-noise formula of Eq.~\eqref{eq:eta_inst_bias} requires an
additional modeling assumption, namely that these yields can be treated as an effective single-run probability. We implicitly made that assumption in the main text.
\\

\noindent Finally, we hypothesize that, since biological magnetosensing likely depends on the robust accumulation of chemical products, either singlet yield, $\langle$\textcolor{InstituteBlue}{\underline{$\left|\ket{S(t, \omega)}\right|^2$}}$\rangle_T$ or $\Phi_{k}(\omega)$, is the best suited magnetometry signal for our analysis. 
\vfill
\newpage

\subsection{The bright-dark picture provides magnetometry insights}
\label{maginsight}

\noindent The bright--dark state decomposition provides a useful way to interpret which initial states are favorable for magnetometry in this toy model. A good magnetometric state is not simply one that is as dark as possible. Although a dark component is useful because it preserves phase memory by avoiding immediate participation in the oscillatory bright-sector dynamics, a purely dark state is generally a poor sensor because it does not efficiently convert field-dependent phase accumulation into a measurable singlet signal. What is needed instead is a state with both dark and bright character: the dark component stores the relative phase, while the bright component provides the readout channel. In other words, we know that the signal is not controlled solely by populations: the field sensitivity comes from how the field changes the relative phase and interference between bright and dark pieces; from this perspective, interestingly, the relevant resource for magnetometry is not dark-state population alone, but field-sensitive bright--dark coherence. This is very close in spirit to Ramsey interferometry: one component stores phase, another component enables readout, and sensitivity comes from the phase-dependent interference between them. The dark--bright basis is therefore not just a mathematical convenience: it identifies the actual interferometric degree of freedom of the sensor. Three key insights follow.
\\

\noindent Firstly, the dark--bright interpretation helps clarify the proper operating regime for sensing under this toy model. The slope in the denominator of Eq.~(\ref{eta}) is taken at the $\omega \to 0$ limit. In dark--bright language, zero field is special because the coherence becomes phase-locked in a way that creates the singular low-field feature, but not a usable linear-response slope for magnetometry: the relevant singlet observables being even in $\omega$, the response slope vanishes at $\omega=0$. Therefore, practical sensing requires a finite bias field $\omega_{\rm E}$, about which the bright--dark relative phase changes linearly with $\omega$, see Eq.~(\ref{eq:D_inst_expand}).
\\

\noindent Secondly, the bright--dark picture clarifies why, as will be shown in Supplementary \linebreak Section~\ref{numden}, the initial states which maximize the denominator of the sensitivity (Eqs.~(\ref{etaT}) and~(\ref{etak})) are concentrated near the $\{|S\rangle,|T_0\rangle\}$ edge of the state space. In the bright--dark basis, both $|S\rangle$ and $|T_0\rangle$ are equal superpositions of bright and dark, differing only in the sign of the bright--dark coherence. In turn, population in $|T_\pm\rangle$ mainly acts as spectator weight: it does not help build the useful interferometric channel, and at finite field it is increasingly trapped by the Zeeman splitting. This is consistent with the fact that the slope of the magnetometry signal is maximal when the initial population in $|T_\pm\rangle$ is minimized, so that the dynamics remain concentrated in the active $\{|S\rangle,|T_0\rangle\}$ manifold. The choice between $|S\rangle$ and $|T_0\rangle$ is therefore not universal, but depends on the ratio $\frac{\omega_{\rm E}}{a}$ and on the interrogation time, since these two states have the same bright--dark populations but opposite coherence sign. Maximizing only the denominator does not guarantee a minimum for the sensitivity; however, these qualitative arguments suggest that optimal states for magnetometry are not pure dark states, but rather states that maximize controllable bright--dark interference while minimizing inactive $|T_\pm\rangle$ population.
\\

\noindent Thirdly, the interrogation time $T$ or lifetime $k^{-1}$ must be long enough for the bright and dark components to accumulate an appreciable relative phase, but not so long that temporal averaging suppresses the oscillatory, coherence-sensitive part of the slope, 
or that the effective duration of each run imposes too severe a metrological penalty $\eta \propto \sqrt{t}$. The onset of useful Earth-field sensitivity is for experimental interrogation times that are on the order of, or larger than, $\sim \frac{1}{\omega_{\rm E}}$ --- this, at its core, is a reflection both of Nyquist–Shannon's sampling theorem, and of the fact that, in order to sense a field of magnitude $|\hat{B}|$, an electron spin must be able to `precess enough about the field' before reaction or decoherence limits the coherent evolution. In the dark--bright picture, this scaling reflects the fact that the dark and bright pieces need enough time to accumulate an $O(1)$ relative phase. Too short a lifetime or interrogation time means no appreciable interferometric phase; in turn, too much averaging washes the coherence-sensitive part of the slope out: the oscillatory, interference-sensitive pieces are suppressed by the time average itself. For example, the bright--dark coherence term in the finite-$T$ average singlet yield signal goes as $\frac{\sin(\cdot)}{T}$, so its contribution decays as $\frac{1}{T}$. 
\\

\noindent Hence, the dark--bright picture suggests concrete magnetometry design principles, which nature may plausibly have exploited: operate around a bias field for which the bright--dark relative phase responds approximately linearly; possibly (or sometimes) choose states on the $\{|S\rangle,|T_0\rangle\}$ edge; and use interrogation windows or lifetimes long enough to accumulate relative phase, but not so long that temporal averaging suppresses the slope and the time cost per shot becomes too large.
\vfill
\newpage

\subsection{Analysis of the sensitivity's numerator and denominator}
\label{numden}

\noindent In the main text, we study the sensitivity for the finite-$T$ average singlet yield, $ \langle \textcolor{InstituteBlue}{\underline{\left|\ket{S(t,\omega)}\right|^2}}\rangle_T$.
\\

\noindent Before anything, note that, for the bounded signal $ \langle \textcolor{InstituteBlue}{\underline{\left|\ket{S(t,\omega)}\right|^2}}\rangle_T \leq 1$, the shot-noise prefactor is itself bounded: 
\begin{equation}
    \sqrt{\langle \textcolor{InstituteBlue}{\underline{\left|\ket{S(t, \omega_{\rm E})}\right|^2}}\rangle_T\bigl(1-\langle \textcolor{InstituteBlue}{\underline{\left|\ket{S(t,  \omega_{\rm E})}\right|^2}}\rangle_T\bigr)} \leq \frac{1}{2} \ .
\end{equation}

\noindent This bound is useful, but it does not imply that the numerator can be neglected. 
\\

\noindent Indeed, the numerator of the sensitivity also depends only on \(\Sigma\) and \(\Delta\), because the yield is affine in those two variables. Writing the signal generically as \(\mathcal S(\Sigma,\Delta)=C+D\Sigma+E\Delta\), one finds
\begin{equation}
    \mathcal S(1-\mathcal S)=\frac14-\left(\mathcal S-\frac12\right)^2 \ ,
\end{equation}
so the numerator is minimized whenever the yield is driven as far from \(\frac{1}{2}\) as possible, i.e., as close to \(0\) or \(1\) as possible. Since \(\mathcal S\) is linear in \((\Sigma,\Delta)\), the quantity \(\mathcal S(1-\mathcal S)\) is a concave quadratic on the allowed \((\Sigma,\Delta)\) triangle, and its minimum is therefore attained at one of the vertices \((1,1)\) (pure initial \(\ket{S}\)), \((1,-1)\) (pure initial \(\ket{T_{0}}\)), or \((0,0)\) (any initial mixture of \(\ket{T_{\pm}}\) adding up to the full population weight). Other states such as \(50\%\ |T_0\rangle, 50\%\ |S\rangle\) cannot minimize the numerator except in accidental degenerate cases. This explains why the bounded shot-noise numerator can still materially reorder the full sensitivity, even when the denominator-based criteria that will be presented below suggest clear trends.
\\

\noindent Let us analyze when the denominator of the sensitivity is at a maximum. 
Examining \linebreak $\left|(2\Sigma-1)A+\Delta B\right|$, we find that there are two qualitative ways of getting a bad (large) sensitivity for fixed $\Sigma$ and $\Delta$. If both \(A\) and \(B\) are small, then all states perform badly. This is what happens, for example, at very short times: there has not been enough time to accumulate a field-dependent phase. If the two slope channels cancel, even if \(A\) and \(B\) are individually large, the combination $\left|(2\Sigma-1)A+\Delta B\right|$ can become small because of destructive \linebreak cancellation (or interference). Hence, $A$ and $B$ should not be optimized individually; rather, their sum should be optimized.
\\

\noindent If \(\Sigma\) and \(\Delta\) are allowed to vary, we conclude that initial population within the \(\ket{T_\pm}\) manifold is usually detrimental to the denominator: decreasing \(\Sigma\) increases the total weight outside the \(\{\ket{S},\ket{T_0}\}\) sector and simultaneously reduces the allowed range of \(\Delta\) (since \(|\Delta|\le \Sigma\), which shrinks the maximum accessible coherence imbalance), thus weakening both slope contributions.
\\

\noindent For a given \(\Sigma\), the allowed range for \(\Delta\) is \(-\Sigma \le \Delta \le +\Sigma\); 
to maximize the slope, \(\Delta\) should be as large as possible in magnitude, i.e., it should be valued at one of the endpoints \(\Delta=\pm\Sigma\). These correspond to the two extreme choices within the coherence-active \(\{\ket{S},\ket{T_0}\}\) manifold: \(\Delta=+\Sigma\) (respectively, \(\Delta=-\Sigma\)) corresponds to all population starting in \(\ket{S}\) (\(\ket{T_0}\)). Hence, at fixed initial loading $\Sigma$ within the \(\{\ket{S},\ket{T_0}\}\) sector, splitting weight between \(\ket{S}\) and \(\ket{T_0}\) is usually not optimal; one of the two endpoints is.
\\

\noindent For the two endpoints, we can rewrite the sensitivity denominator as:
\begin{equation}
\left|(2\Sigma-1)A+\Delta B\right|
=
\left|(2\Sigma-1)A\pm \Sigma B\right| \ ;
\end{equation}

\noindent for improved magnetometry, thus, at fixed $\Sigma$, the choice between a $|S\rangle$-born radical pair and a $|T_0\rangle$-born radical pair is controlled by the full competition between the population-sector contribution $A$ and the coherence contribution $B$. In particular, the preferred endpoint depends on the sign and relative magnitude of $(2\Sigma-1)A$ and $\Sigma B$, so there is no universal statement that $|S\rangle$ is always optimal or that $|T_0\rangle$ is always optimal. The optimum can therefore flip with $T$ or with $\omega_{\rm E}$, because both $A$ and $B$ depend on them. This is also consistent with the fact that, although $|S\rangle$ and $|T_0\rangle$ are closely related by the Hamiltonian dynamics, the singlet readout is not invariant under the swap $|S\rangle\leftrightarrow|T_0\rangle$. Under such a swap, the coherence contribution changes sign, so both the signal and its slope generally change. 
More explicitly, at fixed $\Sigma$, pure $\ket{S}$ gives the larger denominator when
$((2\Sigma-1)A)(\Sigma B)>0$, and pure $\ket{T_0}$ gives the larger denominator when $((2\Sigma-1)A)(\Sigma B)<0$. If $A=0$ or $B=0$, the magnitude of the slope is the same for an initial pure $\ket{S}$ or pure $\ket{T_0}$ state. If 
$B\sim 0$, however, the coherence channel becomes ineffective, and the distinction between $|S\rangle$ and $|T_0\rangle$ correspondingly weakens.
\\

\noindent Note that, if one seeks to improve the sensitivity solely by maximizing the denominator, an optimum always lies on the boundary $\Sigma=1$, that is, with no $\ket{T_\pm}$ initial population at all. In other words, the slope denominator is maximized by putting all the initial population in the $\{\ket{S},\ket{T_0}\}$ manifold. The proof is simple:
\begin{equation}
   \max_{|\Delta|\le \Sigma}
\left|(2\Sigma-1)A+\Delta B\right|
=
|2\Sigma-1|\,|A|+\Sigma|B| \ , 
\end{equation}
\noindent which is maximized at \(\Sigma=1\). Other maxima may exist in special cases like \(B = 0\), where other \(\Sigma\) values can tie with \(\Sigma = 1\).
\\


\vfill
\newpage
\subsection{Derivation of sensitivity for the exponentially-weighted singlet yield as magnetometry signal}
\label{mag2}

\noindent The conclusions here are essentially the same as for the other singlet yield considered in the main text.
\\

\noindent  We take the yield $\Phi_k(\omega)$ for the generic initial spin state $\alpha \ket{T_{+}}, \ \beta  \ket{T_{0}}, \ \gamma  \ket{T_{-}},$ \linebreak  $(1-\alpha-\beta-\gamma) \ket{S}$ (last line of Table~\ref{tab:mfe_k}),
\begin{equation}
    \Phi_k(\omega) =
\frac{\Sigma}{2}
-\frac{(2\Sigma-1)a^2}{8(k^2+\Omega^2)}
+\frac{\Delta}{2}K(\omega)  \ ,
\end{equation}

\noindent and differentiate it: 
\begin{equation}
    \frac{\partial \Phi_k(\omega)}{\partial \omega}
=
\frac{(2\Sigma-1)a^2\omega}{4(k^2+\Omega^2)^2}
+\frac{\Delta}{2}K'(\omega) \ .
\end{equation}

\noindent At the operating point \(\omega=\omega_{\rm E}\),
\begin{equation}
    \Phi_k(\omega_{\rm E}) =
\frac{\Sigma}{2}
-\frac{(2\Sigma-1)a^2}{8(k^2+\Omega_E^2)}
+\frac{\Delta}{2}K(\omega_E) \ ,
\end{equation}
\noindent and
\begin{equation}
     \frac{\partial \Phi_k(\omega)}{\partial \omega}\Bigg|_{\omega=\omega_{\rm E}} \ = \  (2\Sigma-1)A'+\Delta B' \ ,
\end{equation}

\noindent where
\begin{align}
    A' &\equiv \frac{a^2\omega_{\rm E}}{4\bigl(k^2+\Omega_{\rm E}^2\bigr)^2} \ , \\
B' &\equiv \frac12 K'(\omega_{\rm E}) \ .
\end{align}

\noindent Using  \(t=k^{-1}\) as the effective single-run time, the lifetime-weighted sensitivity is
\begin{equation}
    \eta_{\rm E}(k) = \frac{\sqrt{\Phi_k(\omega_{\rm E})\bigl(1-\Phi_k(\omega_{\rm E})\bigr)}}{\left|(2\Sigma-1)A'+\Delta B'\right|}\,\sqrt{k^{-1}} \ .
\end{equation}

\noindent The analysis of the numerator (which is bound), the denominator $\left|(2\Sigma-1)A'+\Delta B'\right|$, and their implications for magnetometry, is identical to that of Supplementary Section~\ref{numden}.
\\

\noindent In particular, the large-\(k\) limit is bad for a clear reason: as \(k\to\infty\), the recombination is so fast that the system is effectively read-out before any appreciable spin evolution takes place. This can be seen from 
\begin{equation}
    \lim_{k \to \infty}\Phi_k(\omega) =
\frac{\Sigma+\Delta}{2} \ , 
\end{equation}
which is exactly the initial singlet population. Hence, in the very fast-recombination limit, the yield just reports the initial singlet content. This is equivalent to saying that the first derivative of the yield goes to zero if $k \to \infty$, and thus the sensitivity diverges.
\\

\noindent The small-$k$ limit is also bad. For fixed nonzero \(\omega_{\rm E}\), when \(k\to 0\), the weighting becomes very broad in time, the coherence term \(K(\omega_{\rm E})\) tends to zero, and the slope contribution from the population sector $A'$ approaches a finite limit. Thus, the signal approaches the long-time branch of the yield, which is fine (and finite!); however, the sensitivity still contains the factor $\sqrt{k^{-1}}$, which blows up as \(k\to0\). In other words, at small \(k\), each run takes too long, so even though the field slope may saturate, the sensitivity worsens.
\\

\noindent In summary, lifetime-weighted magnetometry must also work within an optimum $k$ range. This is the lifetime-broadened analog of choosing an optimal interrogation time; like $T$ previously, the timescale $k^{-1}$ sets the tradeoff between phase accumulation and readout speed.
\vfill
\newpage

\end{document}